\documentclass{aa}  
\usepackage{graphicx}
\usepackage{txfonts}
\usepackage{tabularx}
\usepackage{appendix}
\usepackage{threeparttable}  
\usepackage{dcolumn}
\usepackage{amsmath}
\usepackage{longtable}
\usepackage{float}
\usepackage{hyperref}
\usepackage{multirow}
\usepackage[usenames,dvipsnames]{xcolor}
\usepackage{footmisc}
\usepackage{booktabs} 
\usepackage[export]{adjustbox} 

\newcommand{\degs}{\mbox{$^{\rm o}$}}

\newcommand{\hi}{H{\sc\,i}}
\newcommand{\hii}{H{\sc\,ii}}
\newcommand{\uchii}{UC\,H{\sc\,ii}}
\newcommand{\hchii}{HC\,H{\sc\,ii}}

\newcommand{\choh}{$\rm CH_{3}OH$}

\newcolumntype{d}[1]{D{.}{\cdot}{#1}}

\newcolumntype{.}{D{.}{.}{-1}}

\begin{document} 

\title{A global view on star formation: The GLOSTAR\\ Galactic plane survey. IX.\\
Radio Source Catalog III: $2\degr<\ell<28\degr$, $36\degr<\ell< 40\degr$, $56\degr < \ell < 60\degr$ and $|b|< 1\degr$, VLA B-configuration} 
\author{A.\,Y.\,Yang 
        \inst{1,2,3}\thanks{E-mail: yangay@nao.cas.cn or ayyang@mpifr-bonn.mpg.de},
        S. A. Dzib \inst{3,4}, 
        J. S. Urquhart\inst{5}, 
        A. Brunthaler \inst{3}, 
        S.-N. X. Medina\inst{3,6}, 
        K. M. Menten \inst{3}, 
        F. Wyrowski \inst{3}, 
        G. N. Ortiz-Le\'{o}n\inst{7,3},
        W. D. Cotton\inst{8},
        Y. Gong \inst{3},
        R. Dokara \inst{3},
        M. R. Rugel\inst{3,9,10}, 
        H. Beuther\inst{11},
        J. D. Pandian\inst{12},
        T. Csengeri\inst{13},
        V. S. Veena\inst{3,14}, 
        N. Roy\inst{15},
        H. Nguyen\inst{3},
        B. Winkel\inst{3}, 
        J. Ott\inst{10},
        C. Carrasco-Gonzalez\inst{16},
        S. Khan\inst{3},
        A. Cheema\inst{3} }

\institute{National Astronomical Observatories, Chinese Academy of Sciences, A20 Datun Road, Chaoyang District, Beijing, \\ 100101, P. R. China
\and Key Laboratory of Radio Astronomy and Technology, Chinese Academy of Sciences, A20 Datun Road, Chaoyang District, \\ Beijing, 100101, P. R. China 
 \and 
 Max-Planck-Institut f\"ur Radioastronomie (MPIfR), Auf dem H\"ugel 69, 53121 Bonn, Germany
 \and
IRAM, 300 rue de la piscine, 38406 Saint Martin d'H\`eres, France
\and
Centre for Astrophysics and Planetary Science, University of Kent, Canterbury, CT2\,7NH, UK
\and
 German Aerospace Center, Scientific Information, 51147 Cologne, Germany
\and 
Instituto Nacional de Astrof\'isica, \'Optica y Electr\'onica, Apartado Postal 51 y 216, 72000 Puebla, Mexico
\and 
National Radio Astronomy Observatory,  520 Edgemont Road, Charlottesville, VA 22903, USA
\and Center for Astrophysics | Harvard $\&$ Smithsonian, 60 Garden St., Cambridge, MA 02138, USA
\and National Radio Astronomy Observatory, P.O. Box O, 1003 Lopezville Rd, Socorro, NM 87801, USA
\and Max Planck Institute for Astronomy, Koenigstuhl 17, 69117 Heidelberg, Germany 
\and Department of Earth $\&$ Space Sciences, Indian Institute of Space Science and Technology, Trivandrum 695547, India
\and Laboratoire d'astrophysique de Bordeaux, Univ. Bordeaux, CNRS, B18N, all\'ee Geoffroy Saint-Hilaire, 33615 Pessac, France 
\and I. Physikalisches Institut, Universit\"at zu Köln, Z\"ulpicher Str. 77, 50937 K\"oln, Germany
\and Department of Physics, Indian Institute of Science, Bangalore 560012, India
\and Instituto de Radioastronom\'{i}a y Astrof\'{i}sica (IRyA), Universidad Nacional Aut\'{o}noma de M\'{e}xico  Morelia, 58089, M\'{e}xico 
}

  \abstract
   { As part of the { GLObal view of STAR formation in the Milky Way (GLOSTAR) survey}, we present the high-resolution continuum source catalog for the regions ($\ell = 2\degr-28\degr, 36\degr-40\degr, 56\degr-60\degr$, $\&\,|b|<1.0\degr$), observed with the \emph{Karl G. Jansky Very Large Array (VLA)} in its B-configuration. 
   The continuum images { were} optimized to detect compact sources on angular scales up to 4\arcsec, and have a typical noise level of $\rm 1\sigma \sim 0.08\,mJy\,beam^{-1}$ for an angular resolution of 1\arcsec, which makes GLOSTAR currently the highest resolution as well as the most sensitive radio survey of the northern Galactic plane at 4--8\,GHz. We extracted 13354 sources above a threshold of $5\sigma$ and 5437 sources above $7\sigma$ that represent the { high-reliability} catalog. We determined the in-band spectral index ($\alpha$) for the sources in the 7$\sigma$-threshold catalog. The mean value is $\alpha=-0.6$, which indicates that the catalog is dominated by sources emitting non-thermal radio emission. We identified the most common source types detected in radio surveys:  251 \hii\ region candidates (113 new), 282 planetary nebulae (PNe) candidates (127 new), 784 radio star candidates (581 new), and 4080 extragalactic radio source candidates (2175 new). A significant fraction of \hii\ regions and PNe candidates have $\alpha<-0.1$ indicating that these candidates could contain radio jets, winds or outflows from high-mass and low-mass stellar objects. 
   We identified 245 variable radio sources by comparing the flux densities of compact sources from the GLOSTAR survey and the Co-Ordinated Radio ``N'' Infrared Survey for High-mass star formation (CORNISH), and find that most of them are infrared quiet. The catalog is typically 95\% complete for point sources at a flux density of $\rm 0.6\,mJy$ ( i.e., a typical 7$\sigma$ level) and the systematic positional uncertainty is $\lesssim 0$\farcs1. The GLOSTAR data and catalogs are available online at \emph{https://glostar.mpifr-bonn.mpg.de}. }

  \keywords{catalogs – surveys – radio continuum: general – stars: formation – (ISM:) \hii\ regions - techniques: interferometric }
\titlerunning{The GLOSTAR survey: Continuum Source Catalog}
\authorrunning{A.\,Y.\,Yang et al. }
   \maketitle
%

\section{Introduction}

The  GLObal view of STAR formation in the Milky Way (GLOSTAR)\,\footnote{\label{glostar_web} \url{https://glostar.mpifr-bonn.mpg.de/}}) survey covers 145$\rm\,deg^{2}$ of the northern Galactic plane, namely the region $-2\degs <\ell< 60\degs$ $\&$ $|b|< 1\degs$ and the Cygnus X region ($76\degs < \ell < 83\degs$ $\&$ $-1\degs< b < 2\degs$), where $\ell$ and $b$ are Galactic longitude and latitude, respectively. 
It used the Karl G. Jansky Very Large Array (VLA) in its B and D-configurations as well as the Effelsberg 100-m telescope at C band (4$-$8\,GHz) to observe full polarization continuum emission,  the 4.8 GHz line of formaldehyde (H$_2$CO), the 6.7 GHz maser line of methanol (CH$_3$OH), and several radio recombination lines (RRLs).
Full details of the GLOSTAR survey have been described in an overview paper \citep{Brunthaler2021AA651A85B} that, in particular 
discusses the results for a ``pilot region''($28\degr < \ell < 36\degr$). 
Radio emission of young stellar objects in the Galactic centre region ($-2\degs < \ell < 2\degs\, \&\,|b| < 1.0\degr$) have been discussed in \citet{Nguyen2021AA651A88N}.
\citet{Chakraborty2020MNRAS4922236C} have studied the differential sources count for the pilot region.
The detections and properties of the 6.7 GHz \choh\ maser sources found in  the Cygnus X region have been reported by \citet{Ortiz-Leon2021AA651A87O}, and by \citet{Nguyen2022AA666A59N} for the region $-2\degs < \ell < 60\degs$\,$\&\,|b|< 1.0\degr$. 
\citet{Dokara2021AA651A86D,Dokara2023AA671A145D} have presented the population and properties of Galactic supernova remnants (SNRs) detected in the GLOSTAR survey. 
The 4.8\,GHz H$_2$CO absorption in the Cygnus X region using both the Effelsberg 100-m telescope and VLA are presented by \citet{Gong2023AA678A130G}. 

\begin{table*}
\centering
\caption{\large Summary of the GLOSTAR papers}
\begin{tabular}{p{3.2cm}|p{3.5cm}|p{1.2cm}|p{1.5cm}|p{3.7cm}|p{2.2cm}}
\hline
\hline
  \multirow{2}{*}{The GLOSTAR papers}  &   \multirow{2}{*}{Sensitivity\,($\rm 1\sigma$)}  & \multirow{2}{*}{Beam} &    \multirow{2}{*}{Telescope }& \multirow{2}{*}{Sky coverage}  & 
 \multirow{2}{*}{Context}    \\
   &       &    &    &   \\  
\hline
\multirow{4}{*}{\citet{Brunthaler2021AA651A85B}} &  $\rm \sim 11\,mJy\,@\,4\,km\,s^{-1}$ &  $\sim 1\arcsec$ & VLA-B  & \multirow{2}{*}{$-2\degs<\,\ell<60\degs$, $|b|<1\degs$} & \multirow{4}{*}{Overview paper} \\
   &  $\rm \sim 0.08\,mJy$ & $\sim 18\arcsec$ & VLA-D  &      &   \\
   \cline{2-4}
   & $\rm \sim 80mK\,@\,5km\,s^{-1}$ & \multirow{2}{*}{$\sim 180\arcsec$} & \multirow{2}{*}{Effelsberg} & Cygnus X: $76\degs <\ell<83\degs$   &   \\
      & $\rm \sim 2\,mJy\,beam^{-1}$   &   &    &   \hspace{1.6cm}$-1\degs<b<2\degs$   &   \\

\hline
\multirow{2}{*}{\citet{Medina2019AA627A175M}$^\star$}    &  \multirow{2}{*}{$\rm 0.06-0.15\,mJy\,beam^{-1}$} &  \multirow{2}{*}{ $\sim$18\arcsec}   & \multirow{2}{*}{VLA-D } & $28\degs<\ell<36\degs$  & Continuum    \\
   &  & & & $|b|<1$\degs  & Source Catalog  \\
   \hline
\citet{Dokara2021AA651A86D}   & $\rm 0.06-0.15\,mJy\,beam^{-1}$ & $\sim$18\arcsec & VLA-D & $-2\degs<\ell<60\degs$  & Supernova    \\
\cline{2-4}
\citet{Dokara2023AA671A145D} &  $\rm \sim 2\,mJy\,beam^{-1}$  & $\sim$180\arcsec & Effelsberg &  $|b|<1\degs$  &  Remnants   \\
\hline
\multirow{3}{*}{\citet{Ortiz-Leon2021AA651A87O}} 
   &  $\rm 0.05-0.43\,mJy\,beam^{-1}$   & \multirow{2}{*}{$\sim 1.5\arcsec$}  & \multirow{2}{*}{VLA-B} & \multirow{3}{*}{Cygnus X} & \multirow{2}{*}{Continuum}   \\
    & $\rm 0.1-2.6\,mJy\,beam^{-1}$  & \multirow{2}{*}{$\sim 15\arcsec$} & \multirow{2}{*}{VLA-D} &   & \multirow{2}{*}{\choh\,maser }    \\
     &  $\rm 0.028\,Jy\,@\,0.18km\,s^{-1}$  & &   &  &     \\

\hline
\multirow{2}{*}{\citet{Nguyen2021AA651A88N}}    &  \multirow{2}{*}{$\rm 0.07-1.0\,mJy\,beam^{-1}$ } & \multirow{2}{*}{$\sim$18\arcsec} & \multirow{2}{*}{VLA-D} & $-2\degs <\ell<2\degs$  & \multirow{2}{*}{YSOs in CMZ}   \\
 &    & &  & $|b|<1\degs$  &     \\
\hline
 \multirow{2}{*}{\citet{Nguyen2022AA666A59N}} &  \multirow{2}{*}{$\rm \sim18\,mJy\,@\,0.18\,km\,s^{-1}$ }   & \multirow{2}{*}{$\sim$18\arcsec} & \multirow{2}{*}{VLA-D  } & $-2\degs \leq \ell \leq 60\degs$ & \multirow{2}{*}{ \choh\,maser } \\
   &    & &  & $|b|<1$\degs     &   \\
   \hline
 \multirow{2}{*}{\citet{Dzib2023AA670A9D}}    &   \multirow{2}{*}{$\rm \sim 0.06\,mJy\,beam^{-1}$ } & \multirow{2}{*}{$\sim$1\arcsec}  & \multirow{2}{*}{VLA-B }   & $28\degs< \ell<36\degs$  & Continuum \\
  &   &  & & $|b|< 1\degs$  &  Source catalog   \\
\hline
  \citet{Gong2023AA678A130G}      &   $\rm 0.02\,Jy\,@\,0.5\,km\,s^{-1}$ & $\sim$25\arcsec & VLA-D & \multirow{2}{*}{Cygnus X}  & \multirow{2}{*}{$\rm H_{2}CO$}  \\
 \cline{2-4}
    & $\rm 0.1\,K\,@\,0.5\,km\,s^{-1}$ & $\sim$145\arcsec & Effelsberg &  &     \\
\hline
 \multirow{3}{*}{This work}   
&   \multirow{3}{*}{$\rm 0.05-0.13\,mJy\,beam^{-1}$} & \multirow{3}{*}{$\sim$1\arcsec} & \multirow{3}{*}{VLA-B }   & $2\degs<\ell<28\degs$ &   \multirow{2}{*}{Continuum}  \\
  &   &  & &  $36\degs< \ell< 40\degs$ &  \multirow{2}{*}{Source Catalog}   \\
  &   &  & &  $56\degr< \ell< 60\degs$, $|b|< 1$\degs &     \\
\hline
 \multicolumn{6}{c}{GLOSTAR website: https://glostar.mpifr-bonn.mpg.de/}     \\
 \hline
 \multicolumn{6}{c}{Notes: The paper with $\star$ includes early introduction to the survey.}     \\
\hline
\hline
\end{tabular}
\label{tab:summary_papers}
\end{table*} 
Since the GLOSTAR survey utilizes data from the VLA in two different configurations with quite different angular resolutions, two sub-catalogs have been created -- a B-configuration catalog at high angular resolution ($\sim 1\arcsec$) and a D-configuration catalog at low angular resolution ($\sim 18\arcsec$). 
The sub-catalogs for the Galactic center ($-2\degs < \ell < 2\degs$) and the Cygnus X region in the B-configuration and the D-configuration catalog from $2\degs < \ell < 28\degs$, $36\degs < \ell < 60\degs$, and $|b|< 1.0\degr$ are in preparation (for example, Medina et al.; Ortiz-Leon et al., in prep.), while the catalogs of the pilot region ($28\degs < \ell < 36\degs$) have already been published for both the D- and B-configuration data \citep{Medina2019AA627A175M, Dzib2023AA670A9D}. 
Table\,\ref{tab:summary_papers} summarizes the observations, sky coverage, and context of the papers of the GLOSTAR survey. 
The region $40\degs < \ell < 56\degs$ has not been observed in the B-configuration due to limited observing time allotted for the survey. 
In this paper, we construct and discuss the catalog extracted from the GLOSTAR B-configuration images for the remaining region covering $2\degs< \ell < 28\degs$, $36\degs< \ell < 40\degs$, $56\degs < \ell < 60\degs$, and $|b|< 1.0\degr$. 

This paper is organized as follows: 
Section\,\ref{sect:obs_cal_img} describes the details of the  observations, data reduction, and the noise map of the survey. Section\,\ref{sect:source_cat_construction} presents and describes the extraction of the sources and their properties, 
reliability, completeness level, the in-band spectral index determination, source classification, and clustered/fragmented sources. 
In Section\,\ref{sect:results}, 
we present the catalog and discuss the properties and Galactic distribution of the sources in our high-reliability catalog (these are sources above a 7$\sigma$ detection threshold).    
In Section\,\ref{sect:discussion} we compare the properties of our GLOSTAR catalog with other radio continuum surveys such as CORNISH\,\citep[Co-Ordinated Radio `N' Infrared Survey for High-mass star formation, ][]{Hoare2012PASP,Purcell2013ApJS2051P}, MAGPIS\,\citep[The Multi-Array Galactic Plane Imaging Survey, ][]{White2005AJ,Helfand2006AJ}, and THOR\citep[The \hi\, OH, Recombination line survey of the Milky Way, ][]{Bihr2015AA580A112B,Bihr2016AA588A97B,Beuther2016AA595A32B,Wang2018AA619A124W}. We also discuss the properties of \hii\ region candidates, planetary nebula candidates, extragalactic source candidates, and variable sources detected in the GLOSTAR survey. 
We present a summary of this work and highlight our conclusions in Section\,\ref{sect:conclusion}.



\section{Observations and data reduction}
\label{sect:obs_cal_img}

\subsection{Observations}
\label{observations}

As a part of the GLOSTAR survey, the full Stokes continuum observations were conducted with the VLA in B-configuration and covered two portions the  C-band (4--8\,GHz), namely 4.2--5.2\,GHz and 6.4--7.4\,GHz to avoid strong radio frequency interference (RFI) around 4.1\, GHz and 6.3\,GHz, with 16 spectral windows and 64 channels, each channel having a bandwidth of 2\,MHz. The synthesized beam in B configuration at C-band is $\sim1\farcs0$ and the FWHM primary beam size is $\sim 6\farcm5$ \citep{Brunthaler2021AA651A85B}. 
With a typical integration time of $\sim$15 seconds
per pointing, the total observation time per a $2\degr\times1\degr$ region is $\sim \rm 5\,h$ and the root mean square (rms) noise in the images is expected to be $\rm \sim 0.08\,mJy\,beam^{-1}$. 
The phase calibrators for this work, listed in Table~\ref{tab:obs_info}, were observed every 5--10 minutes to correct the amplitude and phase of the interferometer data for atmospheric and instrumental effects.
The absolute flux density scale was calibrated by comparing the observations of the standard flux calibrators J1331+305 (3C286) and J0137+3309 (3C48) with their models provided by the NRAO \citep{Perley2017ApJS2307P}. 
The observation and instrument parameters of the B-configuration data presented in this work are summarized in Table\,\ref{tab:obs_info}.

\setlength{\tabcolsep}{2pt}
\begin{table}
\centering 
\caption{Summary of VLA continuum observations in B-configuration for the region 
($2\degr < \ell < 28\degr,\,36\degr < \ell < 40\degr, \&\,56\degr < \ell < 60\degr$) of the GLOSTAR survey.
Details of the B-configuration observations of the pilot region  ($28\degr < \ell < 36\degr$) are presented in \citet{Dzib2023AA670A9D}. 
Full observation details of the GLOSTAR survey are described in \citet{Brunthaler2021AA651A85B}. }
\label{tab_obsparms}
\begin{tabular}{lc}\hline \hline
Parameter              &  \\ 
\hline
VLA Proposal ID      & 14A-420/15B-175/16A-174 \\
Frequency (GHz)      & C-band (4-8\,GHz)   \\ 
Array configuration  &   BnA (for $\ell<10\degr$) and B  \\  
Observing mode       & Continuum \\ 
Spectral window      &  16 \\
No. Channels         &  64 \\ 
Bandwidth per channel      & 2 MHz \\ 
Primary beam         & $\sim$ 6$\farcm$5 \\ 
Synthesized beam     & $\sim$ {$1\farcs0$}   \\
Observing dates      & 2014 Apr. $-$ 2016 Oct. \\ 
Integrated time per pointing & $\sim$ 15\,seconds \\ 
Typical sensitivity      & $\rm \sim 0.08\,mJy\,beam^{-1}$  \\
No. pointings  &  $\sim$676 per $2\degr\times1\degr$ \\  
Total observing time & $\sim$5\,h\, per $2\degr\times1\degr$     \\  
Flux density calibrator & 3C286 $\&$ 3C48 \\ 
Phase calibrators & J1820--2528 \,[ $2\degr<\ell\leq10\degr$]\\
& J1811--2055 [$10\degr<\ell\leq12\degr$]\\
& J1825--0737 [$12\degr<\ell\leq28\degr$]\\
& J1907+0127 [$36\degr<\ell\leq46\degr$]\\
& J1925+2106 [$56\degr<\ell\leq58\degr$]\\
& J1931+2243 [$58\degr<\ell\leq60\degr$]\\
\hline
\end{tabular}
\label{tab:obs_info}
\end{table}
\subsection{Data reduction, calibration and imaging}
\label{sect:calibration_and_image}
The data calibration and imaging pipelines were  performed in a semi-automatic manner using the OBIT\,\footnote{\url{https://www.cv.nrao.edu/~bcotton/Obit.html}} package \citep{Cotton2008PASP439C} with scripts written in \emph{python}, which made use of the ObitTalk interface to access  tasks from the Astronomical Image Processing Software package (AIPS)\footnote{\url{https://www.aips.nrao.edu/}} tasks \citep{Greisen2003ASSL109G}. 
The full details of the data reduction pipelines for the whole survey are described in the GLOSTAR overview paper \citet{Brunthaler2021AA651A85B}. 
The continuum data calibration and imaging are presented in \citet{Medina2019AA627A175M} for D-configuration data and in \citet{Dzib2023AA670A9D} for B-configuration data. 
As a summary with additional remarks for the above references, we illustrate the pipeline logic of the calibration and imaging processes in the Appendix Fig. \,\ref{fig:calibration_imaging_pipeline}, which includes the OBIT tasks used for each step.
For completeness, the process is briefly described below.

Raw data collected with the VLA were calibrated and edited following 12 major steps as shown in the left-panel of Fig\,\ref{fig:calibration_imaging_pipeline}. 
The downloaded data in archival science data model (ASDM) format were converted to AIPS format using OBIT task \emph{BDFIn}. 
The initial flagging table consists of online flags (i.e., bad data collected during the observing time) and new flags from shadowed antennas, outliers in the time and frequency domains, as well as the time domain RMS filtering of calibrator data, using OBIT tasks \emph{UVFlag}, \emph{MednFlag}, and \emph{AutoFlag}.   
The following tasks were used to correct the parallactic angle (OBIT/\emph{CLCor}), amplitude variations in system temperature, antenna delays relative to the reference antenna (OBIT/\emph{Calib}\,$\&$\,\emph{CLCal}), amplitude and phase in channels (OBIT/\emph{BPass}\,$\&$\,\emph{BPCal}), amplitude and phase in time (OBIT/\emph{Calib},\,\emph{SNSmo}\,$\&$\,\emph{GetJy}). 
The various calibration steps were each followed by an editing step looking for deviant solutions and flagging 
the corresponding data. After the first calibration pass, the corrected flags were kept and the calibration redone applying the previous flags.
Diagnostic plots of various stages were made for each IF (spectral window as defined by \citealt{Cotton2008PASP439C}), polarization, antenna, 
and baseline, including amplitude/phase/delay vs time for all sources and amplitude/phase vs frequency for calibrators.  
Each plot was visually inspected to mark unflagged bad data such as large phase scatters, errant amplitudes, system-temperature spikes, and large delays. 
These flags were manually added to the special editing lists that were automatically applied 
in the pipeline. 
Final flagging tables and calibration solutions were produced and applied to all the data. 
The calibrated uv datasets in AIPS format were converted to uvfits format before proceeding to the next step of imaging.

As shown in the right panel of Fig.\,\ref{fig:calibration_imaging_pipeline}, \emph{OBIT/MFImage} was used on the  calibrated uvfits data for wide-band and wide-field imaging.  
The imaging procedure first did a shallow CLEAN to get a crude sky model to use for flagging RFI, and then did a deeper, multifrequency CLEAN with potential self calibration.
The wideband imaging (\emph{OBIT/MFImage}) in Stokes I for each pointing was made by forming a weighted combined image of the sub-band images, with the field of view of the primary beam at a given frequency.
The bandpass was divided into 9 sub-bands, and a narrow sub-band image was made for each sub-band.
A pixel-by-pixel spectrum was fitted to the sub-band images to determine the spectral index. 
A weighted fit to the spectrum of each pixel was used to estimate the flux density at the central frequency of 5.8 GHz.
To improve the dynamic range, self-calibration was performed for fields with peak brightnesses that exceed a given threshold.
During the imaging/
deconvolution process of the B-configuration data, the baseline range was restricted to be larger than 50\,$k\lambda$ (corresponding to an  angular size < 4\arcsec), in order to reject emission from poorly mapped extended structures, which are numerous in the Galactic plane. This has a minor impact on the overall sensitivity, as discussed in \citet{Dzib2023AA670A9D}.
Each field of view consists of multiple facets and all facets are cleaned in parallel. The CLEAN facet images are combined into a single plane for each sub-band.  
A final $1\degr \times 2\degr$ image was made by the mosaic process \citep{Brunthaler2021AA651A85B}. Here, a circular restoring Gaussian beam with an FWHM of 1$\farcs$0 and a pixel size of 0$\farcs$25 was used. Here the first few rows of pointings from the two neighboring images were also used.

\begin{figure*}
 \centering
    \includegraphics[width = 1.0\textwidth]{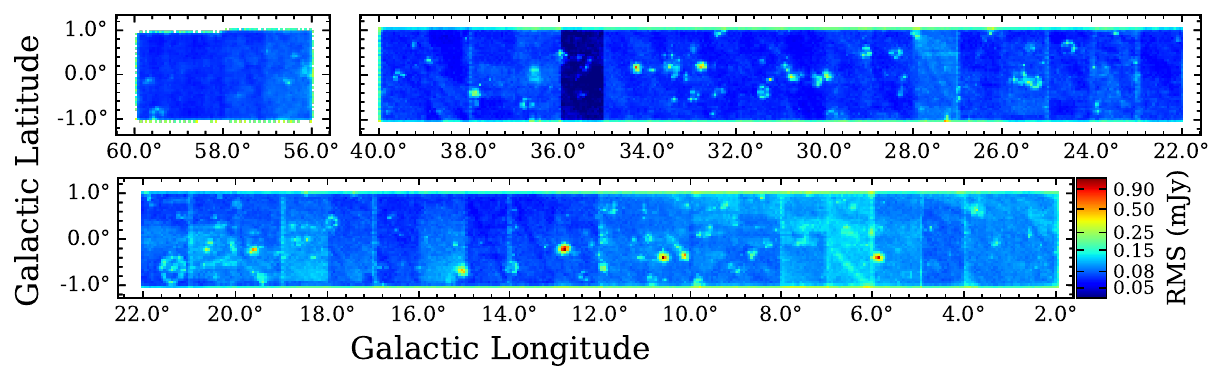} 
  \caption{RMS noise map of the $2\degr<\ell<40\degr, 56\degr<\ell<60\degr,\&\, |b|<1\degr$ region observed in B-configuration of the GLOSTAR survey. Each field is made by sampling on a size of $3\farcm25$. High noise levels are found in the fields associated with bright emissions (the star-forming regions, \hii\ regions, and other bright radio-emitting sources), observed at low declinations (close to the Galactic center), observed in bad weather conditions, and located at the edge area of the survey. High noise striping in Galactic Latitude is due to changes in observing conditions between scans. For B-configuration data of the ``pilot region'' of  the GLOSTAR survey ($28\degr<\ell<36\degr, |b|<1\degr$), see \citet{Dzib2023AA670A9D}. 
  }
 \label{fig:rms_noise_map}
 \end{figure*}
 
\subsection{The noise level}
\label{sect:noise_level}
  \begin{figure}
 \centering
    \includegraphics[width = 0.5\textwidth]{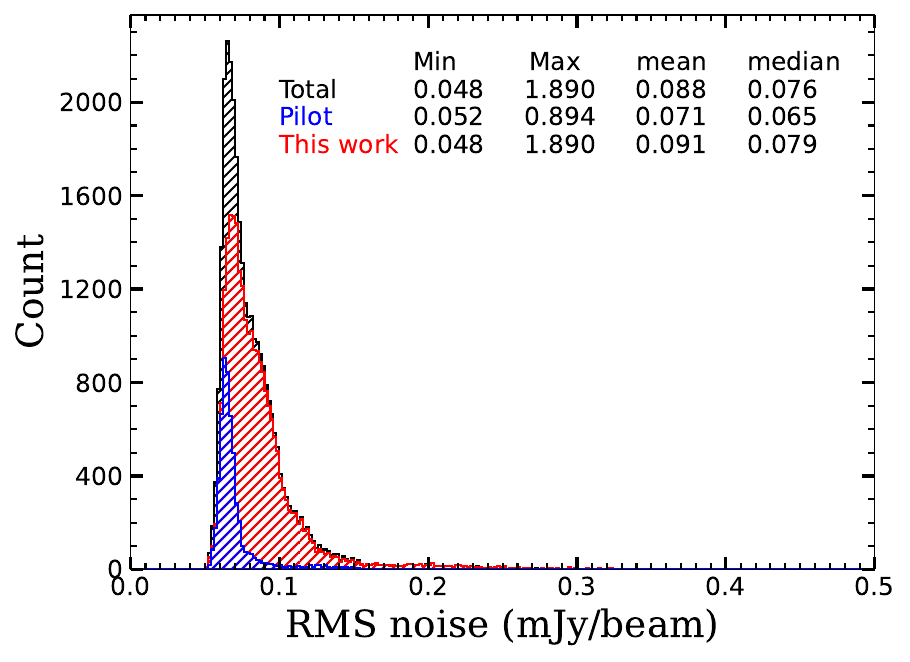} 
  \caption{Distributions of RMS noise level of the GLOSTAR B-configuration images (Fig.\,\ref{fig:rms_noise_map}). 
  The bin width is $\rm 0.002\,mJy\,beam^{-1}$. The red, blue  and black histograms represent values determined for the region discussed here ($2\degr<\ell<28\degr, 36\degr<\ell<40\degr,\,\&\,56\degr<\ell<60\degr$), the pilot region ($28\degr<\ell<36\degr$) published in \citet{Dzib2023AA670A9D}, and the combined area $2\degr<\ell<40\degr,56\degr<\ell<60\degr$ area.  The median RMS noise level is $\rm \sim\rm 0.08\, mJy\,beam^{-1}$ and 95\% of the fields have a noise level $\rm <0.13\,mJy\,beam^{-1}$. }
 \label{fig:rms_distribution}
 \end{figure}
The root mean square (RMS) noise map determined from the B-configuration image of the region covered by the GLOSTAR survey, (i.e., the total area minus the Galactic center region and the Cygnus X region) is presented in Fig.\,\ref{fig:rms_noise_map}. 
The data points represent RMS noise values determined over a
quadratic areas with a side length of $3\farcm25$ \citep[i.e., half of the average primary beam FWHM, see ][]{Brunthaler2021AA651A85B}.  
As expected, the noise is found to be high in fields associated with bright and extended emission (e.g., star-forming complexes, \hii\ regions, and other bright radio-emitting sources), close to the edge of the survey regions, and those observed at low elevations or in bad weather \citep[e.g.,][]{Helfand2006AJ,Hoare2012PASP,Purcell2013ApJS2051P,Bihr2015AA580A112B}. 
For instance, three of the fields show noise levels of $\rm 1\,\sigma>0.9\,mJy$, all of which are observed at relatively low declinations\,($\delta<-17\degr$) and located in massive star-forming complexes such as W28\,A\,(G005.89$-$0.39), W31\,C\, (G010.6$-$0.4), and W33\,(G012.8$-$0.2), with associated masers, compact and \uchii\ regions and outflows \citep[e.g.,][]{Liu2010ApJ7252190L,Beuther2011AA531A26B,Qiu2012ApJ756170Q,Immer2014AA572A63I,Wyrowski2016AA585A149W,Urquhart2018MNRAS4731059U,Yang2022AA658A160Y}.

Fig.\,\ref{fig:rms_distribution} 
presents the distributions of the
RMS noise level for the B-configuration images ($2\degr<\ell<40\degr\,\&\,56\degr<\ell<60\degr$), which include pilot region ($28\degr<\ell<36\degr$) data published by \citet{Dzib2023AA670A9D}. 
The $1\,\sigma$ noise level varies spatially over the fields, from $\sim \rm 0.05\,mJy\,beam^{-1}$ to $\sim \rm 1\,mJy\,beam^{-1}$, with a median RMS noise of $\rm \sim 0.08\,mJy\,beam^{-1}$.  
About 95\% of the fields show RMS values $\rm <0.13\,mJy\,beam^{-1}$, and only $\sim$1\% of the fields have noise levels exceeding the median value by a factor of 3 (i.e.\,$\rm >0.24\,mJy\,beam^{-1}$). 

As presented in Fig.\,\ref{fig:rms_distribution}, the region covered in this work shows a statistically higher noise level than the already published pilot region, and the median RMS values are $\rm \sim 0.079\,mJy\,beam^{-1}$ and $\rm \sim 0.065\, mJy\,beam^{-1}$, respectively. 
This could be due to the fact that this work covers a larger sky area (68$\rm\,deg^{2}$ or 79\% of the total fields) with more high-noise fields than the pilot region (16$\rm\,deg^{2}$). 
For instance, this paper covers the inner part of the GLOSTAR area that has  lower declinations ($\delta<-4\degr$) compared to the pilot region ($\delta>-4\degr$), and has been observed at lower elevations, resulting in higher levels of noise. 
This is also seen in the CORNISH survey that shows a significantly higher noise for lower declination fields \citep{Purcell2013ApJS2051P}. 
From Fig.~\,\ref{fig:rms_distribution}, we can see the noise distribution of this work is bimodal with a second peak of smaller amplitude at the high noise side, which is mainly attributed to the high noise level in the inner regions with lower declinations.

In summary, the B-configuration continuum images of the GLOSTAR survey show a spatially varying noise level of $\rm \sim0.05-0.13\, mJy\,beam^{-1}$ for 95\% of the covered area. The typical RMS noise $\rm \sim 0.08\,mJy\,beam^{-1}$ is consistent with the theoretical prediction. 

\section{Source catalog construction}
\label{sect:source_cat_construction}

The source catalog presented in this work is constructed following the same strategy that was used by \citet{Dzib2023AA670A9D} and \citet{Medina2019AA627A175M} for source extraction and estimation of physical parameters.

\subsection{Source extraction, fluxes, and sizes}
\label{sect:source_extract}

The radio continuum sources are initially extracted using a $5\,\sigma$ local noise threshold determined with the software BLOBCAT\,\citep{Hales2012MNRAS425979H}. 
BLOBCAT produces a catalog of BLOBS that contains the properties for every detected source, including peak pixel coordinates, peak and integrated flux, RMS noise levels, number of pixels comprising each source. 
The effective sizes can be determined from the total number of pixels comprising each source $\rm N_{pix}$ and the pixel size $\rm 0\farcs25$ as $R_{eff}=\sqrt{A/\pi}$, where the source area in arc seconds, $A$, is $N_{pix}\times 0\farcs25 \times 0\farcs25$. 

 \begin{figure}
 \centering
    \includegraphics[width = 0.45\textwidth]{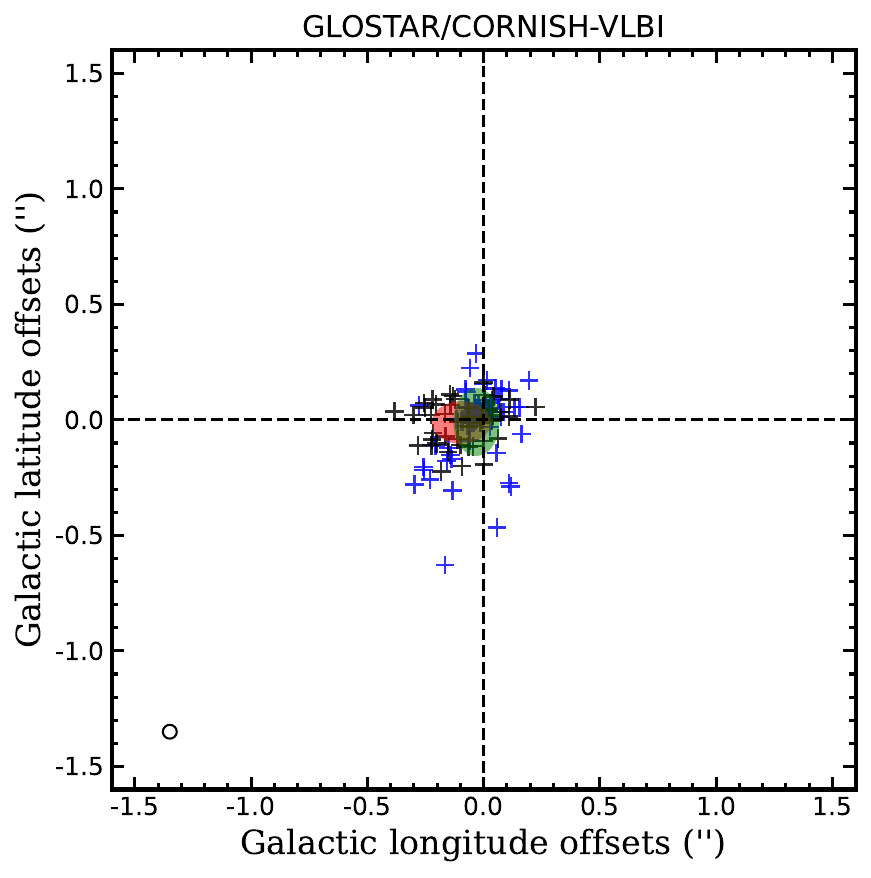} \\
 \includegraphics[width = 0.45\textwidth]{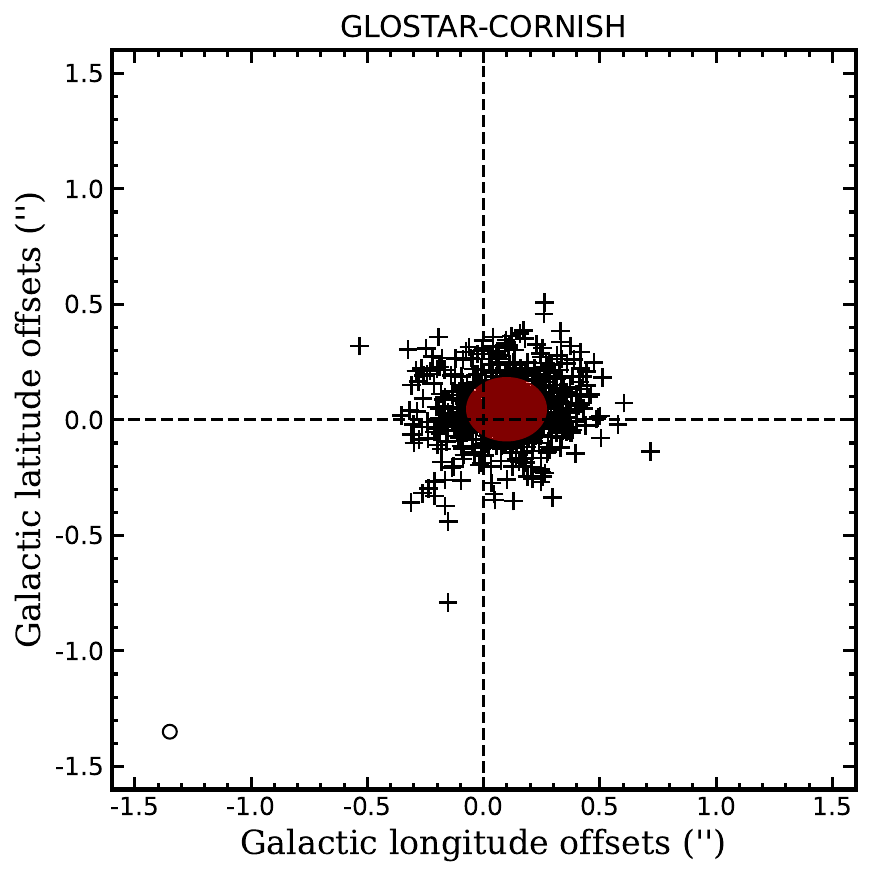} \\
    \caption{Comparison of GLOSTAR  astrometry with CORNISH and VLBI results. Top panel:  Offsets between GLOSTAR B-configuration positions (this work) and the positions of VLBI calibrators (black crosses, red filled ellipse) for 44 counterparts, and the offsets between CORNISH and VLBI calibrators for 75 counterparts (blue crosses, green filled ellipse). 
    The VLBI calibrators have a (sub)milliarcsec positional accuracy \citep{Petrov2021AJ16114P}. The median and standard deviation of the offsets are  $-0\farcs06$ and $-0\farcs11$, suggesting a systematic positional difference of $\lesssim$0\farcs1 for the GLOSTAR and Very Long Baseline Array (VLBA) calibrators.  
  Bottom panel: Position offsets between the GLOSTAR  B-configuration and CORNISH positions for 669 counterparts. The mean and standard deviation of the offsets are 0\farcs08 and 0\farcs16, suggesting a systematic positional difference of $\lesssim$0\farcs1 for the GLOSTAR and CORNISH. }
 \label{fig:position_offset}
 \end{figure}
\subsection{Astrometry}
\label{sect:astrometry}

In order to estimate the astrometric accuracy of our data, we determined the position offsets of VLBA calibrators between GLOSTAR and \citet{Petrov2021AJ16114P}  which have (sub)milliarcsecond astrometic accuracy. 
We also compared the GLOSTAR with the CORNISH positions \citep[]{Hoare2012PASP} which has an angular resolution of 1\farcs5 and a position accuracy of 0\farcs1.
As seen in the upper-panel of Fig.\,\ref{fig:position_offset}, we find no significant position offset between the GLOSTAR B-configuration and VLBA calibrator positions for the 44 common sources with mean offsets $\pm$ standard deviations  of -0\farcs09 $\pm$ 0\farcs13 in Galactic longitude and -0\farcs01 $\pm$ 0\farcs09 in Galactic latitude, respectively.
Similarly, as shown in the bottom panel of Fig.\,\ref{fig:position_offset}, we find no significant offsets between GLOSTAR and CORNISH for 699 compact common sources, with 0\farcs09 $\pm$ 0\farcs17 and 0\farcs05 $\pm$ 0\farcs14 in the Galactic longitude and latitude directions, respectively. 
Based on the above analysis, we conclude that the systematic positional uncertainty of our GLOSTAR B-configuration positions is $\lesssim$0\farcs1, which is consistent with that determined for the pilot region in \citet{Dzib2023AA670A9D}.
 

  \begin{figure}
 \centering
    \includegraphics[width = 0.45\textwidth]{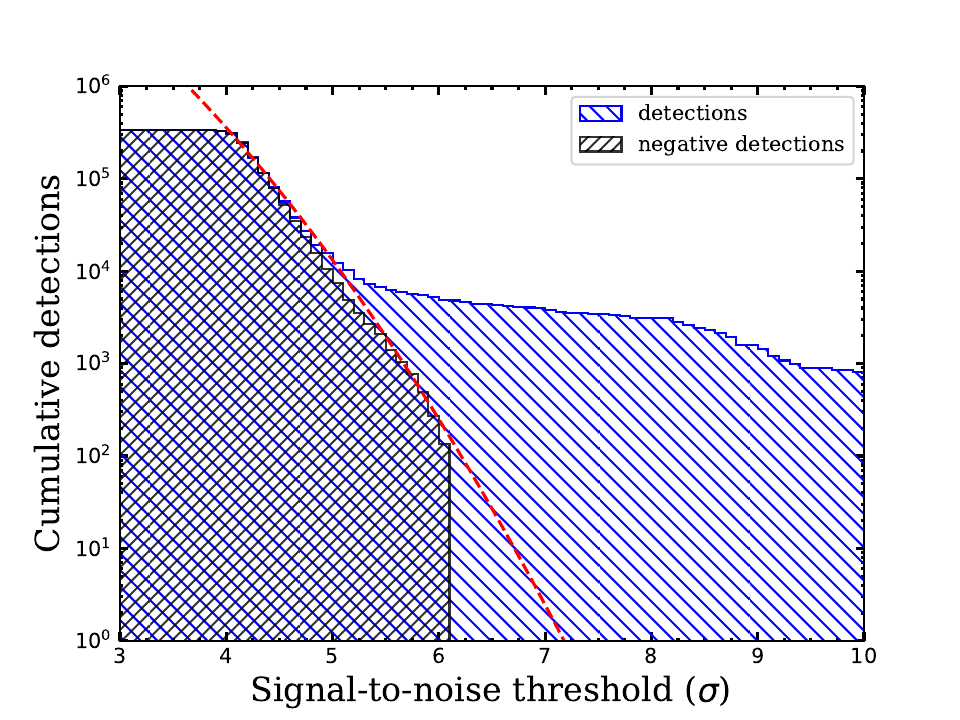} 
  \caption{Cumulative distribution of spurious sources (black histogram) and all detections (blue histogram) as a function of the signal-to-noise ratio ($\rm 1\sigma\sim0.08\,mJy$), to estimate the spurious sources expected in the GLOSTAR survey. 
  The red dashed line fitted to the negative detections indicates the expected spurious sources. 
  Below $4.5\,\sigma$, the detections are dominated by spurious sources (occupying 93\%) and the false detections decrease to 186 at $6.1\,\sigma$. Above $7\,\sigma$, fewer than 5 sources are expected to be false. }
 \label{fig:false_detection_distribution}
 \end{figure}
\subsection{Spurious source estimation}
\label{sect:spurious_sources}
A total of 13689 sources are detected above $5\,\sigma$ in the B-configuration images of the GLOSTAR region discussed in this paper ($2\degr<\ell<28\degr,36\degr<\ell<40\degr, \&\,56\degr<\ell<60\degr$). 
It is to be noted that the number of false detections is strongly related to the signal-to-noise ratio (S/N) of sources and a significant fraction of $5\,\sigma$ detections are likely to be spurious or caused by residual sidelobes of nearby strong sources \citep{Helfand2006AJ,Purcell2013ApJS2051P}. 
To estimate the number of the spurious sources expected in the GLOSTAR survey, we follow the strategy used by CORNISH  \citep{Purcell2013ApJS2051P} and run a source finder on the inverted data to extract negative detections (i.e., spurious sources), which can be done by using the tool Aegean\,\citep{Hancock2012MNRAS4221812H}.  
As the cumulative distribution of spurious sources shown in  Fig.\,\ref{fig:false_detection_distribution}, the false detections decrease rapidly for increasing signal-to-noise ratios. For sources below $5\,\sigma$, more than two-thirds are likely spurious sources, whereas the fraction of false detections falls to about half for sources between $5\,\sigma$ and $6\,\sigma$. 
The false detections decrease to 186 at $6.1\sigma$.  
The red dashed line in Fig.\,\ref{fig:false_detection_distribution} shows that the population of false detections is close to what is expected from Gaussian statistics, by fitting the negative detections with $f(\sigma)=1-erf(\sigma/\sqrt{2})$ (where $erf(\sigma/\sqrt{2})$ is the Gaussian error function as outlined in \citealt{Purcell2013ApJS2051P}).
Based on this, for sources above $7\,\sigma$, less than 5 spurious sources are expected to be detected. 
Therefore, we use $7\,\sigma$ as a threshold to split the catalog into two: a high-reliability catalog with 5497 sources above $7\,\sigma$ and a low-reliability catalog with 7917 sources between $5\,\sigma$ and $7\,\sigma$. 
The low-reliability $5-7\sigma$ threshold catalog contains many real sources as well as spurious sources (nearly half), which is listed in Appendix table\,\ref{tab:total_5to7sigma}.  
The full catalog will be available in the GLOSTAR website \footref{glostar_web} and CDS. 
In the following section, we present, analyze, and discuss the high-reliability $7\,\sigma$ catalog, as listed in Table\,\ref{tab:total_7sigma}.  

  \begin{figure}
 \centering
    \includegraphics[width = 0.45\textwidth]{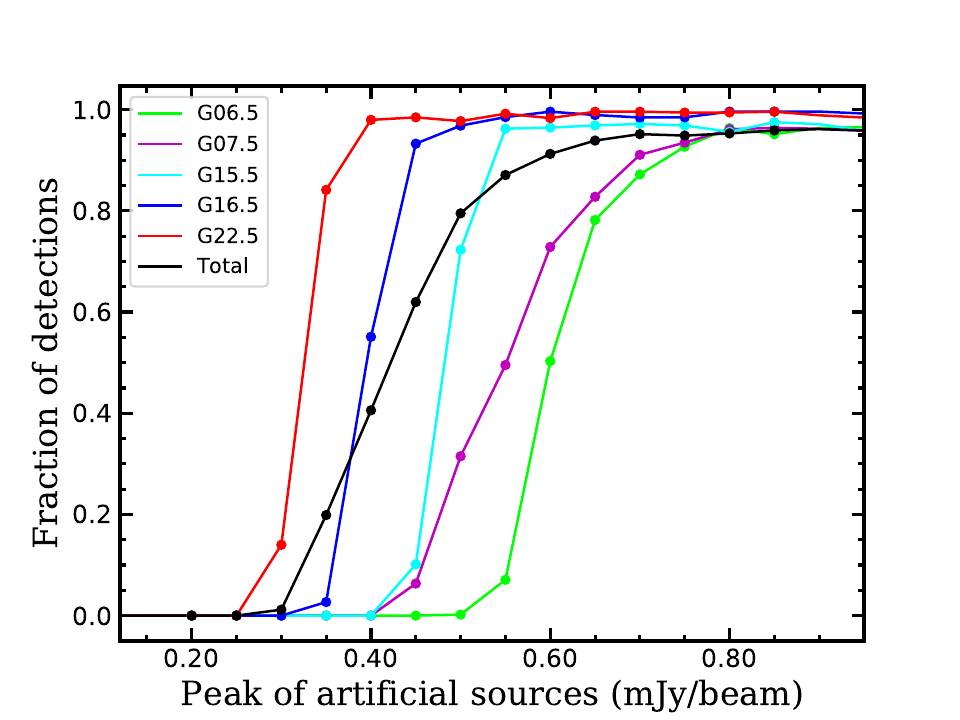}
  \caption{Completeness fraction of recovered artificial sources as a function of the peak flux density of the total added artificial sources. The completeness of the selected fields with different typical noises are shown in different colors. The mean RMS values for the high, low, and typical noise regions are  $\rm 0.12\,mJy\,beam^{-1}$, $\rm 0.06\,mJy\,beam^{-1}$, and $\rm 0.08\,mJy\,beam^{-1}$, respectively. 95\% completeness limit is reached at a flux density of $\sim$0.35\,mJy for the best case G22.5 (red line) and $\sim$0.8\,mJy for the worst case G06.5. Typically, the GLOSTAR survey in B-configuration is 95\% complete to point sources at the $\sim$0.6\,mJy level (i.e., the chosen $7\,\sigma$ threshold for the catalog in Sect.\,\ref{sect:spurious_sources}), as shown in black line for all the selected fields. See the completeness in Sect.\,\ref{sect:completeness} for details.} 
 \label{fig:completeness}
 \end{figure}
\subsection{Completeness}
\label{sect:completeness}
The completeness limit of the catalog can be investigated by running a source finder for added artificial point sources in the empty sky, that is, fields with few detections above $5\,\sigma$. 
The noise distributions vary spatially as outlined in Sect.\,\ref{sect:noise_level}, and hence to investigate the completeness limit of the whole survey, we chose several regions with a size of $0.5\degr\times 0.5\degr$ that sampled all noise regimes  such as high noise ($\rm \sim\,0.12\,mJy\,beam^{-1}$), low noise ($\rm \sim\,0.06\,mJy\,beam^{-1}$) and typical noise ($\rm\sim0.08\,mJy\,beam^{-1}$). 
Then, one thousand artificial point sources are injected into each selected region. The position and peak flux densities of the injected sources are randomly produced from a uniform distribution, with a flux density range of 0.05\,mJy\,$-$\,1.0\,mJy. 
Following the same source extraction strategy described in Sect.\,\ref{sect:source_extract}, BLOBCAT was used to extract these added artificial sources. 
After 10 iterations of injection and extraction processes, the extracted results are compared with the injected source parameters to estimate the completeness. 
Fig.\,\ref{fig:completeness} shows 
the fraction of the recovered artificial sources as a function of the peak intensity of the added artificial sources for fields with different noise levels. The completeness fraction varies among these fields, and fields with lower noise levels tend to reach the same completeness limit at  lower flux density values. We find that a 95\% completeness limit corresponds to $\sim$0.35\,mJy (red line in Fig.\,\ref{fig:completeness}) for the best case of G22.5 and $\sim$0.8\,mJy (lime line in Fig.\,\ref{fig:completeness}) for the worst case of G06.5. 
Typically, above $\sim$0.6\,mJy (i.e., the $7\,\sigma$-threshold detection limit for the high-reliability catalog in Sect.\,\ref{sect:spurious_sources}), the survey is 95\% complete for point sources, as shown in the black line in Fig.\,\ref{fig:completeness} for all the selected fields. 

\subsection{Spectral index determination}
\label{sect:spetral_index}
\begin{figure}
 \centering
    \includegraphics[width = 0.45\textwidth]{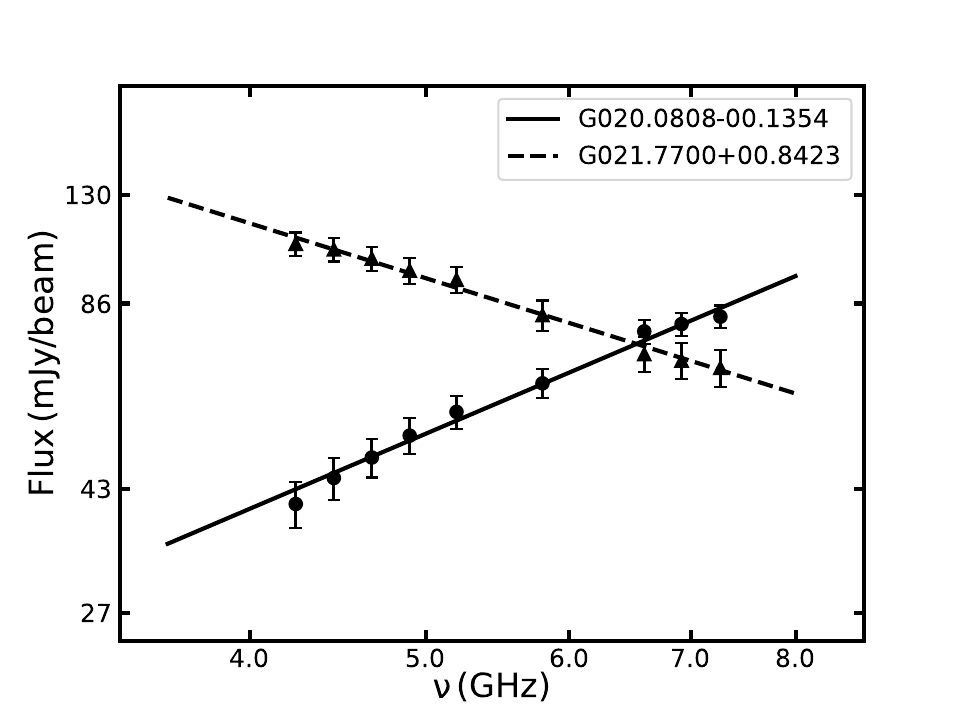} \\
    \includegraphics[width = 0.45\textwidth]{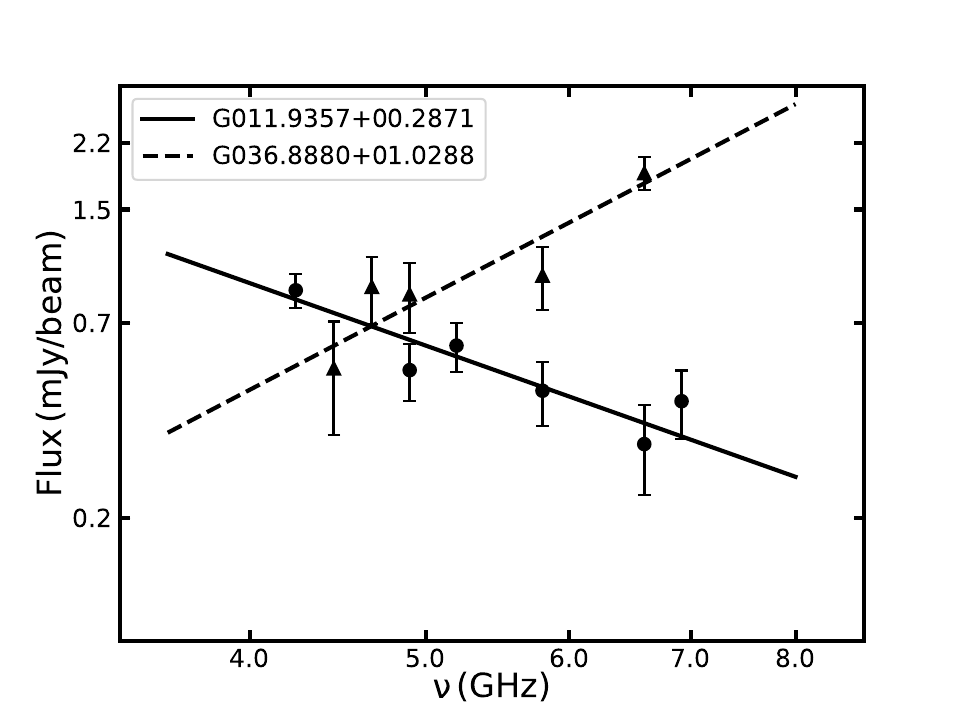} 
  \caption{Example of the peak flux as a function of the in-band frequency for bright sources (top panel) and faint sources (bottom panel).
  Each data point (circle or triangle) refers to the peak intensity at each sub-band, with the error of the peak measurement. 
The solid and dashed lines in the top panel show the best fit to the peak flux densities from  9 sub-bands for two bright sources: G020.0808-00.1354 ($\alpha = 1.1\pm 0.07$) and G021.7700+00.8423 ($\alpha = -0.92\pm 0.04$), respectively. 
  The solid and dashed lines in the bottom panel present the spectral fitting of 5 and 6 sub-bands for two faint sources: G011.9357+00.2871 ($\alpha = -1.7\pm 0.36$) and G036.8880+01.0288 ($\alpha = 2.5\pm 0.6$), respectively.
  }

 \label{fig:spectral_fit_example}
 \end{figure}

As mentioned in Sect.\,\ref{sect:obs_cal_img}, the GLOSTAR survey covers a wide bandwidth from 4-8\,GHz, which was split into 9 sub-bands and each sub-band was imaged separately. 
We are thus able to determine the in-band spectral indices $\alpha$ by (1) extracting the peak flux density of each source within the sub-bands and (2) fitting the peaks to the formula $S_{\rm \nu}\propto \nu^{\alpha}$ based on the scipy function \emph{$curve\_fit$}. 
The uncertainties in the peak flux densities are taken into account in
the fitting process to estimate the uncertainty for the spectral index. 
When we are not able to extract the peak flux densities in all 9 sub-bands for some sources due to high noise or RFI in some sub-bands, the spectral indices are determined by fitting the remaining data points. 
More than 98\% of the $7\,\sigma$ sources have reliable data suitable for spectral fitting in at least five sub-bands.
Fig.\,\ref{fig:spectral_fit_example} gives an example of the spectral index ($\alpha$) fitting process for two bright sources with all 9 sub-bands and two faint sources with 5 and 6 sub-bands. 
With this strategy, we measured in-band spectral indices for 5430 sources in the 7$\sigma$-threshold catalog. 
The in-band spectral index of the 5430 sources ranges from -2.85 to 2.72, as listed in Table\,\ref{tab:summary_phy_param}.  
The uncertainty in spectral index $\sigma_{\alpha}$ ranges from 0.01 to 2.63 with mean and median values of $\sim$0.2 for both, and is strongly correlated with the signal-to-noise ratio of the sources.
Given that 97\% of the 7-$\sigma$ sources are compact with $Y_{\rm factor}<2.0$ (defined as the ratio between integrated flux density and peak flux density), the measured spectral index is not expected to be affected significantly by spatial filtering of the interferometer (due to missing short spacings). 

\subsection{Source classification}
\label{sect:classification}

  \begin{figure}
 \centering
    \includegraphics[width = 0.45\textwidth]{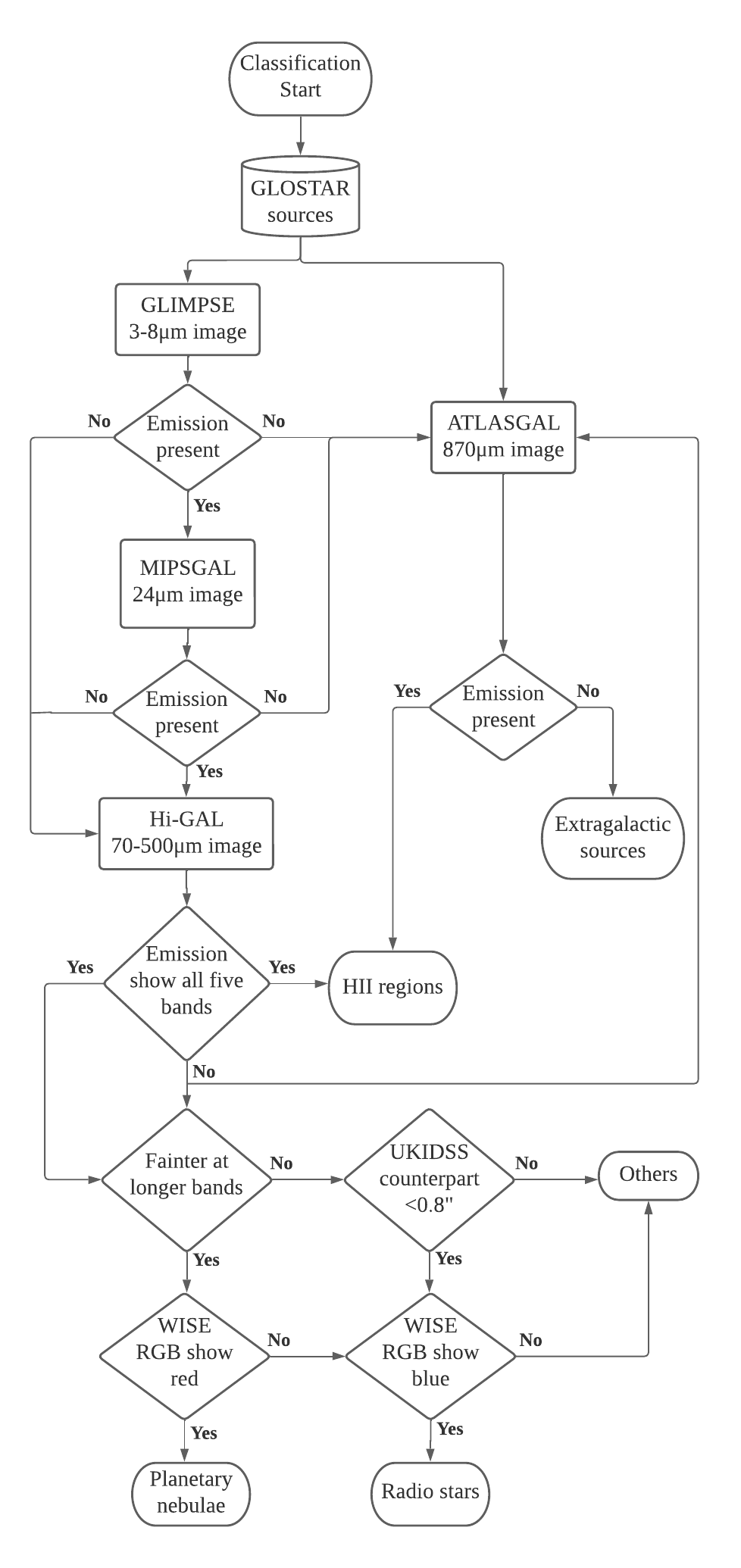} 
  \caption{Flowchart illustration of the radio source classification process. See the text of Sect.\,\ref{sect:classification} for further details.}
 \label{fig:classification_process}
 \end{figure}
 
  \begin{figure*}
 \centering
 (a) an \hii\ region candidate G002.6150+00.1338 \\
\includegraphics[width = 0.95\textwidth]
{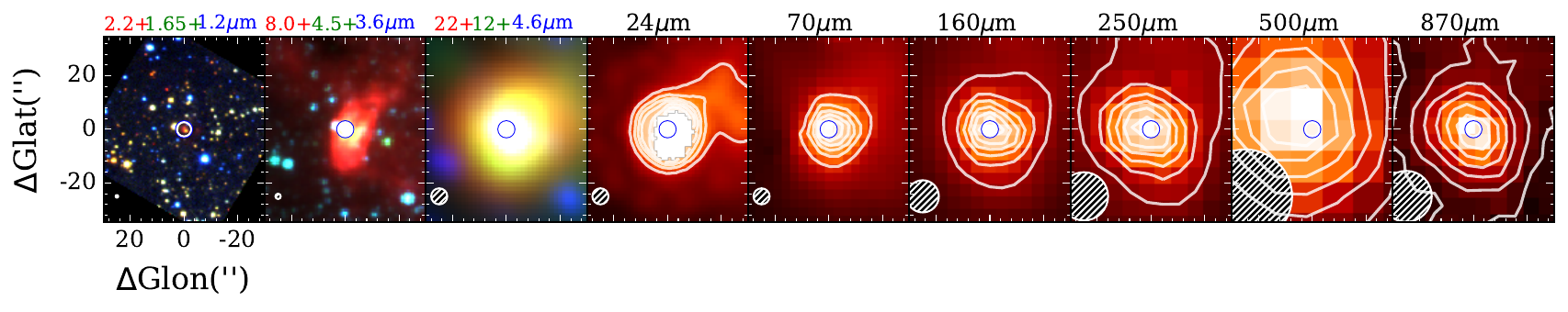} \\
(b) a planetary nebula candidate G003.3298-00.8693 \\
\includegraphics[width = 0.95\textwidth]{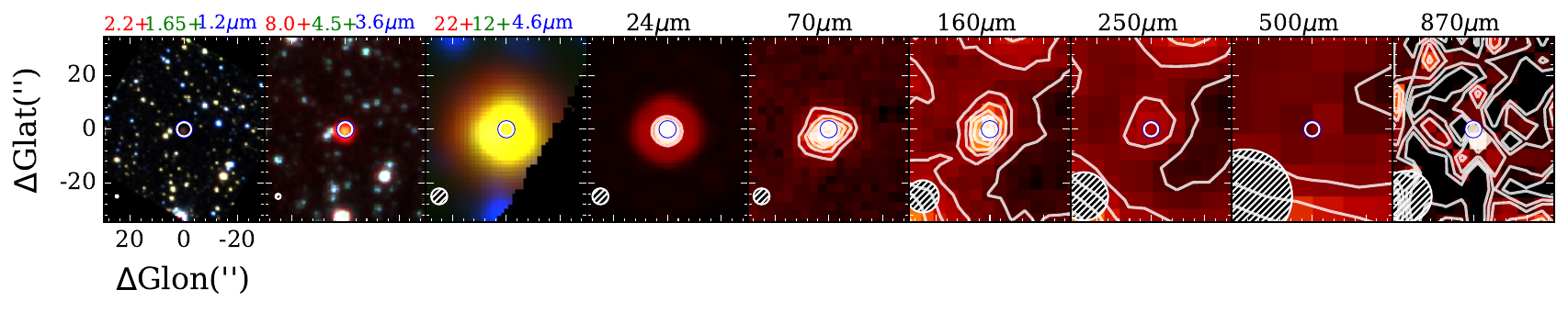} \\
(c) a radio star candidate G012.7919+00.0310 \\
\includegraphics[width = 0.95\textwidth]
{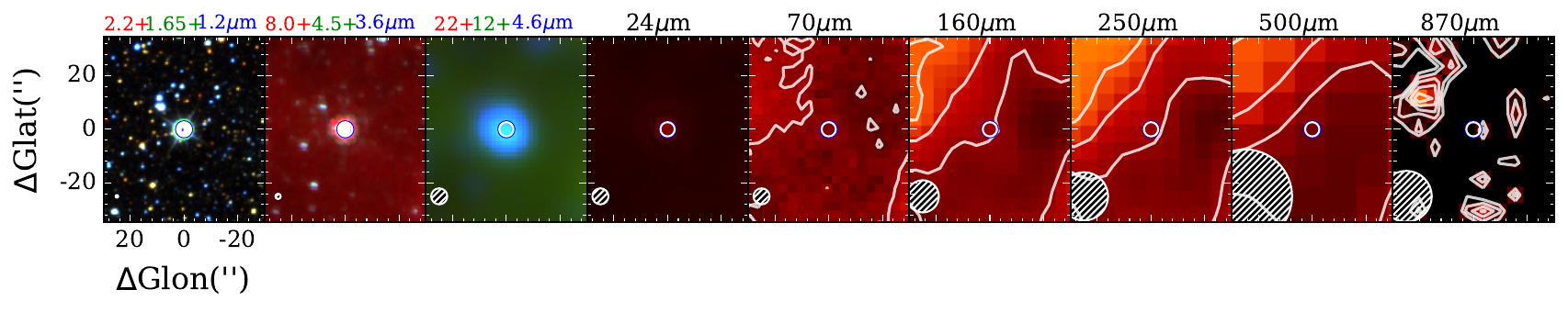} \\
(d) an extragalactic candidate G004.3528+00.2119 \\
\includegraphics[width = 0.95\textwidth]{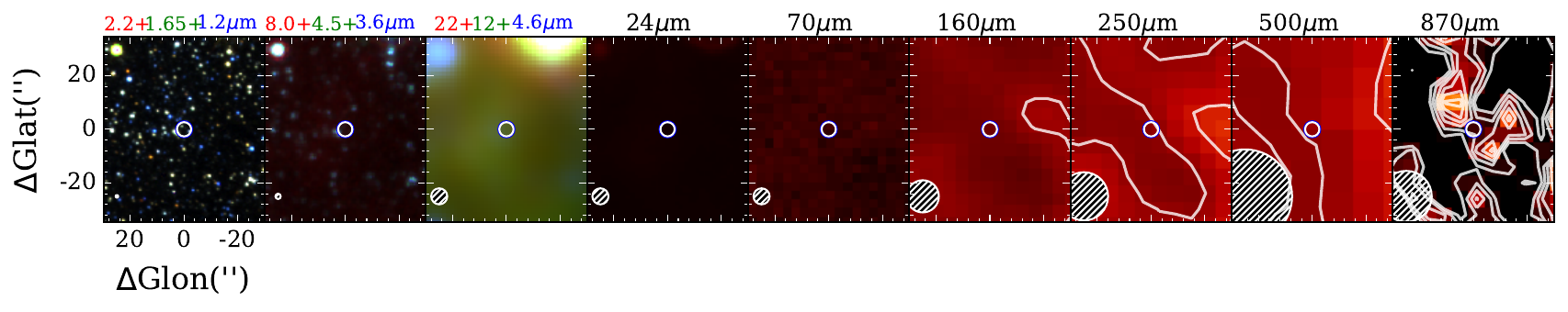}
\caption{Illustration of source types in the classification process of Sect.\,\ref{sect:classification}. 
From top to bottom: Typical multiband images of a radio source classified as an \hii\ region, a planetary nebula, a radio star and a extragalactic source. From left to right: three-colors composition image of UKIDSS (Red = 2.2\,$\mu m$, Green = 1.65\,$\mu m$, Blue = 1.2\,$\mu m$), three-color composition
image of GLIMPSE (Red = 8\,$\mu m$, Green = 4.5\,$\mu m$, Blue = 3.6\,$\mu m$), three-color composition of WISE (Red = 22\,$\mu m$, Green = 12\,$\mu m$, Blue = 4.6\,$\mu m$), MIPSGAL 24\,$\mu m$, Hi-GAL 70\,$\mu m$, Hi-GAL 160\,$\mu m$, Hi-GAL 250\,$\mu m$, Hi-GAL 500\,$\mu m$, and ATLASGAL
870\,$\mu m$. The blue and white circle in the center show the position of radio emission. 
The FWHM beams of UKIDSS\,(0.8\arcsec), GLIMPSE\,(2\arcsec), WISE\,(6\arcsec\, at 12\,$\mu m$),
MIPSGAL\,(6\arcsec\,at 24\,$\mu m$), Hi-GAL\,(6\arcsec$-$35\arcsec), ATLASGAL\,(19\arcsec) are indicated by the white circles with black hatched lines shown in the lower-left corner of each image.}
 \label{fig:example_class}

 \end{figure*}

As outlined in the surveys of CORNISH \citep{Purcell2013ApJS2051P}, THOR \citep{Wang2018AA619A124W}, and the GLOSTAR pilot \citep{Dzib2023AA670A9D}, the classification processes of the GLOSTAR sources were conducted based on multiwavelength counterparts and/or emission properties from the Galactic plane surveys such as the near-infrared (NIR) UKIDSS survey at J\,H\,K bands \citep{Lucas2008MNRAS391}, the GLIMPSE survey at $3-8\,\mu m$ \citep{Churchwell2009PASP}, the mid-infrared (MIR) WISE survey at $4-22\,\mu m$ \citep{Wright2010AJ1401868W} and the MIPSGAL survey $24\,\mu m$ \citep{Carey2009PASP76C}, the far-infrared (FIR) Hi-GAL at $70-500\,\mu m$ \citep{Molinari2010PASP}, and the submillimeter (Submm) ATLASGAL survey at $870\,\mu m$ \citep{Schuller2009AA,Contreras2013AA549A45C,Csengeri2014AA565A75C,Urquhart2014AA568A41U,Urquhart2018MNRAS4731059U,Urquhart2022MNRAS3389U}. 
To search for counterparts, we adopt the beam size of these surveys as the search radius, such as $0.8\arcsec$ for UKIDSS, $2\arcsec$ for GLIMPSE, $6\arcsec$ for WISE, and $6\arcsec$ for MIPSGAL at $24\,\mu m$. 
We also carried out a visual inspection of the emission in different bands to look for association with extended structure. 
This was done by constructing three-color images from UKIDSS (red K-band $2.2\,\mu m$, green H-band $1.65\,\mu m$ and
blue J-band $1.2\,\mu m$), GLIMPSE (red $8.0\,\mu m$, green $4.5\,\mu m$, and blue
$3.6\,\mu m$), and WISE (red $22.0\,\mu m$, green $12.0\,\mu m$ and blue $4.6\,\mu m$) surveys, as well as images from the MIPSGAL, Hi-GAL, and ATLASGAL surveys. 

The GLOSTAR sources are then classified as five types of candidates: \hii\ regions, radio stars, planetary nebulae, extragalactic sources, and others, using the criteria below: 
\begin{itemize}

 \item[] \emph{\hii\ regions:} As ionized gas regions around massive stars are located in dense molecular clouds, \hii\ regions are bright in the Submm, FIR, and MIR \citep{Churchwell2002ARAA,Anderson2012AA537A1A,Urquhart2013MNRAS435,Thompson2016mksconfE15T,Yang2019MNRAS4822681Y}.  \hii\ regions are also bright in the GLIMPSE three-color images due to emission from the associated polycyclic aromatic hydrocarbons (PAHs)  \citep{Purcell2013ApJS2051P,Tsai2009AJ1374655T}. Since they are deeply embedded in molecular clouds, young \hii\ regions may still be dark or weak in NIR and even in some MIR bands \citep{Hoare2012PASP,Murphy2010MNRASa2,Yang2021AA645A110Y}. Therefore, radio sources associated with Submm and/or FIR emission are classified as \hii\ regions. 

    \item[] \emph{Planetary nebulae (PNe):} 
    As ionized gas regions around young white dwarf stars \citep{Bobrowsky1998Natur469B}, PNe and \hii\ regions have similar emission properties at infrared and radio wavelengths \citep{Anderson2012AA537A1A}. 
    Compared to \hii\ regions, the spectral energy distributions (SEDs) of PNe tends to peak at shorter MIR wavelengths and fall off steeply at FIR \citep{Anderson2012AA537A1A,Purcell2013ApJS2051P}, which often makes PNe undetectable in the submm range  \citep{Urquhart2013MNRAS435,Urquhart2018MNRAS4731059U}. 
    Although some nearby PNe are detectable in the FIR (Hi-GAL) and the submm (ATLASGAL), their emission typically tends to be fainter at longer wavelengths \citep{Anderson2012AA537A1A,Purcell2013ApJS2051P}. 
    Since the SEDs typically peak in the MIR range, PNe tend to appear red in WISE and GLIMPS three-color images.
    Moreover, PNe are likely to be isolated point-like sources in GLIMPSE and UKIDSS images due to the absence of  molecular clouds \citep{Zhang2009ApJ706252Z,Hoare2012PASP}.

    \item[] \emph{Radio stars:} Both thermal and non-thermal radio emission from radio stars have been observed, that could arise from evolved OB stars, active stars, or active binaries, as discussec by \citet{Hoare2012PASP}. In the submm and FIR ranges, emssion from radio stars tends to be weak or absent. 
    In the WISE, GLIMPSE, and UKIDSS three-color images of , radio stars tend to appear as blue  and point-like sources  \citep{Hoare2012PASP,Urquhart2013MNRAS435}. Also, a GLOSTAR source is classified as a radio star if it has an NIR UKIDSS counterpart (offset $<0\farcs8$) that is suggested to be a star in the UKIDSS catalog \citep{Lucas2008MNRAS391}. 
    
    \item[]  \emph{Extragalactic sources:} extragalactic sources are usually not seen in the Submm, FIR, MIR, and NIR Galactic plane surveys \citep{Hoare2012PASP,Urquhart2013MNRAS435}. Some of them are associated with very faint diffuse and/or point-like counterparts (offset < 0\farcs8) in the NIR UKIDSS, and the counterparts are classified as galaxies by \citet{Lucas2008MNRAS391}. 
    
    \item[] \emph{Others:} Radio sources that cannot be categorized as one of the above four types are classified as type ``others''. One type in this class is the photodissociation region (PDR), namely, the interface between the ionized region and the molecular cloud, which are normally extended and bright at GLIMPSE, usually with weak or no emission in the FIR and Submm \citep{Hoare2012PASP}.  
    
\end{itemize}

An overall flow chart of the classification process is shown in Fig.\,\ref{fig:classification_process} and the example sources are shown in Fig.\,\ref{fig:example_class}. The presence or absence of emission at other wavelengths, along with the source classification is indicated in Table\,\ref{tab:total_7sigma}.
In total, among the 5437 $7\,\sigma$-threshold sources, we identify candidates of 251 \hii\ regions, 784 radio stars, 282 PNe, 4080 extragalactic sources, and 29 Others. 
Among sources classified as the type ``Others'', 11 are likely to be PDR region candidates  \citep[i.e., normally associated with extended emission at MIR, and only weak or no emission at FIR and SMM wavelengths;][]{Hoare2012PASP}  while the rest are unidentified. As expected, most sources ($\sim$75\%) are classified as extragalactic candidates. 

To quantify the quality of the classification, we use a matching radius of 2" for determining SIMBAD counterparts  \citep{Wenger2000AAS_simbad_database}, as outlined in the pilot paper of GLOSTAR \citep{Dzib2023AA670A9D}. 
Among the 5437 classified sources, 1010 are found to have SIMBAD counterparts, and more than half (542/1010) are classified as the SIMBAD type `Radio'. 
For the 251 \hii\ regions candidates, 182 show SIMBAD counterparts, 95\% (173/182) of which have SIMBAD types that are consistent with the expected properties of \hii\ regions at different wavelengths, such as SIMBAD types of NIR/MIR/IR sources, molecular clouds, YSOs, star formation regions, dense cores, millimetric/submillimetric  sources, and \hii\ regions. 
Similarly, 166 PNe, 119 radio stars and 416 extragalactic sources have SIMBAD counterparts, with our classification being consistent with the SIMBAD type for 95\% (158/166), 92\% (111/119) and 97\% (406/416) of PNe, radio stars and extragalactic sources respectively. 
In addition, the consistency of classification between this work and the CORNISH survey are 100\% for \hii\ regions and PNe, and 98\% for extragalactic sources.  
This supports the validity of our classification criteria. 

We noted that our classification criteria may misclassify a radio source if it displays the same multiband emission properties as the above source types but does not belong to any of these categories. For example, three extragalactic source candidates are found to be associated with three pulsars within a 1.1\arcsec radius from the SIMBAD database, such as G016.8052-01.0011 \citep[e.g., PSR J1825-1446,][]{Hobbs2004MNRAS3531311H,Wang2020AA644A73W},  G023.2721+00.2979\citep[e.g., PSR J1832-0827,][]{Wang2001MNRAS328855W,Yao2017ApJ83529Y}, and G023.3856+00.0631 \citep[e.g., PSR J1833-0827,][]{Tian2007ApJ657L25T,Jankowski2019MNRAS4843691J}. Therefore, further investigation is required to understand the nature of these candidates.

\subsection{Clustered sources}
\label{sect:cluster_source}

  \begin{figure}
 \centering
     \includegraphics[width = 0.43\textwidth]{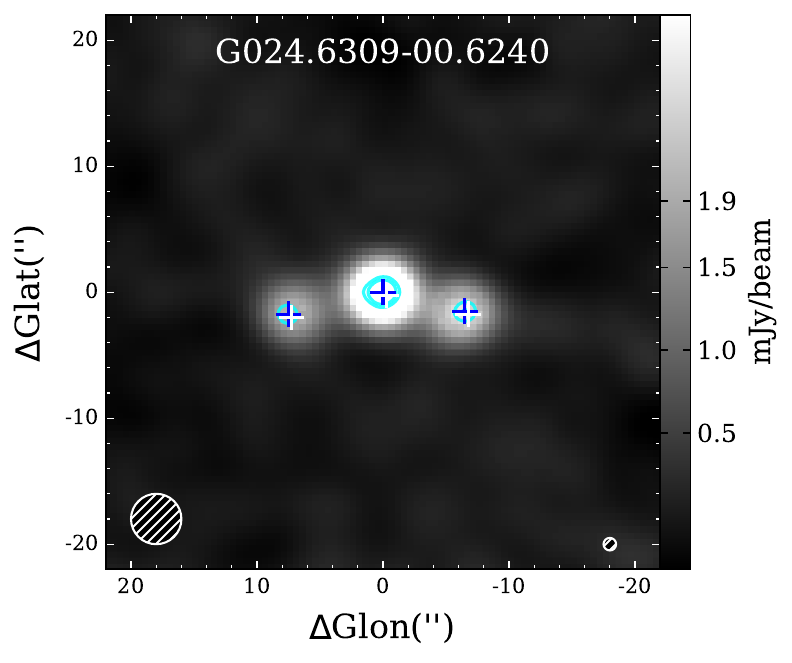} \\
    \includegraphics[width = 0.43\textwidth]{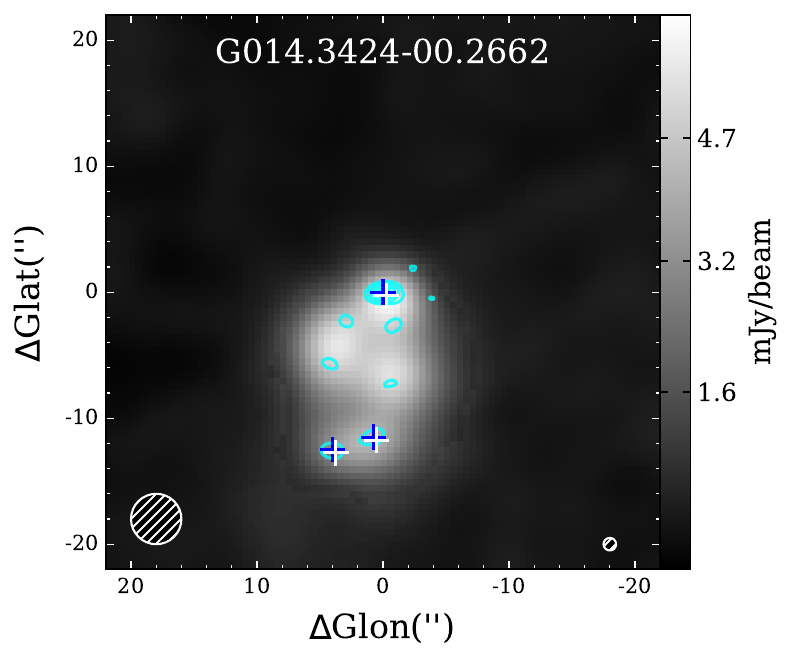} \\
    \includegraphics[width = 0.43\textwidth]{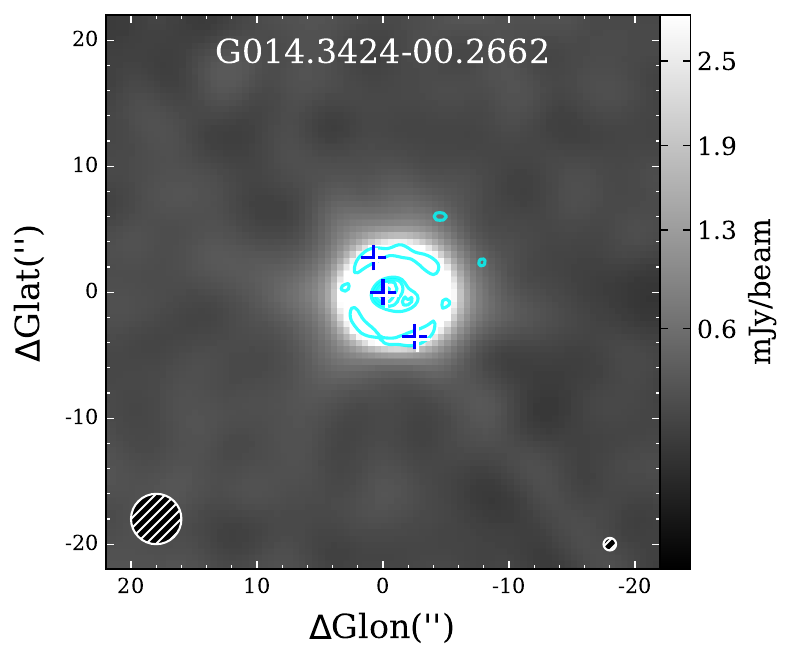} \\
      \caption{Example of cluster sources in three cases. Top panel: the cluster sources in normal case. Middle-panel: example of clusters consists of over-resolved sources are listed in Table\,\ref{tab:over_resolved_source}. 
      Bottom panel: example of clusters with semi-ring-like fragments. In this case, the sources with semi-ring-like structures have been removed from the catalog.  
      The positions of 7$\sigma$ cluster sources are noted by blue and white pluses. The Galactic name of each figure refers to the source in the center position.
      The cyan contours {are emission from} B-configuration images (this work), starting at 5$\sigma$ with 5$\sigma$ increment. 
      The background of each image presents the B+D configuration of GLOSTAR \citep{Brunthaler2021AA651A85B}. 
      The FWHM beams of B ($1\farcs0$) and B+D configuration ($4\farcs0$) are presented by the white circles in the lower-right and lower-left corners of each image.} 
 \label{fig:cluster_sources}
 \end{figure}
  
Small clusters of radio sources are identified in the catalog using a friends-of-friends method \citep[e.g.,][]{Purcell2013ApJS2051P}, that is, a source is associated with a cluster if it is located within 12\arcsec\ of any other member in the catalog. Among the total of 5497 sources,  60 sources are likely to be artifacts or semi-ring-like fragmented components that have been removed from the catalog. 
570 sources are found to be associated with 258 clusters. Among the 258 clusters, the majority ($\sim$84\%;\,216/258) harbors two radio sources. 
About 16\% (42/258) clusters harbor more than two radio sources, with $\sim$13\% (33/258) harboring three radio sources and $\sim$3\% (9/258) having more than three sources. 
All cluster members of the B-configuration are usually detected as one source in the D-configuration (beam=18\arcsec) of the GLOSTAR with the D-configuration names from \citet{Medina2019AA627A175M} and Medina et al., in prep., being shown in Column 12 of Table\,\ref{tab:total_7sigma}.  

As mentioned in Sect.\,\ref{sect:calibration_and_image}, the imaging process was restricted to have baselines of more than $50\,k\lambda$, due to which
the survey is optimized to detect emission with angular size scales
up to 4\arcsec. 
Hence, the extended sources with complex structures tend to be resolved and decomposed into multiple compact components, presented as group sources in clusters. 
To distinguish between adjacent but unrelated sources (top panel of Fig.\,\ref{fig:cluster_sources}) and resolved sources, we manually inspected each of the clustered sources. 
About 155 clustered members are associated with 56 clusters that are likely to trace resolved/fragmented sources, as shown in the middle-panel of Fig.\,\ref{fig:cluster_sources}. 
All the clustered sources that trace resolved sources are listed in Table\,\ref{tab:over_resolved_source}.
Some clustered sources show one or two semi-ring-like structures around the central compact bright component, suggesting that a single source that is fragmented as seen in the bottom panel of Fig\,\ref{fig:cluster_sources} and Figure 3 of \citet{Dzib2023AA670A9D}. 
The semi-ring-like fragments and the associated central compact component are regarded as one single radio source. 
It should be noted that the flux density is trustworthy only for the detected compact sources, and the flux of the extended sources, especially the fragmented cluster sources are underestimated due to the lack of short baselines. 
In the future, the combined B$+$D images from the GLOSTAR survey will be used for the reliable measurement of both compact and extended emissions of these resolved or fragmented sources.
The imaging of the  combined D+B configuration data of the entire survey is ongoing and will be published in a subsequent paper. 

\section{Results}

\label{sect:results}
We found 5497 sources above a $7\,\sigma$ threshold in the region ($2\degr<\ell<28\degr, 36\degr<\ell<40\degr\,\&\,56\degr<\ell<60$) of GLOSTAR in B-configuration. 
We visually inspected the $7\,\sigma$ detections and exclude 18 artifacts and 42 semi-ring-like fragmented components (see Sect.\,\ref{sect:cluster_source}), giving a final catalog of 5437 sources. 
Among these 5437 sources, 251 are likely to be \hii\ regions, 784 are candidates of radio stars, 282 are PNe candidates, and 4080 are likely to have extragalactic origins. The remaining 40 that cannot be classified into any of the above four types are regarded as ``Other'', including 11 PDRs and 29 ``unclear'', as mentioned in Sect.\,\ref{sect:classification} and listed in Table\,\ref{tab:total_7sigma}. 

\subsection{Catalog description}
\label{sect:catalog_description}
The catalog contains 17 columns for each source, as presented in Table\,\ref{tab:total_7sigma}. 
Columns 1-8 are determined or derived by the source finder tool BLOBCAT as described in Sect.\,\ref{sect:source_extract}, and correspond to the  Galactic name of the GLOSTAR source,  Galactic longitude $\ell$ and latitude $b$, the signal-to-noise ratio (S/N), peak flux density $S_{\rm peak}$ and its uncertainty $\Delta S_{\rm peak}$, integrated flux density $S_{\rm int}$ and its uncertainty $\Delta S_{\rm int}$. 
The $Y_{\rm factor}$ (defined as the ratio between the integrated flux density
and the peak flux density, $Y_{\rm factor}$=$S_{\rm int}/S_{\rm peak} $) is listed in Column 9.
The source effective radius $R_{\rm eff}$, determined by the pixel size and the total number of pixels composing the source as outlined in Sect.\,\ref{sect:source_extract}, and the spectral index $\alpha$,  obtained by fitting the peak flux densities of the sub-band images (see Set.\,\ref{sect:spetral_index}), are presented in Columns 10-11. 
Column 12 gives the corresponding D-configuration name from the GLOSTAR survey \citep[][Medina et al., in prep.]{Medina2019AA627A175M}, which includes cluster sources as discussed in Sect.\,\ref{sect:cluster_source}. 
The information about the presence or absence of counterparts in the NIR, MIR, FIR and Submm (see Sect.\,\ref{sect:classification}) are displayed in Columns 13-16.
Column 17 gives the source classification based on the multiwavelength properties as outlined in Sect.\,\ref{sect:classification}. 
All the columns are displayed in Table\,\ref{tab:total_7sigma} for a small portion of the catalog, with full catalog being available at CDS and the GLOSTAR website \footref{glostar_web}. 

\setlength{\tabcolsep}{3.2pt}
\begin{table*}
\centering
\caption[]{ \it \rm GLOSTAR B-configuration catalog for $2\degr<\ell<28\degr, 36\degr<\ell<40 \,\& \,56\degr<\ell<60\degr $ and $|b| < 1\degr$. }
 \scriptsize
 \begin{tabular}{p{2.0cm}p{0.9cm}p{0.9cm}p{0.6cm}p{0.5cm}p{0.6cm}p{0.6cm}p{0.6cm}p{0.6cm}p{0.6cm}p{1.3cm}p{2.2cm}p{0.5cm}p{0.5cm}p{0.5cm}p{1.0cm}p{0.7cm}}
\hline
\hline
GLOSTAR B-conf. & $\ell$ & $b$ & $\rm S/N$ & $S_{\rm peak}$ & $\Delta S_{\rm peak}$ & $S_{\rm int}$ & $\Delta S_{\rm int}$ & $Y_{\rm factor}$ & $R_{\rm eff}$ & $\alpha\pm \Delta\alpha$ & GLOSTAR D-conf. & \multicolumn{3}{c}{Infrared counterpart} & Submm & Class \\
Gname & $\degr$ & $\degr$ &   & \multicolumn{2}{c}{$mJy/beam$}    & \multicolumn{2}{c}{$mJy$} &   & $\arcsec$ &   & Gname &  NIR &   MIR &   FIR & counterpart  &  \\
\hline
(1) & (2) & (3) & (4) & (5) & (6) & (7) & (8) & (9) & (10) & (11) & (12) & (13) & (14) & (15) & (16) & (17)  \\
\hline
G002.0122+00.7438 & 2.01218 & 0.74384 & 32.6 & 5.73 & 0.36 & 8.19 & 0.45 & 1.43 & 1.1 & -0.56$\pm$0.23 & G002.012+00.744 &  $\checkmark$  &  $\checkmark$  &  $\checkmark$  & $-$ & PN  \\
G002.0174+00.6687 & 2.01743 & 0.66867 & 9.1 & 1.13 & 0.14 & 0.94 & 0.13 & 0.83 & 0.6 & -0.25$\pm$0.49 & G002.018+00.669 & $-$ &  $\checkmark$  & $-$ & $-$ & Unclear \\
G002.0217+00.0011 & 2.0217 & 0.00111 & 8.0 & 0.93 & 0.13 & 0.78 & 0.12 & 0.84 & 0.6 & 0.18$\pm$0.37 & G002.022+00.001 &  $\checkmark$  & $-$ & $-$ & $-$ & EgC  \\
$\vdots$ & $\vdots$ & $\vdots$ & $\vdots$ & $\vdots$ & $\vdots$ & $\vdots$ & $\vdots$ & $\vdots$ & $\vdots$ & $\vdots$ & $\vdots$ & $\vdots$ & $\vdots$ & $\vdots$ & $\vdots$ & $\vdots$  \\
G025.6392+00.5310 & 25.63925 & 0.53101 & 49.6 & 4.35 & 0.25 & 5.46 & 0.29 & 1.26 & 1.2 & -1.40$\pm$0.11 & G025.639+00.531 & $-$ & $-$ & $-$ & $-$ & EgC  \\
G025.6497+01.0495 & 25.64966 & 1.04951 & 10.0 & 2.35 & 0.27 & 2.57 & 0.27 & 1.09 & 0.8 & -0.23$\pm$1.00 & $-$ &  $\checkmark$  &  $\checkmark$  &  $\checkmark$  & $-$ & HII  \\
G025.6523+00.7887 & 25.65232 & 0.78866 & 7.7 & 0.53 & 0.08 & 0.52 & 0.07 & 0.98 & 0.6 & 0.24$\pm$0.57 & G025.652+00.789 & $-$ & $-$ & $-$ & $-$ & EgC  \\
$\vdots$ & $\vdots$ & $\vdots$ & $\vdots$ & $\vdots$ & $\vdots$ & $\vdots$ & $\vdots$ & $\vdots$ & $\vdots$ & $\vdots$ & $\vdots$ & $\vdots$ & $\vdots$ & $\vdots$ & $\vdots$ & $\vdots$  \\
G059.9412-00.0416 & 59.94116 & $-$0.04161 & 8.1 & 0.86 & 0.12 & 1.32 & 0.12 & 1.54 & 0.8 & 0.05$\pm$0.58 & G059.941-00.042 &  $\checkmark$  & $-$ & $-$ & $-$ & EgC \\
G059.9665+00.0971 & 59.96647 & 0.0971 & 7.2 & 1.17 & 0.17 & 0.65 & 0.17 & 0.55 & 0.5 & 1.46$\pm$0.58 & $-$ & $-$ & $-$ & $-$ & $-$ & EgC  \\
G059.9697+00.5393 & 59.96966 & 0.53934 & 7.4 & 1.38 & 0.2 & 1.35 & 0.2 & 0.98 & 0.6 & -0.41$\pm$0.90 & G059.970+00.540 & $-$ & $-$ & $-$ & $-$ & EgC \\

\hline
\hline
\end{tabular}
\tablefoot
{The description of each column is displayed in Sect.\,\ref{sect:catalog_description}.
Please noted that the flux measurements might be unreliable for sources which have $S_{\rm int}<S_{\rm peak}$, that is, $Y_{\rm factor}=S_{\rm int}/S_{\rm peak}<1.0$, and please see Sect.\,\ref{sect:peak_int_flux} for details.  
Column (12) lists the corresponding D-configuration counterparts (Medina et al., In. Prep.), which indicates the cluster sources as discussed in Sect.\,\ref{sect:cluster_source}. 
Symbol $\checkmark$ in Column 13-16 indicates that the sources show counterparts and/or emission at NIR\,(i.e., the UKIDSS and/or GLIMPSE surveys), MIR\,(i.e., the WISE and/or MIPSGAL surveys), FIR\,(i.e., the Hi-GAL survey), Submm\,(i.e., the ATLASGAL survey) as discussed in Sect.\,\ref{sect:classification}. 
Symbol $-$ refers to no association of counterpart and/or emission. 
Column (17) for source classification as described in Sect.\,\ref{sect:classification}: EgC = Extragalactic source candidate, HII = \hii\ region candidate, Radio star, PN = planetary nebula candidate, Other = source that cannot be classified as one of the above four types, such as PDR (Photodissociation region) or source with no clear classification. 
 Only a small portion of the data is provided here. 
 And the full version will be available in electronic form at the CDS via anonymous ftp to cdsarc.cds.unistra.fr (130.79.128.5)
or via https://cdsarc.cds.unistra.fr/cgi-bin/qcat?J/A+A/. }
\label{tab:total_7sigma}
\end{table*}

\setlength{\tabcolsep}{1.6pt}
\begin{table}
\centering
\caption[]{ \it \rm GLOSTAR D-configuration sources that are detected as fragmented sources (e.g.,  multiple components or over-resolved of an extended
source) in the B-configuration images. }
 \footnotesize
 \begin{tabular}{p{2.7cm}p{2.7cm}p{1.1cm}p{1.1cm}p{0.6cm}}
\hline
\hline
GLOSTAR B$-$conf.  & GLOSTAR D$-$conf. & Num. & $S_{int}$ & Class \\
Gname &  Gname & frags. & $\rm(mJy)$ &  \\
(1) &  (2) & (3) & (4) &  \\
G003.3511$-$00.0774 & G003.350$-$00.077 & 3 & 100.98 & HII \\
G005.4751$-$00.2430 & G005.476$-$00.244 & 3 & 15.32 & HII \\
G008.1397$-$00.0271 & G008.140$-$00.027 & 2 & 4.88 & HII \\
G008.6693$-$00.3560 & G008.669$-$00.356 & 3 & 231.16 & HII \\
$\vdots$ & $\vdots$ & $\vdots$ & $\vdots$ & $\vdots$ \\
G027.1858$-$00.0817 & G027.186$-$00.081 & 2 & 7.76 & HII \\
G027.2799$+$00.1446 & G027.280$+$00.144 & 4 & 51.44 & HII \\
G027.7018$+$00.7040 & G027.701$+$00.705 & 2 & 6.98 & PN \\
G038.8757$+$00.3080 & G038.876$+$00.308 & 2 & 52.67 & HII \\
G056.6162$+$00.1707 & G056.616$+$00.171 & 3 & 11.25 & Egc \\
\hline
\hline
\end{tabular}
\tablefoot{Column (1) shows the GLOSTAR B-configuration name of the brightest member in the clusters. 
Column (2) shows the GLOSTAR D-configuration name for the over-resolved/fragmented clusters, as shown in the middle panel of Fig.\,\ref{fig:cluster_sources}.  
Cols(3)-(4) show the number of fragments and the integrated fluxes of the brightest component. 
Column\,(5) shows the source type as classified in Sect.\,\ref{sect:classification}. 
Other properties can be found in Table\,\ref{tab:total_7sigma}.
 Only a small portion of the data is provided here. The full table is available in electronic form at the CDS via anonymous ftp to cdsarc.cds.unistra.fr (130.79.128.5)
or via https://cdsarc.cds.unistra.fr/cgi-bin/qcat?J/A+A/. }
\label{tab:over_resolved_source}
\end{table}

\subsection{Galactic distribution}
\label{gal_distribution}
\begin{figure}
 \centering
    \includegraphics[width = 0.45\textwidth]{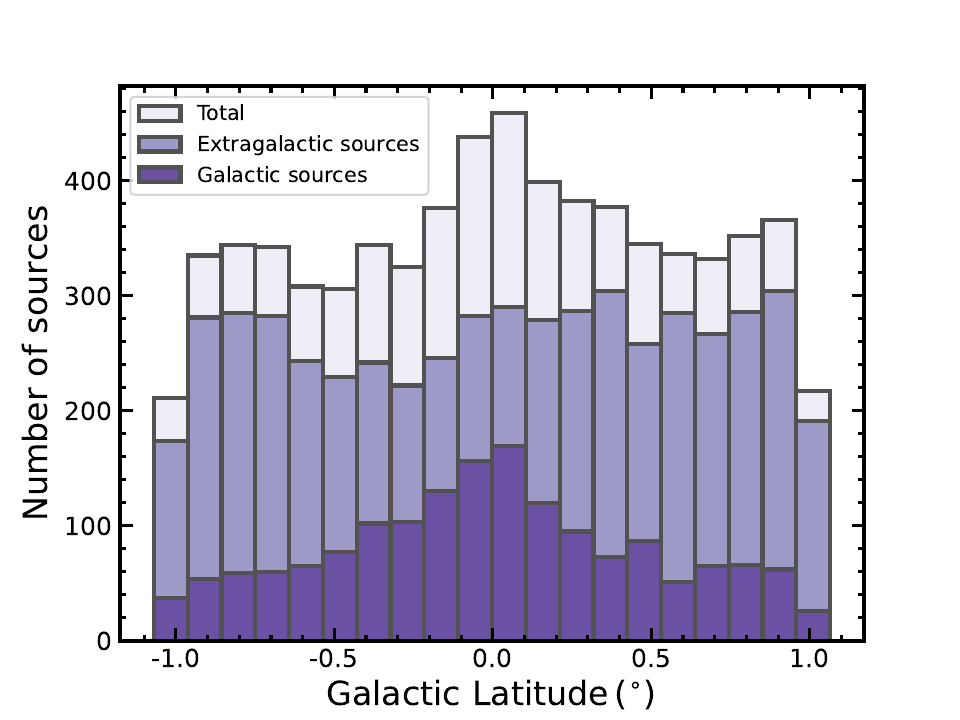} \\
    \includegraphics[width = 0.45\textwidth]{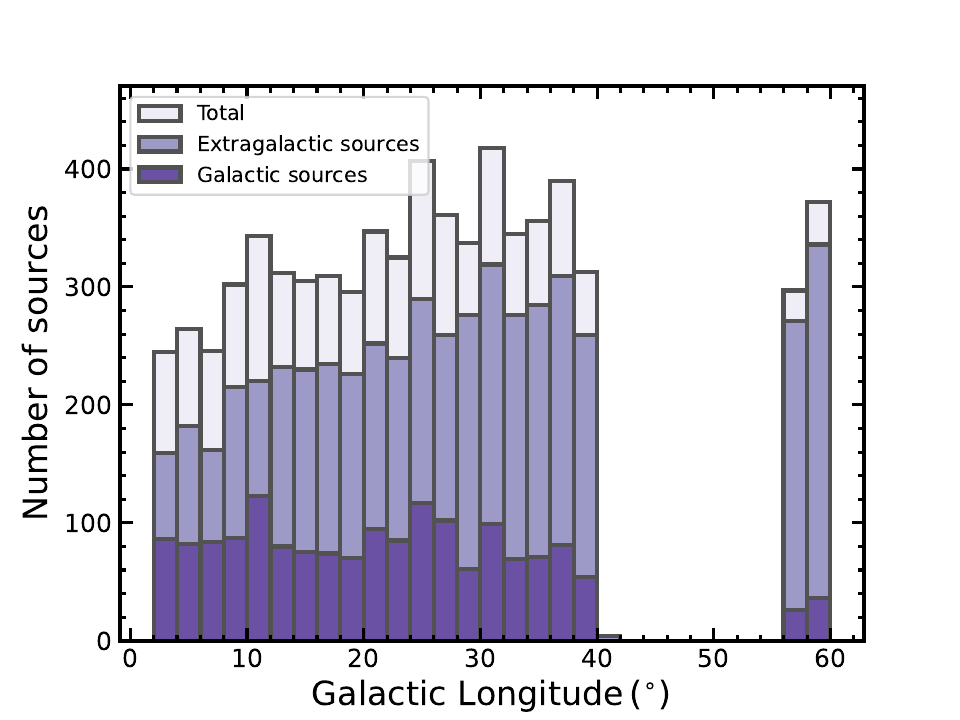} \\
      \caption{Distribution of 6894 GLOSTAR B-configuration $7\,\sigma-$threshold sources as a function of Galactic Latitude (upper panel) and Longitude (lower panel) for the B-configuration catalog, including the catalog of this work ($2\degr<\ell<28\degr\, 36\degr<\ell<40\&\,56\degr<\ell<60$ and $|b| < 1\degr$), as listed in Table\,\ref{tab:total_7sigma} and the published pilot catalog ($28\degr<\ell<36\degr$ and $|b| < 1\degr$) in \citet{Dzib2023AA670A9D}. The bin sizes are 0.1\degr and 2.0\degr for the upper and lower panel, respectively. The blank region in Galactic longitude $40\degr<\ell<56\degr$ is not covered by the GLOSTAR survey in B-configuration.}
 \label{fig:distr_glon_lat}
 \end{figure}

To show the Galactic distributions for the whole B-configuration catalog, we combined our work with the catalog published in \citet{Dzib2023AA670A9D}. Fig.\,\ref{fig:distr_glon_lat} presents the distributions of GLOSTAR sources as a function of Galactic Longitude ($\ell$) and Latitude ($b$). The full sample, the extragalactic, and Galactic sources are shown in light purple, purple, and dark purple, respectively. 

As shown in the upper-panel of Fig.\,\ref{fig:distr_glon_lat}, the number of sources per 0.1\degr latitude bin increase gradually toward zero latitude for the total sample (light purple), which is due to the expected peak at $b\sim 0.0\degr$ of the Galactic sources (dark purple). 
The extragalactic sources (purple) show a relatively flat distribution with Galactic latitude, which is also seen in \citet{Hoare2012PASP} for the CORNISH survey. 
We note that the distribution of Galactic sources in latitude is slightly asymmetric and skewed toward $b>0\degr$, which is also seen for the Galactic SNR distribution in GLOSTAR \citep{Dokara2021AA651A86D} and in THOR \citep{Anderson2017A&A605A58A}. 
This asymmetry is supported by a Shapiro-Wilk test of the Galactic source distribution in latitude that gives a statistical confidence of $p$-value$\ll$0.001. 
We found that the source counts of \hii\ regions peaks at latitude $b\sim 0.0\degr$, as noted by \citet{Urquhart2013MNRAS435}, while the distributions of radio stars and planetary nebulae are relatively flat along latitude from $-$1.0\degr\ to 1.0\degr. 

The lower-panel of Fig.\,\ref{fig:distr_glon_lat} shows the source counts in 2\degr\ bins of Galactic longitudes. The source counts are seen to have peaks at certain longitudes such as $\ell\approx 10\degr$ and $\ell\approx 26\degr$. These correspond to locations of star formation regions such as W31, SFC1-4 \citep{Thompson2006AA4531003T,Murray2011ApJ729133M} and W42 \citep{Gao2019AA623A105G,Liu2019ApJS24014L,Feng2021PASJ73467F}.
The Kolmogorov$-$Smirnov (K-S) tests between the Galactic sources and the extragalactic sources give a statistical confidence $p$-value $\ll$ 0.001, which confirms that their distributions are significantly different in both Galactic longitude and latitude.
The source count of extragalactic sources is seen to drop in regions with high noise such as $|b|>0.9\degs$ at the survey edges and 2\degr$<\ell<$8\degr\ at low declination, as shown in Fig.\,\ref{fig:rms_noise_map}. 
Considering the high noise at 2\degr$<\ell<$8\degr, the number of extragalactic sources per $2\degr$ longtitude bin still decrease gradually toward longitude zero, which is also seen in the distribution of non-classified sources with extragalactic origin in THOR survey (Figure 13 of  \citealt{Wang2018AA619A124W}). The number of Galactic sources per $2\degr$ longtitude bin increase toward longitude zero, which is also seen for the distributions of the resolved sources with galactic origin in the CORNISH survey (Figure 18 of \citet{Purcell2013ApJS2051P}) and the Galactic sources (PNe and \hii\ regions) in the THOR survey (Figure 13 of  \citealt{Wang2018AA619A124W}).

\subsection{Source Properties}
\begin{table}
\centering
\caption {\rm Summary of source properties of the $7\,\sigma-$threshold catalog. In Columns (2-5) we give the minimum ($x_{min}$), maximum ($x_{max}$), $\rm mean\pm standard\,deviation$ ($x_{mean}\pm x_{std}$), and median values ($x_{med}$) of these parameters for the total sample, the extragalactic sample and the Galactic sample. 
}
\begin{tabular}{lllll}
\hline
\hline
Parameter  &  $x_{min}$ & $x_{max}$ & $x_{mean}\pm x_{std}$ & $x_{med}$\\
\hline
\multicolumn{5}{c}{Total 5437 sources}     \\
\hline
$S_{\rm int}\rm \,(mJy)$ & 0.34 & 1453.77 & $5.41\pm29.87$& 1.26 \\
$S_{\rm peak}\rm \,(mJy)$ & 0.39 & 1175.55 & $4.05\pm22.09$& 1.16 \\
$R_{\rm eff}\rm \,(arcseconds)$ & 0.45 & 2.69 & $0.88\pm0.26$& 0.80 \\
$\alpha$ & $-$2.85 & 2.72 & $-0.54\pm0.69$& $-$0.52 \\
$Y=S_{\rm int}/S_{\rm peak}$ & 0.40 & 7.06 & $1.16\pm0.42$& 1.07 \\

\hline
\multicolumn{5}{c}{ 4080 Extragalactic sources}     \\
\hline
$S_{\rm int}\rm \,(mJy)$ & 0.39 & 1453.77 & $4.41\pm29.06$& 1.20 \\
$S_{\rm peak}\rm \,(mJy)$ & 0.39 & 1175.55 & $3.83\pm24.36$& 1.11 \\
$R_{\rm eff}\rm \,(arcseconds)$ & 0.45 & 2.65 & $0.86\pm0.23$& 0.80 \\
$\alpha$ & $-$2.85 & 2.72 & $-0.58\pm0.67$& $-$0.56 \\
$Y=S_{\rm int}/S_{\rm peak}$ & 0.40 & 4.98 & $1.11\pm0.28$& 1.06 \\

\hline
\multicolumn{5}{c}{ 1357 Galactic sources}     \\
\hline
$S_{\rm int}\rm \,(mJy)$ & 0.34 & 553.56 & $8.43\pm32.00$& 1.49 \\
$S_{\rm peak}\rm \,(mJy)$ & 0.39 & 249.90 & $4.73\pm13.07$& 1.31 \\
$R_{\rm eff}\rm \,(arcseconds)$ & 0.51 & 2.69 & $0.94\pm0.32$& 0.83 \\
$\alpha$ & $-$2.75 & 2.49 & $-0.44\pm0.75$& $-$0.41 \\
$Y=S_{\rm int}/S_{\rm peak}$ & 0.49 & 7.06 & $1.32\pm0.68$& 1.11 \\

\hline
\hline
\end{tabular}
\label{tab:summary_phy_param}
\end{table} 
In this section, we present the source properties for the high reliability ($7\,\sigma-$threshold) catalog.
Table\,\ref{tab:summary_phy_param} displays a statistical summary of the source properties for the total catalog in Table\,\ref{tab:total_7sigma}, the extragalactic sources, and the Galactic sources. 
The typical values of flux density, effective radius, and spectral index of the sources in the catalog are $S_{\rm int}\sim \rm1.0\,mJy$, $R_{\rm eff}\sim 0\farcs8$, and $\alpha\,\sim -0.52$, indicating that the detected sources are typically compact and weak, and are dominated by extragalactic sources with non-thermal emission. 

\subsubsection{Source effective size}
\label{sect:source_size}

\begin{figure}
 \centering
    \includegraphics[width = 0.45\textwidth]{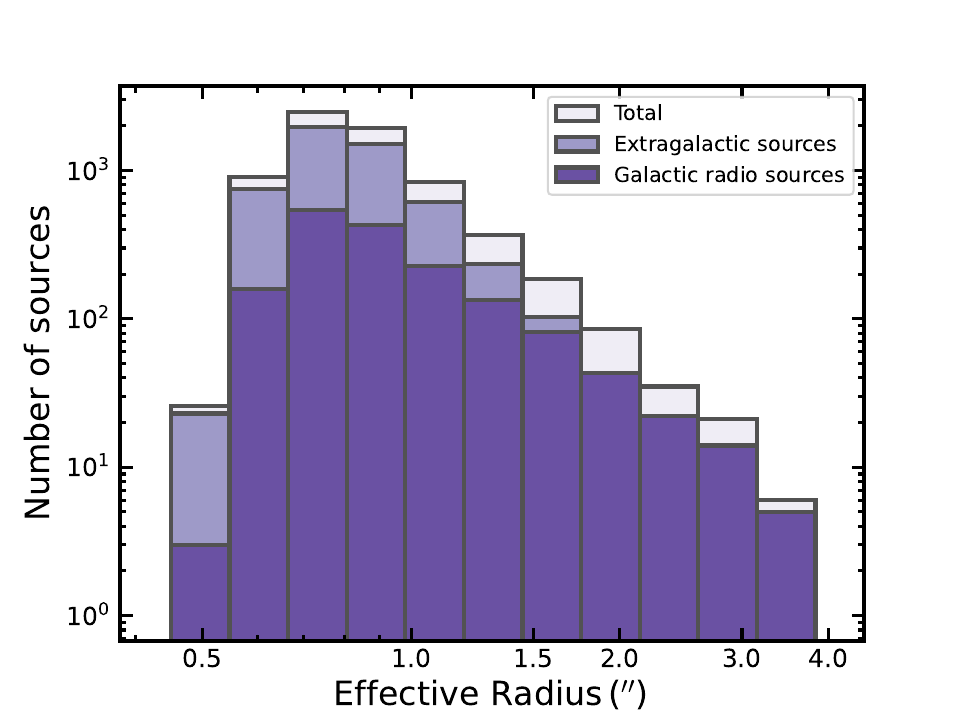} 
      \caption{Distribution of source effective radius for the 6894 sources in the $7\,\sigma-$threshold catalog of this work (Table\,\ref{tab:total_7sigma}) and the published pilot catalog in \citet{Dzib2023AA670A9D}}. 
 \label{fig:distr_angle_size}
 \end{figure}

Fig.\,\ref{fig:distr_angle_size} shows the source counts as a function of source effective sizes (see Sect\,\ref{sect:source_extract}) for the high reliability catalog of GLOSTAR B-configuration (light purple), with the extragalactic and Galactic sources being shown in purple and dark purple respectively. 
As shown in Table\,\ref{tab:summary_phy_param},  the effective radius of the sample ranges from 0.45\arcsec\ to 2.69\arcsec, with a mean and median value of 0.86\arcsec\  and 0.80\arcsec\ respectively. 
As expected, the Galactic sources have larger sizes with a higher mean $R_{\rm eff}$ compared to the extragalactic sources, and the two distributions are significantly different as suggested by the K-S test ($p$-value$\ll$0.001). 

The effective source sizes are calculated from the total number of pixels comprising each source obtained from the BLOBCAT software (see Sect.\,\ref{sect:source_extract}).  
As discussed for the THOR survey \citep{Bihr2016AA588A97B,Wang2018AA619A124W}, the effective radius is not a good parameter to distinguish between resolved and unresolved sources.
Hence, we use the parameter $Y_{\rm factor}$\,(i.e., $S_{\rm int}/S_{\rm peak}$ in Table\,\ref{tab:total_7sigma}), namely, the ratio between the integrated (in units of mJy) and peak flux density (in units of mJy/beam), to divide sources into three subsamples: extended sources ($Y_{\rm factor}>2.0$), compact sources ($1.1<Y_{\rm factor}\,\leq\,2.0$) and unresolved/point-like sources ($Y_{\rm factor}\,\leq 1.1$). This is identical to what was used in the pilot region of the GLOSTAR B-configuration \citep{Dzib2023AA670A9D} and the D-configuration \citep{Medina2019AA627A175M}, as well as in previous work \citep[e.g.,][]{Bihr2016AA588A97B, Yang2019MNRAS4822681Y}. 
As expected, the majority of the detected sources (97\%=5255/5437 with $Y_{\rm factor}\,\leq\,2$) are compact (2053 with $1.1<Y_{\rm factor}\,\leq\,2.0$) and unresolved/point-like sources (3202 with $Y_{\rm factor}\,\leq\,1.1$), due to the fact that the images are optimized for detection of compact emission (see Sect.\ref{sect:calibration_and_image}). 
In the above three categories of unresolved, compact, and extended sources, 79\%, 73\%, and 32\% of them are extragalactic sources candidates, respectively.
About 98.6\% of the extragalactic sources and 91\% of the Galactic sources are compact and unresolved with $ Y_{\rm factor}\,\leq 2.0$. 
The remaining 1.4\% of extragalactic sources are extended with $Y_{\rm factor}>2.0$, which are expected to be associated with radio galaxy lobes, while the extended Galactic sources (9\% of the Galactic sources) are found to be preferentially associated with \hii\ regions and PNe. 
%

 \subsubsection{Peak and integrated flux density}
\label{sect:peak_int_flux}
\begin{figure}
 \centering
    \includegraphics[width = 0.45\textwidth]{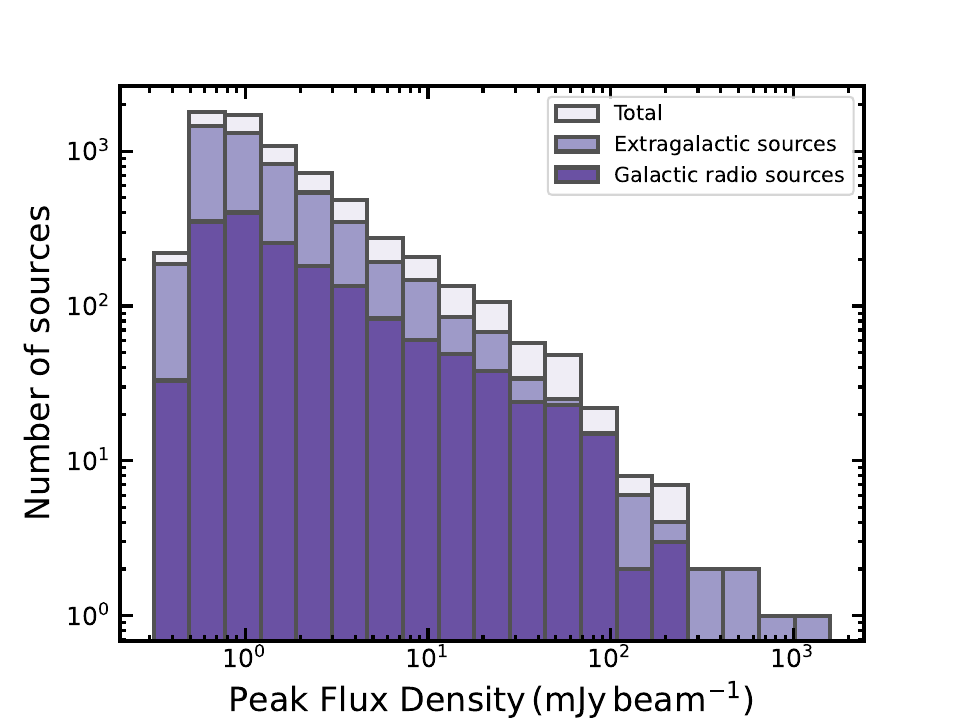} \\
    \includegraphics[width = 0.45\textwidth]{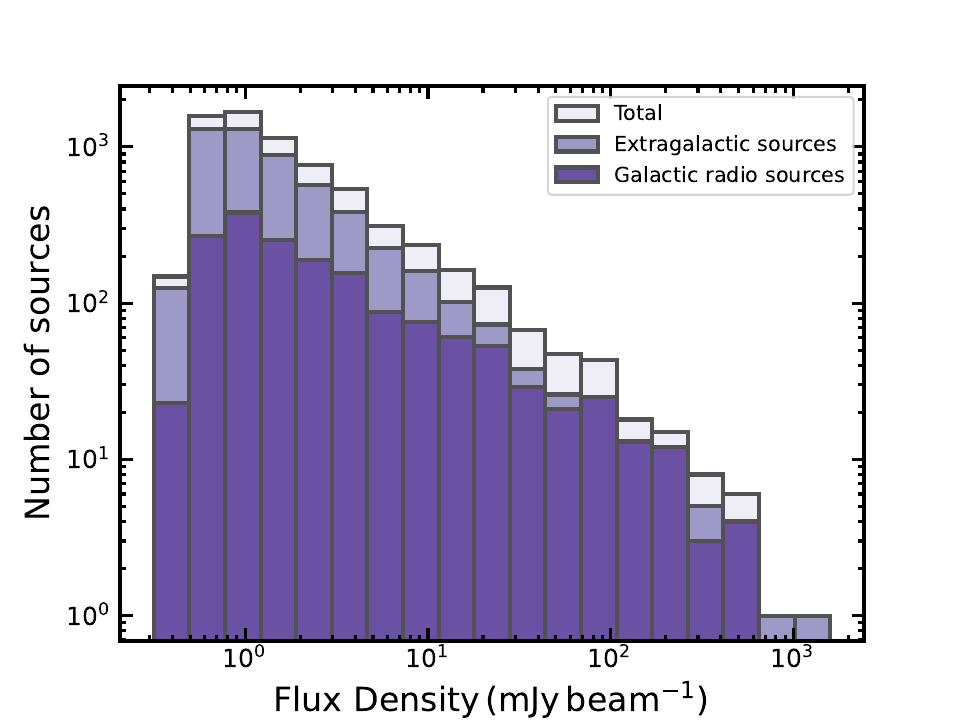} \\ 
  \caption{Distributions of peak (upper panel) and integrated (lower panel) flux density for the 6894 $7\,\sigma-$threshold sources of GLOSTAR B-configuration, including Table\,\ref{tab:total_7sigma}  and the published pilot catalog in \citet{Dzib2023AA670A9D}}.  
 \label{fig:distr_peak_int}
 \end{figure}

The peak and integrated flux densities are obtained as outputs from the source extraction tool, BLOBCAT (see Sect.\,\ref{sect:source_extract}).
Fig.\,\ref{fig:distr_peak_int} shows the distributions of peak and integrated flux density for the high reliability catalog (light purple), the extragalactic sources (purple), and the Galactic sources (dark purple).  
The decline in the source count for flux densities below 0.65 mJy is due to the non-uniform noise distribution and the resulting variation in the $7\,\sigma$ detection limit over the survey region. 

It is to be noted that there is a significant population of unresolved sources with $Y_{\rm factor}<1.0$, which means that their integrated flux densities ($S_{\rm int}$) are lower than their peak intensities ($S_{\rm peak}$). While for most of the sources, the integrated flux densities and the peak intensities are still consistent within 3$\sigma$, there are a few sources where the discrepancy is significant. This could happen when the unresolved sources are (1) not fitted with enough pixels by BLOBCAT (i.e., fitted area less than beam); (2) located in negative side lobes from nearby bright sources; and (3) not cleaned properly. These are also seen in other survey catalogs generated using BLOBCAT, such as that of THOR \citep{Bihr2016AA588A97B,Wang2018AA619A124W}  and the GLOSTAR pilot region \citep{Dzib2023AA670A9D, Medina2019AA627A175M}. Therefore, for these unresolved sources with a $Y_{\rm factor}<1.0$, we used their peak flux densities for further analysis discussed in the following sections. 

\subsubsection{Spectral index}
\label{sect:secptral_index_results}
\begin{figure}
 \centering
    \includegraphics[width = 0.45\textwidth]{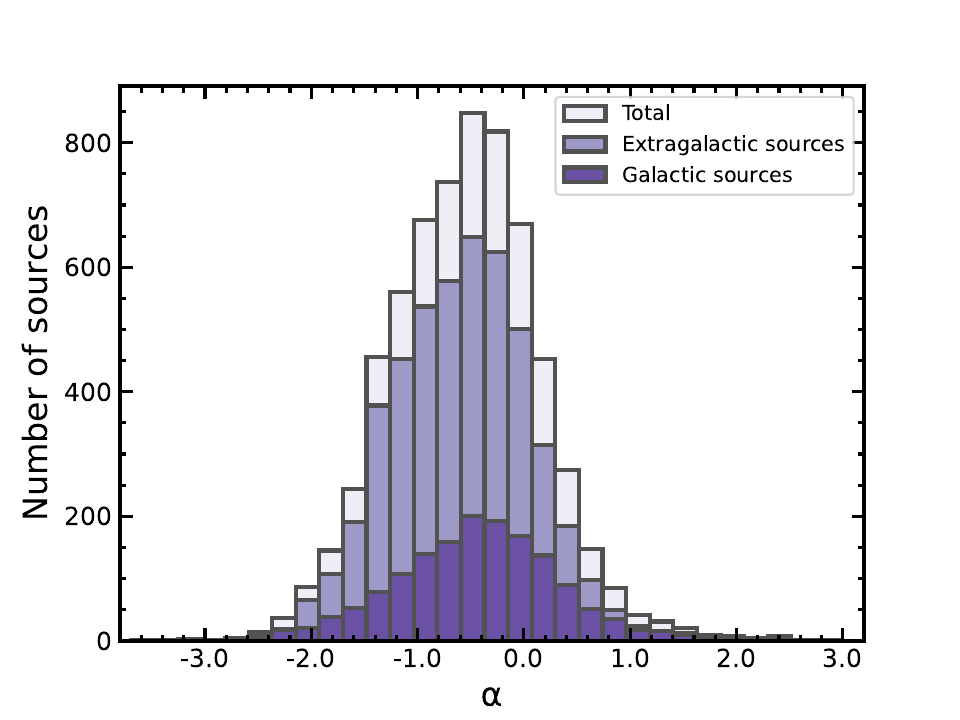} 
      \caption{Distribution of spectral index $\alpha$ for the $7\,\sigma-$threshold catalog of the GLOSTAR B-configuration, including Table\,\ref{tab:total_7sigma}  and the published pilot catalog in \citet{Dzib2023AA670A9D}}.
 \label{fig:distr_alpha}
 \end{figure}

Fig.\,\ref{fig:distr_alpha} shows the distribution of spectral index for the full high reliability sample of this work (light purple), extragalactic (purple), and Galactic sources (dark purple).  
The measured in-band spectral index of this work is consistent with that measured in THOR for compact sources that are detected in both surveys, as discussed later in Sect.\,\ref{sect:compare_thor}.
The K-S test of the spectral index between the Galactic sources and the extragalactic sources ($p$-value$\ll$0.001) suggests that they are significantly different. 

The spectral index is a common and useful tool for distinguishing between thermal and non-thermal radio emission, broadly corresponding to positive and negative spectral indices, respectively. 
The catalog is dominated by non-thermal radiation, as it includes 74\% of sources with $\alpha\,< -0.1$.  
The spectral index of extragalactic sources (purple) and the total sample (light purple) peaks at $\alpha\sim -0.6$, as seen in the pilot paper \citep{Dzib2023AA670A9D}. 
Among the 4080 extragalactic sources, about 77\% have negative spectral indices with $\alpha\,< -0.1$, while 23\% show spectral indices from thermal emission with $\alpha\,> -0.1$. 
The spectral index of Galactic sources (dark purple in Fig.\,\ref{fig:distr_alpha}) peaks at $\sim -0.2$, with 67\% having $\alpha<-0.1$ (mainly radio stars) and the remaining 33\% having $\alpha>-0.1$ (mainly \hii\ regions and planetary nebulae).
This shows that the radio emission from both Galactic and extragalactic sources is dominated by non-thermal radiation. 
It is to be noted that the uncertainty in the spectral index is strongly correlated with the logarithm of single-to-noise ratio (S/N), with a correlation coefficient of $\rho=$-0.89 and $p$-value $\ll$ 0.001. 
Hence, the in-band spectral indices of weak sources just above the detection limit of 7$\sigma$ are less reliable, with there being some cases of two adjacent pixels having very different spectral indices. A similar observation was made by \citet{Rosero2016ApJS22725R} for weak and compact radio emission in high-mass star-forming regions.

\section{Discussion}

\label{sect:discussion}
\subsection{Comparison with other radio surveys}
 The comparison between GLOSTAR and other radio surveys allows us to discuss the consistency of peak and integrated flux densities, and positions of the detected sources. 
 Due to the differences in spatial filtering, angular resolution, and $uv$ coverage, the comparisons of radio properties are valid for compact sources detected in these surveys. 
 In this section, we compare the properties of compact sources in our catalog with other surveys such as CORNISH \citep[e.g.,][]{Hoare2012PASP,Purcell2013ApJS2051P}, MAGPIS \citep[e.g.,][]{White2005AJ,Helfand2006AJ}, and THOR \citep[e.g.,][]{Bihr2016AA588A97B,Beuther2016AA595A32B,Wang2018AA619A124W}. 

\subsubsection{The CORNISH survey}
\label{sect:compare_cornish}

 \begin{figure}
 \centering
    \includegraphics[width = 0.45\textwidth]{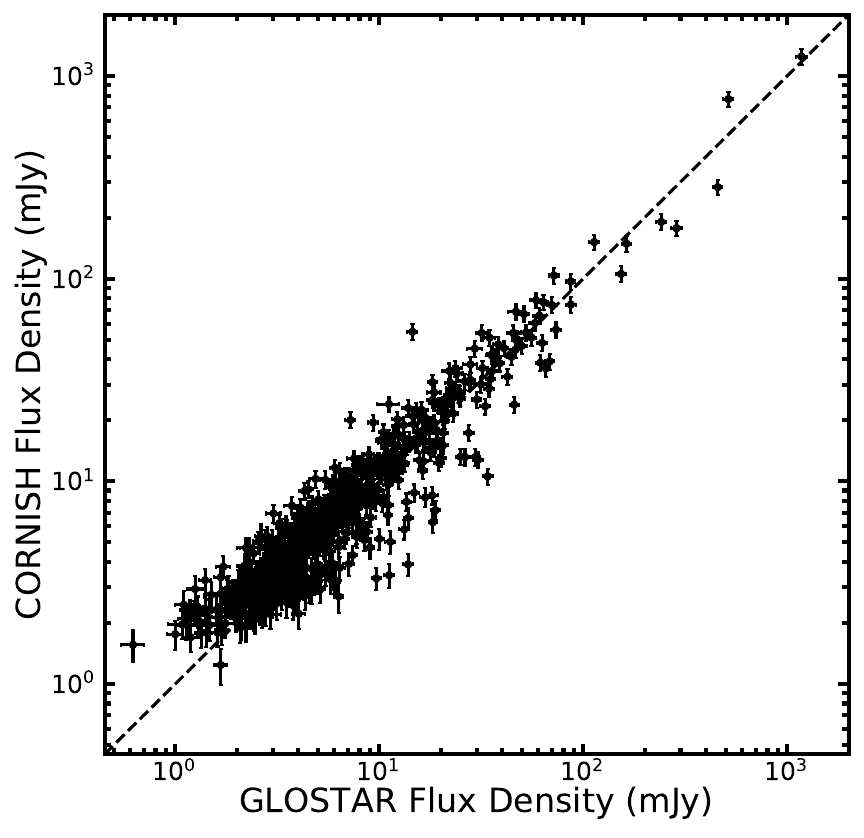} \\
 \includegraphics[width = 0.45\textwidth]{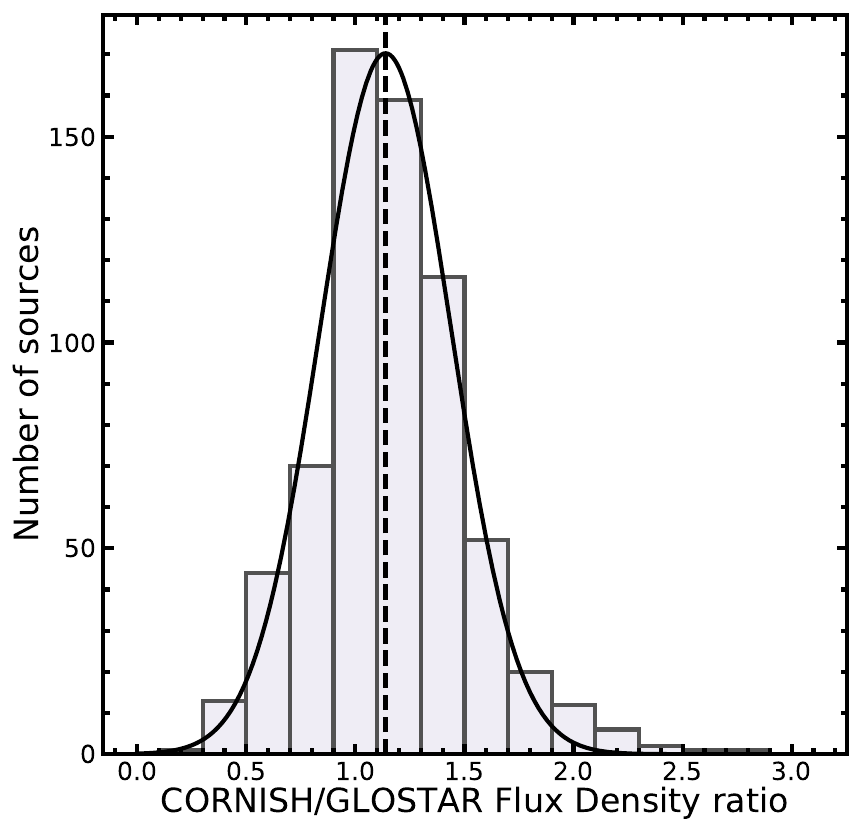} \\
  \caption{Comparison of flux densities between GLOSTAR and CORNISH. Top panel: The comparison of flux densities for 669 compact sources detected by both CORNISH and GLOSTAR catalogs. The error bar of each point shows the uncertainty of flux.
  Bottom panel: The histogram of the flux density ratios of compact sources between CORNISH and GLOSTAR. The black line displays the Gaussian fit to the histogram. The dashed line presents the mean value of the distribution, with mean and standard deviation of the flux ratio from the Gaussian fit are 1.14 and 0.3, respectively. }
 \label{fig:flux_position_cornish_glostar}
 \end{figure}
The CORNISH survey has the most similar observation setup and sky coverage to the B-configuration of the GLOSTAR survey, making it an excellent resource for a comprehensive inspection of the source properties in this study.
CORNISH \citep{Hoare2012PASP,Purcell2013ApJS2051P} used the VLA in B and BnA configurations at 5 GHz to conduct a Galactic plane survey from $10\degr<\ell<65\degr$ and $|b|<1\degr$, with a resolution of 1.5\arcsec\, and a median RMS noise of $\rm \sim\,0.4\,mJy\,beam^{-1}$. A total of 2638 high-reliability CORNISH sources are detected above the $7\sigma$ limit ($\sim \rm 2.5\,mJy\,beam^{-1}$), containing extended emission on scales no greater than 14\arcsec. 
Compared to CORNISH, GLOSTAR in B-configuration has a similar angular resolution of 1.0\arcsec\ and better sensitivity of $\rm \sim\,0.08\,mJy\,beam^{-1}$ (i.e., the $\rm 7\sigma$ detection limit $\sim \rm 0.56\,mJy\,beam^{-1}$), but a poorer sampling of extended emission with the images being mostly insensitive to emission on scales $>$4\arcsec\,(see Sect.\,\ref{sect:calibration_and_image}). Despite their differences in the $uv$ coverage, the properties of compact and unresolved sources are expected to be common in both catalogs. 
The comparison between GLOSTAR and CORNISH for the pilot region ($28\degr<\ell<36\degr$, $|b|<1\degr$) has discussed in \citet{Dzib2023AA670A9D}.

Within the overlap region of this work ($10\degr<\ell<28\degr$, $36\degr<\ell<40\degr$,  $56\degr<\ell<60\degr$, and $|b|<1\degr$), CORNISH detected 1210 high-reliability sources above 7$\sigma$, including 742 compact sources (defined as sources with angular sizes$\leq$1.8\arcsec\ in \citealt{Purcell2013ApJS2051P}). 
In the overlap region, GLOSTAR detected 4381 sources above 7$\sigma$, highlighting the improvement in sensitivity.
Among these 742 compact CORNISH sources, we find a match of 669 GLOSTAR sources, using a circular matching threshold of 1.8\arcsec, giving a match rate of $\sim 90\%$. The match rate is similar (90\%=259/290) even if we only consider sources that are unresolved by CORNISH (i.e., sizes $\leq 1\farcs5$).  
In spite of the improved sensitivity of the GLOSTAR survey, there are 73 sources that are detected by CORNISH but not by GLOSTAR. An examination of the location of these sources reveals that about 54\% (40/73) are located at the survey edges with $|b|>1.0$ that are not covered by GLOSTAR. 
The remaining sources are likely to be variable radio sources as discussed later in Sect.\,\ref{sect:variable_sources}. 
The catalog discrepancies between CORNISH and GLOSTAR have also been discussed in the GLOSTAR pilot papers by \citet{Dzib2023AA670A9D} and \citet{Medina2019AA627A175M}.

Fig.\,\ref{fig:flux_position_cornish_glostar} shows the comparison of flux densities of CORNISH and GLOSTAR for the 669 compact sources that have been detected in both surveys. We measured a mean$\pm$ standard deviation of the flux ratio of $1.14 \pm 0.3$, demonstrating a good agreement between the two surveys. 
The consistency of flux measurements between CORNISH and GLOSTAR was also seen in the D-configuration catalog of the pilot region \citep[see Fig. 11 in][]{Medina2019AA627A175M}. 
Because CORNISH (beam=1\farcs5) and GLSOTAR (beam=1\farcs0) have similar angular resolutions, we have also used these compact matching sources to infer the astrometric accuracy of the GLOSTAR survey, indicating a position uncertainty of $\lesssim 0\farcs1$ for this catalog (see Fig.\,\ref{fig:position_offset} and Sect.\,\ref{sect:astrometry} for details).

Among the compact sources detected in both surveys, CORNISH has classified 17 sources as \hii\ regions, 32 as planetary nebulae, and 577 as extragalactic sources. Among these CORNISH sources with types, GLOSTAR found all 17 \hii\ regions and 32 planetary nebulae, as well as 569 extragalactic sources. This gives a classification match rate of 100\% for \hii\ regions and PNe, and 98\% for extragalactic sources. 
This demonstrates a high level of agreement in classification between the two surveys.

\subsubsection{The MAGPIS survey}
\label{sect:compare_magpis}
 \begin{figure}
 \centering
    \includegraphics[width = 0.45\textwidth]{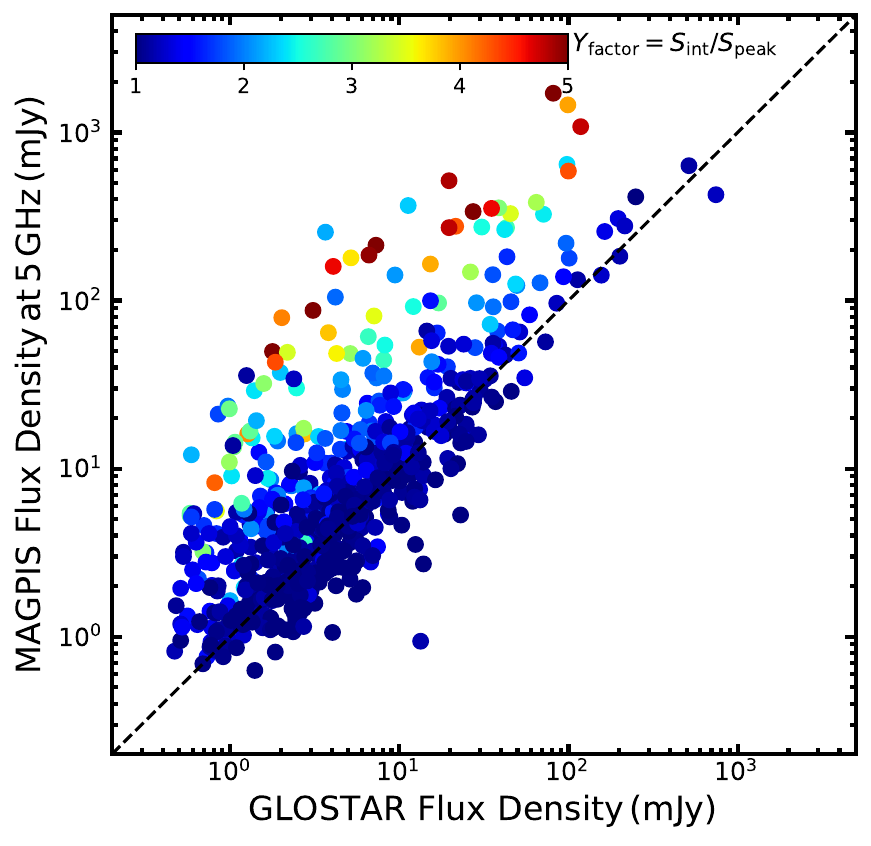} \\
    \includegraphics[width = 0.45\textwidth]{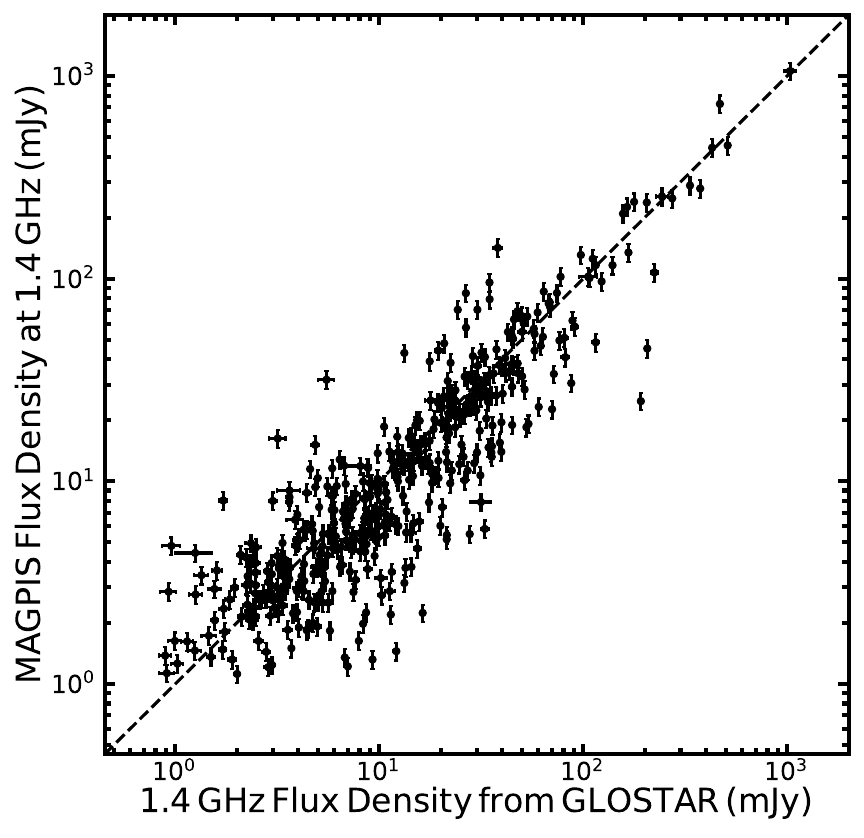} \\
  \caption{Comparison of flux densities between GLOSTAR and MAGPIS. Top panel: The comparison of measured flux densities for 663 MAGPIS compact sources at 5 GHz \citep{White2005AJ} common to GLOSTAR. These compact sources are defined as sources with angular sizes $<6\arcsec$ in MAGPIS. The dashed line means the flux densities are the same in MAGPIS and GLOSTAR. At the top, we show the color bar for the $Y_{\rm factor}$ (defined as $S_{int}/S_{peak}$) of the GLOSTAR detections, indicating the emission of sources in the GLOSTAR image are unresolved (defined as $Y_{\rm factor}<1.1$ in Sect.\,\ref{sect:source_size}), compact ($\rm 1.1<Y_{\rm factor}<2.0$) or extended ($\rm Y_{\rm factor}>2.0$). 
  Bottom panel: The comparison of flux densities of 484 compact MAGPIS at 1.4 GHz that are also detected by GLOSTAR at 5\,GHz. The 1.4 flux densities from GLOSTAR is extrapolated from the 5 GHz flux densities and spectral indices of the GLOSTAR catalog.
  }
 \label{fig:flux_comparison_magpis_glostar}
 \end{figure}

 The Multi-Array Galactic Plane Imaging survey (MAGPIS \footnote{\label{magpis_web} \url{https://third.ucllnl.org/gps/index.html}}) observed the Galactic plane at 5\,GHz between $-10\degr<\ell<42\degr$ and $|b|<0.4\degr$ \citep{White2005AJ} and at 1.4\,GHz between $5\degr<\ell<32\degr$ and $|b|<0.8\degr$ \citep{Helfand2006AJ}, with a resolution of 6\arcsec\ and a median $\rm 1\sigma$ noise of $\sim \rm 0.29\,mJy\,beam^{-1}$. The detection threshold of MAGPIS was chosen to be $\sim 5.5\,\sigma$ ($\sim \rm 1.4\, mJy\,beam^{-1}$) compared to 7$\sigma$($\sim \rm 0.56\, mJy\,beam^{-1}$) for GLOSTAR in this paper. Due to the differences in \emph{uv} coverage, we only examine the properties of compact and unresolved sources in the catalogs of MAGPIS and GLOSTAR B-configuration (including this work and \citet{Dzib2023AA670A9D}). 

For the 5 GHz catalog in \citet{White2005AJ}, within the overlapping region of the GLOSTAR B-configuration ($2\degr<\ell<40\degr$ and $|b|<0.4\degr$), MAGPIS detected 2345 sources above 5.5\,$\sigma$ with angular size measurements, including 935 compact sources \citep[defined as sources with angular sizes $<6\arcsec$ in MAGPIS][]{White2005AJ}. 
In the same region, GLOSTAR detected 6216 sources above 7$\sigma$.
Among the 935 compact MAGPIS sources, we find a match of 684 GLOSTAR sources above 5$\sigma$ using a matching radius of 5\arcsec. The remaining sources that are detected by MAGPIS but not by GLOSTAR are extended with angular sizes greater than 4\arcsec\ (i.e., the largest angular scale structure detected by GLOSTAR in B-configuration).
Due to the differences in the adopted detection thresholds and the \emph{uv} coverage between MAGPIS and GLOSTAR, among the 684 5$\sigma$ matches, 663 are above 7$\sigma$ in the high-reliability catalog of GLOSTAR. 
From the top panel of Fig.\,\ref{fig:flux_comparison_magpis_glostar}, we can see that the measured flux densities of the unresolved sources (i.e., $Y_{\rm factor}<1.1$) in GLOSTAR and MAGPIS are in good agreement, as was also seen in  \citep[][]{Medina2019AA627A175M} for the D-configuration catalog of the pilot region.  
The outliers that have $Y_{\rm factor}<2$ could be from the variable radio source sample such as G031.0777$+$00.1703 in \citet{Dzib2023AA670A9D} which is the outlier point located at the bottom-right of Fig.\,\ref{fig:flux_comparison_magpis_glostar}.
The extended sources with $Y_{\rm factor}>2$ are responsible for the outliers that show higher flux densities in MAGPIS compared to GLOSTAR, which is mainly attributed to differences in \emph{uv} coverage between the two surveys. 

For the 1.4\,GHz catalog in \citet{Helfand2006AJ}, within the overlapping region ($5\degr<\ell<32\degr$ and $|b|<0.8\degr$), MAGPIS detects 3149 sources above $5\,\sigma$, 1153 of which are compact with angular sizes < 6\arcsec sources. Using a matching radius of 5\arcsec, GLOSTAR detects 860 sources above $7\,\sigma$ at 5\,GHz. 
To compare the flux densities at the same observing frequency with MAGPIS 1.4\,GHz, we extrapolated the GLOSTAR 5 GHz flux densities to the 1.4\,GHz flux densities according to the spectral indices of GLOSTAR catalog. To make the 1.4\,GHz extrapolated flux densities reliable, we select the 484 compact sources that have low uncertainties in their spectral indices (i.e., $\sigma_{\alpha}<0.2$, where 0.2 is the typical value of $\sigma_{\alpha}$ as outlined in Sec.\,\ref{sect:spetral_index}). 
The bottom panel of Fig.\,\ref{fig:flux_comparison_magpis_glostar} shows the comparison of 1.4\,GHz flux densities of MAGPIS and GLOSTAR for the 484 compact sources detected by both surveys. 
This suggests that the extrapolated 1.4 GHz flux densities from GLOSTAR agree with the 1.4\,GHz MAGPIS fluxes. 
Considering the differences in \emph{uv} coverage and the observing frequency, the flux measurements of the two surveys are consistent.

\subsubsection{The THOR survey}
\label{sect:compare_thor}
 \begin{figure}
 \centering
    \includegraphics[width = 0.45\textwidth]{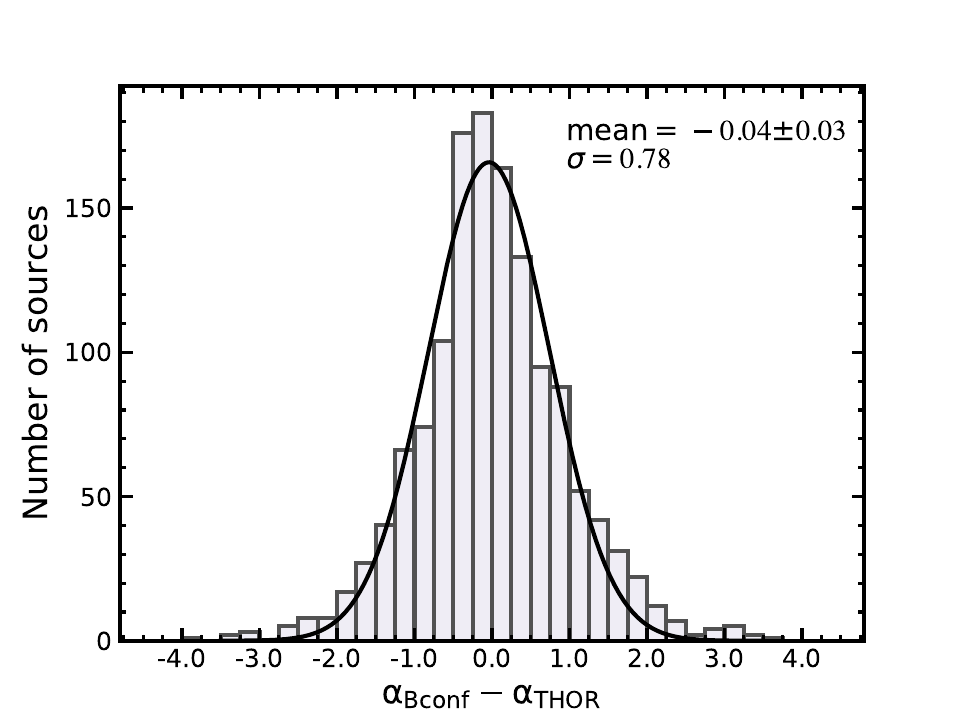} \\
  \caption{The distribution for the difference ($\alpha_{\rm Bconf}-\alpha_{\rm THOR}$) in the measured spectral indices for 1390 compact THOR sources common to GLOSTAR in B-configuration.  
  The black solid line means the Gaussian fit for the distribution, giving a mean value of $-0.04\pm0.03$ (the error of the mean is from the Gaussian fit) and a standard deviation of 0.78. }
 \label{fig:spectral_index_thor_glostar}
 \end{figure}
The THOR survey \citep{Bihr2015AA580A112B,Bihr2016AA588A97B,Beuther2016AA595A32B,Wang2018AA619A124W}) observed HI, OH, recombination lines, and continuum of the Galactic plane between $14.5\degr<\ell<67.4\degr$ and $|b|<1.25\degr$, using the VLA in C-configuration at 1-2\,GHz. 
The continuum images of THOR have a $\rm 1\sigma$ noise level of $\rm 0.3-1.0\,mJy\,beam^{-1}$ and a typical angular resolution of $\sim 25\arcsec$. The detection threshold of THOR is set as 5$\sigma$, and a total catalog of 10387 sources is detected above the threshold. 
Given the large 
differences in the observation setup and \emph{uv} coverage between THOR and GLOSTAR in B-configuration, we can only roughly discuss the similarities and differences between the two surveys. 
The comparison between GLOSTAR and THOR for the pilot region ($28\degr<\ell<36\degr$, $|b|<1\degr$) can be found in \citet{Dzib2023AA670A9D}.

Within the region of overlap between THOR and the GLOSTAR area presented in this work ($14.5\degr<\ell<28\degr$, $36\degr<\ell<40\degr$, $56\degr<\ell<60\degr$ and $|b|<1.0\degr$), 
THOR detected 4083 sources above 5$\sigma$ as reported in \citet{Wang2018AA619A124W}. 
In the overlapped region, GLOSTAR detected 3650 sources above 7$\sigma$ detection level. 
Using a matching radius of 5\arcsec, we found 2363 common sources above 5$\sigma$ and 2001 matches above 7$\sigma$ in the GLOSTAR survey. 
Among these 7$\sigma$ matches, 1764 are regarded as point sources in the THOR survey, 1390 of which have had spectral indices measured by both surveys. 
Fig.\,\ref{fig:spectral_index_thor_glostar} shows the distribution of the difference in the measured spectral index ($\alpha_{\rm Bconf}-\alpha_{\rm THOR}$) for these sources.
The mean difference in spectral index between the two surveys is $-0.04 \pm 0.03$ (the error is estimated from a Gaussian fit to the distribution in Fig.\,\ref{fig:spectral_index_thor_glostar}), with a standard deviation of 0.78. 
The mean value reduces to $-$0.02 if the matched sources are also compact ($Y_{\rm factor}<2$) in GLOSTAR, and down to zero if the sample is restricted to be sources that are classified as extragalactic in Sect.\,\ref{sect:classification}.
The spectral indices of compact sources measured in the two surveys are thus consistent. A similar result was found in the catalog of the pilot region \citep{Dzib2023AA670A9D}. 
Fig.\,\ref{fig:spectral_index_thor_glostar} also shows the presence of a number of sources that have significant differences in spectral index measurement between THOR and GLOSTAR, such as $\rm abs(\alpha_{\rm Bconf}-\alpha_{\rm THOR})>2$. 
This could be due to: (1) the big differences in beam sizes of THOR ($\sim 25\arcsec$) and GLOSTAR in B-configuration ($\sim 1\arcsec$); (2) the measurement uncertainties of spectral index measured in THOR \citep{Bihr2016AA588A97B} and in this work; (3) a turnover in the radio spectra between THOR and GLOSTAR. 

Among the 1764 7$\sigma$ compact matches, only 130 sources have been classified by THOR, including 45 \hii\ regions and 57 PNe candidates. 
The classification in THOR is consistent with our classification for 38 out of the 45 \hii\ regions and 53 out of the 57 PNe. 
This gives a classification match rate of  84\% (38/45) for \hii\ regions and 92\% (53/57) for PNe. 
Differences in classification can be caused by the differences in the matching radius used for comparison in IR surveys between the two surveys, as discussed in the GLOSTAR catalog paper of the pilot region \citep{Dzib2023AA670A9D}.

\subsection{\hii\ region candidates}
\label{sect:hii_candidate}
 \begin{figure}
 \centering
   \includegraphics[width = 0.45\textwidth]{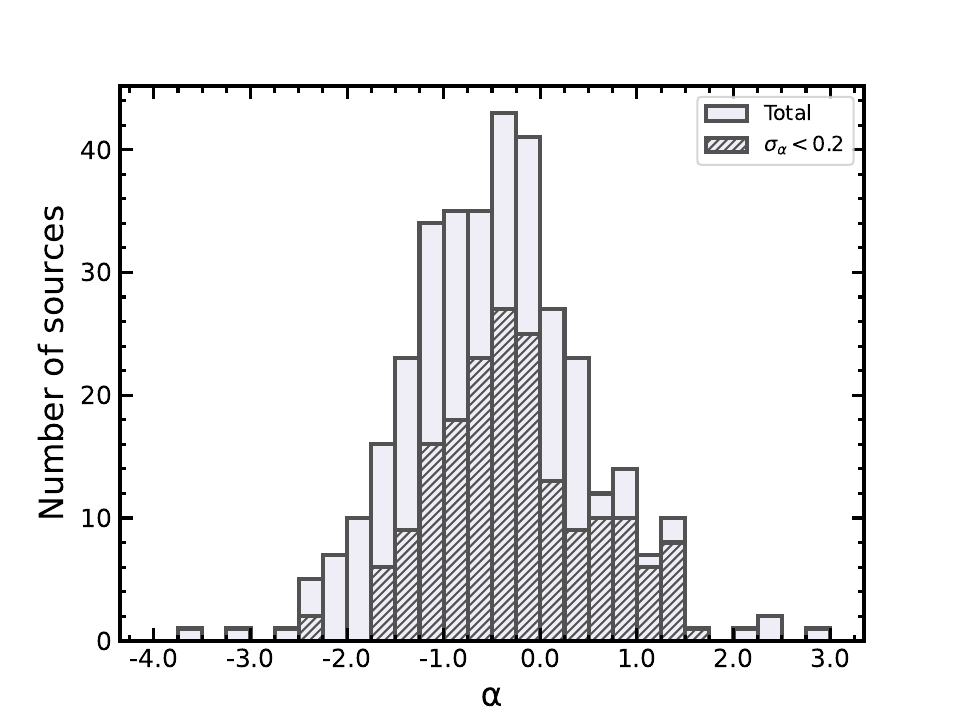} \\
 \includegraphics[width = 0.45\textwidth]{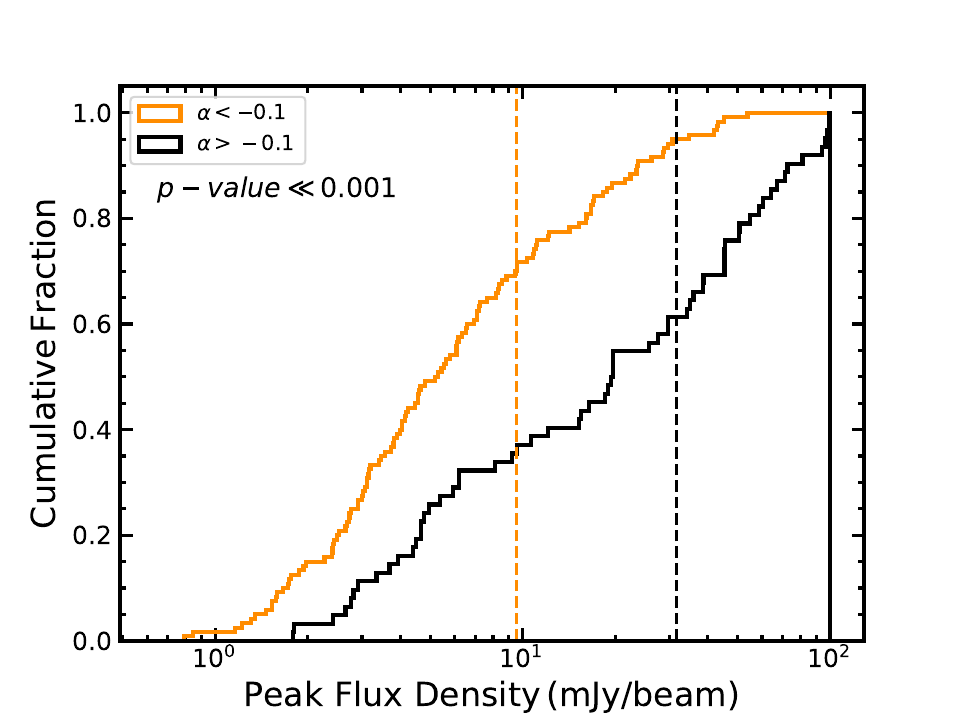} \\
  \caption{Distributions of spectral index and peak flux densities of \hii\ region candidates in GLOSTAR B-configuration. Top panel: The histogram of the measured spectral index for the total 390 \hii\ region candidates of GLOSTAR B-configuration. The hatched histogram represents the 183 candidates with $\sigma_{\alpha}<0.2$. The bin size is 0.25. The 183 candidates are divided into two subsamples: (1) 120  \hii\ regions with $\alpha$ and (2) 63 \hii\ regions with $\alpha>-0.1$, indicating non-thermal and thermal emissions, respectively. Bottom panel: The cumulative distribution of the peak flux densities for the $\alpha<-0.1$ sample (orange line) and the  $\alpha>-0.1$ sample (black line). The orange and black vertical dashed lines show the mean values of 9.5$\rm\,mJy\,beam^{-1}$ and 31.6$\rm \,mJy\,beam^{-1}$ for the two subsamples, respectively.}
 \label{fig:hii_candidates_alpha_flux}
 \end{figure}

 \begin{figure*}
 \centering
 \includegraphics[width = 0.85\textwidth]{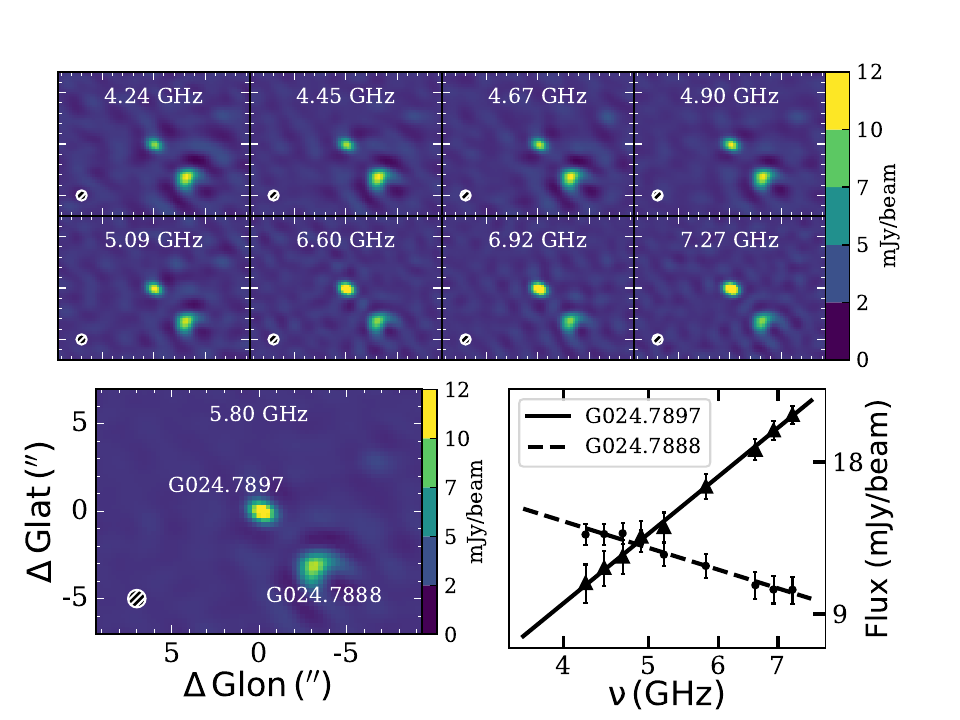} \\
  \caption{Example of two \hii\ region candidates with positive and negative in-band spectral index, as discussed in Sect.\,\ref{sect:hii_candidate}.  The top two rows show the 8 sub-bands of the GLOSTAR image. The bottom-left panel shows the averaged image at 5.8\,GHz used to extract the source, and the bottom-right panel shows the in-band spectral index fitting for the two compact \hii\ regions candidates: the solid line for G024.7897 ($Y_{\rm factor}=1.02$) with $\alpha=1.43\pm0.02$ and the dashed line for G024.7888  ($Y_{\rm factor}=1.97$) with $\alpha=-0.55\pm0.04 $. The FWHM beam of GLOSTAR in B-configuration (1.0\arcsec) is indicated by the white circles in the lower-left corner of each image. There is a clear trend that the fluxes increase and decrease as the increasing frequencies in the sub-bands for G024.7897 and G024.7888 respectively.} 
 \label{fig:hii_candidates_alpha_positive_negative}
 \end{figure*}

As discussed in Sect.\,\ref{sect:classification}, radio sources with submm and FIR emission are classified as \hii\ region candidates. 
These \hii\ region candidates trace radio emission in star formation regions (SFR). 
In this paper, we identified 251 \hii\ region candidates\,\footnote{\label{hii_figure_label} In Appendix\,\ref{appendix_sect:MIR_HII_glimpse}, we present the MIR emission around the \hii\ region candidates.} 
Among these \hii\ regions, 138 are identified/detected by previous work using the CORNISH, THOR, and the SIMBAD database. 
Therefore, 113 \hii\ regions are newly identified in this work. 

The \hii\ region candidates of this work are compact and show a mean effective angular size of 1.2\arcsec, ranging from 0.55\arcsec\ to 2.69\arcsec. 
This indicates that the majority of them belong to the category of the most compact \hii\ regions \citep[e.g,][]{Hoare2007prplconfH}, such as hyper-compact \hii\ (\hchii) regions and ultra-compact \hii\ (\uchii) regions \citep[e.g.,][]{Kurtz2005IAUS,Yang2021AA645A110Y,Liu2021MNRAS5052801L,Patel2023MNRAS5244384P}. 
Combined with the compact \hii\ region candidates in Table 1 of \citet{Dzib2023AA670A9D}, we obtain a sample of 390 \hii\ region candidates in GLOSTAR B-configuration.

The distribution of the spectral index $\alpha$ for the 390 \hii\ region candidates is shown in the top panel of Fig.\,\ref{fig:hii_candidates_alpha_flux}, and $\alpha$ ranges from $-3.70$ and 2.96, with a mean value of $-0.53$. 
Considering the uncertainties of the spectral indices $\sigma_{\alpha}$ as outlined in Sec.\,\ref{sect:spetral_index}, we choose to discuss the \hii\ region candidates that have a reliable spectral index, that is, $\sigma_{\alpha}<0.2$ (where 0.2 refers to the typical value of $\sigma_{\alpha}$ in the catalog). This gives a sample of 183 \hii\ region candidates, with -2.33<$\alpha$<1.58, as the hatched histogram shown in the top panel of Fig.\,\ref{fig:hii_candidates_alpha_flux}.
Previous studies have reported positive and negative spectral indices for \hii\ region candidates in the CORNISH survey \citep{Kalcheva2018AA615A103K} and \hii\ region candidates in the GLOSTAR pilot region \citep{Dzib2023AA670A9D}.

Theoretically, the radio continuum emission of an \hii\ region is thermal and has a spectral index $\alpha$ ($S_{\nu}\propto\nu^{\alpha}$) varying from $ +2$ (optically thick) at low frequency to $-0.1$ (optically thin) at high frequency.  
We use the spectral index $\alpha$ to divide \hii\ region candidates into two subsamples: the $\alpha<-0.1$ group (66\% = 120/183) and the $\alpha>-0.1$ group (34\% = 63/183 ), with mean values of $\alpha\sim-0.7$ and $\alpha\sim0.6$, respectively.
Considering 2 times of the uncertainties in  spectral index ($2\,\sigma_{\alpha}$) for the two samples, the fraction of the $\alpha>-0.1$ sample increase to about 50\%.
The mean spectral index of 0.6 for the $\alpha>-0.1$ group refers to the thermal emission from \hii\ regions, which is similar to the mean of $\alpha\sim 0.6$ observed at 5\,GHz for the young \hii\ regions (\hchii\ and \uchii) sample in \citet{Yang2019MNRAS4822681Y} who suggest the existence of \hii\ regions with a mix of optically thin and thick components along the line of sight. 

Intriguingly, the majority ($\sim 66\%$) of the sample belongs to the $\alpha<-0.1$ group with a mean spectral index of $-$0.7, which indicates that a substantial portion of radio emission in these \hii\ region candidates is non-thermal. 
A number of observational studies have reported the existence of \hii\ regions with a mixture of thermal and non-thermal radiation  \citep[e.g.,][]{Wang2018AA619A124W,Meng2019AA630A73M,Padovani2019AA630A72P} and dominated non-thermal emission  \citep[e.g.,][]{Wilner1999ApJ513775W,vanderTak2005AA437947V,Rosero2019ApJ88099R}. 
The \hii\ region with radio continuum $\alpha < -0.5$ are considered to be dominated by non-thermal emission \citep{Kobulnicky1999ApJ527154K}.
Considering that 74 out of the 120 \hii\ region candidates have $\alpha<-0.5$, we suggest these are dominated by non-thermal emission, while the remaining 46 candidates with $-0.5<\alpha<-0.1$ are likely to be associated with a mixture of thermal and non-thermal radiation. 

Given that the \hii\ region candidates are identified by radio emission in star formation regions (see Sect.\,\ref{sect:classification}), it is possible that the non-thermal radio emission originates from the processes such as radio jets, shocks and outflows from high-mass \citep{vanderTak2005AA437947V} and low-mass stars \citep{Gomez2005MNRAS364738G}. 
We find that the $\alpha<-0.1$ group is more likely to be located in clusters (see Sect.\,\ref{sect:cluster_source}) and be associated with molecular outflows in \citealt{Yang2018ApJS2353Y,Yang2022AA658A160Y} compared to the $\alpha>-0.1$ sample, implying that the non-thermal emission arises from localized spots that are seen only when the large scale emission is filtered out by the interferometer.
From Fig.\,\ref{fig:hii_candidates_alpha_flux}, we can see that the $\alpha<-0.1$ sample shows significantly lower values of peak flux density compared to the $\alpha>-0.1$ sample, indicating that the $\alpha<-0.1$ sample are relatively compact and weak. Thus, these clustered non-thermal emission spots are likely to be radio jets and outflows located in the vicinity of \hii\ regions and in star formation regions \citep[e.g.,][]{Wang2012ApJ745L30W,Purser2016MNRAS4601039P,Rosero2016ApJS22725R,Liu2017ApJ84925L,Qiu2019ApJ871141Q}. 
This is consistent with the findings of \citet{Wang2022ApJ927185W} who suggested that most of the radio sources in the Cygnus region are radio jets and winds originating from massive young stellar objects. 
In Fig.\,\ref{fig:hii_candidates_alpha_positive_negative}, we displayed an example of the two \hii\ region candidates (G024.7897 and G024.7888) that shows positive and negative in-band spectral index derived by fitting the 8 sub-bands radio images of the GLOSTAR. The \hii\ region candidate G024.7897 ($Y_{\rm factor} =1.02$) with $\alpha=1.43\pm0.02$ is supposed to be a ``real'' \hii\ region,  which shows extended green emission \citep[as defined by ][]{Cyganowski2008AJ136.2391C} in Fig.\,\ref{appendix_sect:MIR_HII_glimpse}, and is associated with a maser-emitting \uchii\ region in \citet{Hu2016ApJ83318H}.
In contrast, the nearby \hii\ region candidate G024.7888 ($Y_{\rm factor} = 1.97$) with $\alpha=-0.55\pm0.04$ is likely to be the non-thermal emission from radio jets or outflows in massive star-forming regions. 
Some of the sources in the catalog were confirmed as non-thermal sources through VLBI observations by \citet{Dzib2016ApJ826201D}. 
We note that there are many \hii\ regions like G024.7897 that are surrounded with at least one non-thermal source like G024.7888, as shown in Fig.\,\ref{fig:HII_MIR_images}, which is consistent with the findings in   \citet{Gomez2005MNRAS364738G} who detected a cluster of non-thermal sources around a young and compact \uchii\ region with VLA observations and suggested these non-thermal clusters are originated from low-mass, pre–main-sequence stars. 

In summary, from the $\alpha>-0.1$ sample, we confirm the existence of the \hii\ regions with a mixture of optically thin and thick thermal emission components. 
From the $\alpha<-0.1$ sample, we find that a large fraction of compact \hii\ region candidates are associated with non-thermal emission, suggesting that these candidates can be radio jets, winds and outflows from high-mass and low-mass young stellar objects. This further indicates that there is a significant amount of relativistic electrons that exist in star-forming regions. 
Further investigation is required to confirm the nature of non-thermal emissions in massive star-forming regions.
 
\subsection{ Planetary nebula candidates}
\label{setc:discuss_pne}
We identified 282 planetary nebulae (PNe) based on the classification process described in Sect.\,\ref{sect:classification}. 
Among these, 155 are identified/detected by previous work and 127 PNe are new detections.

These PNe candidates are compact and their effective size range from 0.53\arcsec\ to 2.46\arcsec with a mean and standard deviation of 1.0\arcsec and 0.3\arcsec respectively. 
About 54\% of the sources have effective sizes less than 1.0\arcsec. 
The in-band spectral indices $\alpha$ of the PNe candidates range from $-$2.75 to 1.9, with a mean value of $-0.52$. 
Combining with the 68 PNe candidates identified in the Table 1 of  \citet{Dzib2023AA670A9D} for the pilot region, we obtained a sample of 350 PNe candidates for the GLOSTAR B-configuration catalog. 

As in \hii\ regions, from the 162 PNe candidates with $\sigma_{\alpha}<0.2$, two subsamples are obtained: (1) PNe candidates with $\alpha<-0.1$ (62\%=100/162) and (2) PNe candidates with $\alpha>-0.1$ (38\%=62/162). 
 This is similar to observations from previous surveys such as CORNISH  \citep{Irabor2018MNRAS4802423I} and the pilot region of GLOSTAR \citep{Dzib2023AA670A9D}, who found that PNe were found with radio emission showing both positive and negative spectral indices.
 This suggests that both thermal and non-thermal emission components are associated with PNe candidates. 
 Theoretically, PNe are thus expected to have radio continuum from thermal free-free emission with $-0.1<\alpha<2$ \citep[e.g.,][]{Gomez2005MNRAS364738G,Tafoya2009ApJ691611T,Qiao2016ApJ81737Q}. 
Thus, similar to the \hii\ region candidates, the PNe candidates with spectral index greater than and less than -0.1 are expected to be associated with thermal and non-thermal emission, respectively. 
 
Observationally, a low spectral index threshold of $\alpha<-0.5$ between 1\,GHz and 20\,GHz was suggested to identify non-thermal emission from  planetary nebulae detected in the AT20G survey \citep[i.e., the Australia Telescope 20 GHz survey,][]{Chhetri2015MNRAS4513228C}. 
Planetary nebulae with both non-free-free emission components \citep[e.g.,][]{Casassus2007MNRAS3821607C,Irabor2018MNRAS4802423I} and even non-thermal dominated emission \citep[e.g.,][]{Suarez2015ApJ806105S,Cerrigone2017MNRAS4683450C,Anglada2018AARv263A} have been reported before. 
Non-thermal emission is found in young PNe that are associated with masers \citep{Cohen2006MNRAS369189C} or that are in the formation stage \citep{Cerrigone2017MNRAS4683450C}. 
 In our sample of sources with $\alpha<-0.1$,  59 PNe candidates have $\alpha<-0.5$ and thus are likely to be dominated by non-thermal emission, while the remaining 41 PNe candidates with $-0.5<\alpha<-0.1$ are thought to have both thermal and non-thermal components.
The $\alpha<-0.5$ candidates show significantly lower $Y_{\rm factor}$ and flux densities compared to sources of the $\alpha>-0.5$ sample. This suggests that the $\alpha<-0.5$ sources are denser and weaker and thus likely associated with radio jets.  
 In short, we found a large fraction of PNe are associated with compact non-thermal emissions likely from radio jets, suggesting that these PNe candidates are very young or in the formation stage.
  
\subsection{Extragalactic source candidates}
In this paper, we identified 4080 extragalactic sources based on the classification process outlined in Sect.\,\ref{sect:classification}. 
This is consistent with the expected number of $\sim 4137$ estimated from the FIRST survey \citep{White1997ApJ475479W} and $\sim 4100$ calculated from Equation (A2) of \citet{Anglada1998AJ1162953A} using the 5 GHz source counts of \citet{Condon1984ApJ287461C} and the typical detection level ($\rm \sim 0.6\,mJy\,beam^{-1}$) of this work.
Among these extragalactic sources, 1905 have been identified/detected previously in the CORNISH and THOR surveys and the SIMBAD database. 
Therefore, 2175 extragalactic sources are newly identified by this work. 

Combining with the 1157 extragalactic source candidates in the Table 1 of \citet{Dzib2023AA670A9D} for the pilot region, we have a sample of 5237 extragalactic source candidates for the GLOSTAR B-configuration catalog. These extragalactic source candidates have effective sizes ranging from 0.45\arcsec to 3.24\arcsec, with a mean value of 0.86\arcsec and a standard deviation of 0.25\arcsec. 
As expected, the effective size of the extragalactic source candidates is systematically smaller than that of the Galactic sources (i.e., \hii\ region and PNe candidates). 
Some extragalactic source candidates show extended radio emissions that are likely to be associated with radio galaxy lobes as suggested in the CORNISH survey \citep{Purcell2013ApJS2051P}. 
The in-band spectral index $\alpha$ of the 4836 extragalactic source candidates {are measured, ranging} from $-3.24$ to 2.72, with a mean value of $-0.58$. 
As expected, the majority of extragalactic source candidates (77\%=3725/4836) have $\alpha<-0.1$, indicative of non-thermal synchrotron emission.  
The remaining 23\% extragalactic candidates show $\alpha>-0.1$, which could be due to the spectral turnover due to synchrotron self-absorption and free-free absorption \citep[e.g.,][]{Bicknell2018MNRAS4753493B,Shao2022AA659A159S} of gigahertz peaked spectrum sources \citep[GPS, e.g.,][]{Bicknell1997ApJ485112B} or high-frequency peakers \citep[HFP, e.g.,][]{Dallacasa2000AA363887D}, as discussed in the pilot field GLOSTAR article \citep{Dzib2023AA670A9D}. 
 Given that extragalactic radio sources with $\alpha<-1.3$ are expected to be high-redshift galaxies with ultra steep spectra \citep{Wang2018AA619A124W}, 13\% (698/5237) of extragalactic candidates with $\alpha<-1.3$ in this work are thus likely to be  in this category. 
We note that there are more than 200 extragalactic candidates associated with NIR $\sim 2\mu m$ UKIDSS counterparts that are suggested to be galaxies by \citet{Lucas2008MNRAS391}, which are likely to be radio galaxies in the Zone of Avoidance \citep[e.g.,][]{Marleau2008AJ136662M} or quasars behind the Galactic plane \citep[e.g.,][]{Fu2022ApJS26132F}.

  \begin{figure}
 \centering
    \includegraphics[width = 0.45\textwidth]{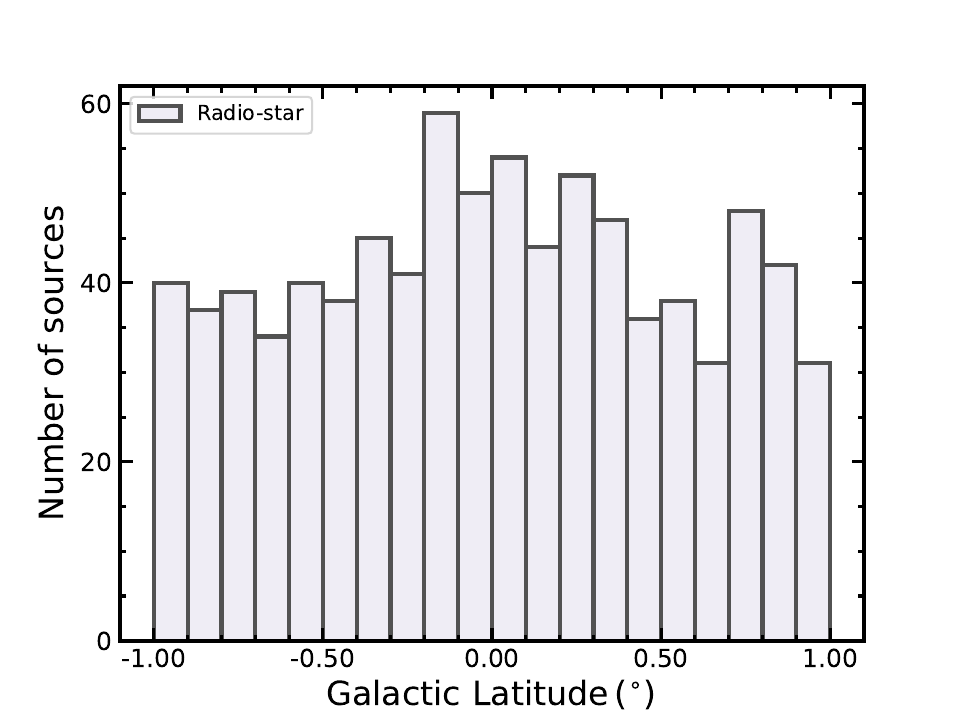} \\
    \includegraphics[width = 0.45\textwidth]{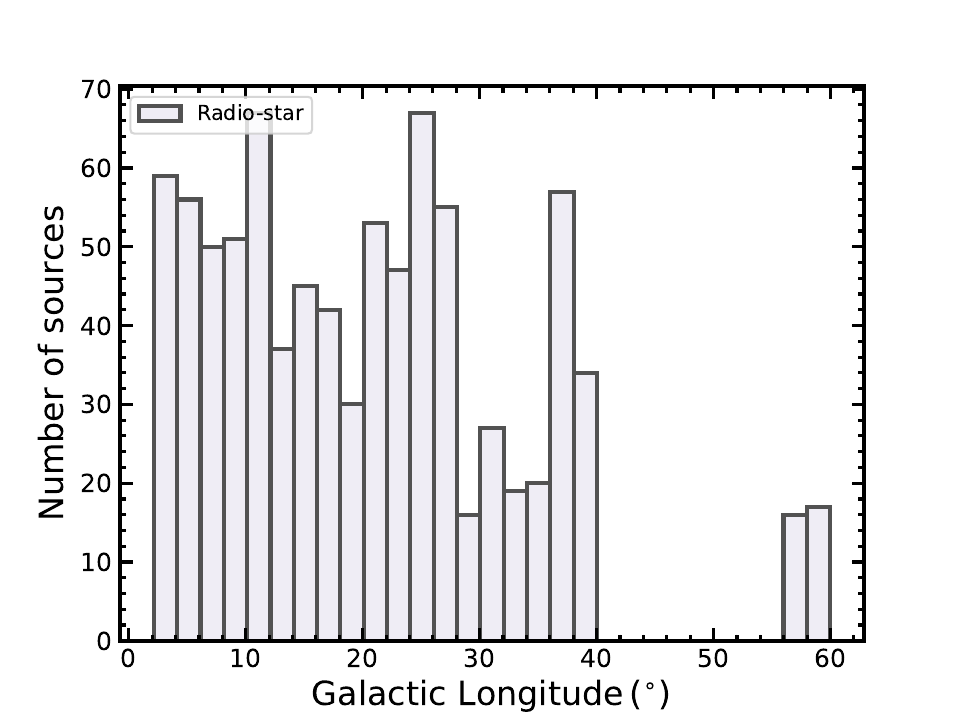} \\
      \caption{Distributions of radio star candidates as a function of Galactic Latitude (upper panel) and Longitude (lower panel) for the GLOSTAR B-configuration catalog, including the catalog of this work ($2\degr<\ell<28\degr\, 36\degr<\ell<40\&\,56\degr<\ell<60$ and $|b| < 1\degr$), as listed in Table\,\ref{tab:total_7sigma} and the published pilot catalog ($28\degr<\ell<36\degr$ and $|b| < 1\degr$) in Table 1 of \citet{Dzib2023AA670A9D}. The bin sizes are 0.1\degr and 2.0\degr for the upper and lower panel, respectively. The blank region in Galactic longitude $40\degr<\ell<56\degr$ is not covered by the GLOSTAR survey in B-configuration.}
 \label{fig:distr_glon_lat_RS}
 \end{figure}

\subsection{Radio star candidates}
\label{sect:radio_stars}
In this paper, we have identified 784 radio star candidates based on the classification process in Sect.\,\ref{sect:classification}, which refers to a group of radio sources that are  point-like and blue in the three-color images at near infrared UKIDSS, GLIMPSE, and WISE, but with weak or no emission at submm and FIR wavelength (due to the dispersion of natal molecular clouds), as shown in panel (c) of Fig.\,\ref{fig:example_class}.
Among these radio star candidates, 203 have been identified/detected by previous work in the CORNISH, THOR, and/or the SIMBAD database. 
Therefore, 581 radio stars are newly detected and identified by this work. 

Combining the 81 radio star candidates in Table 1 of \citet{Dzib2023AA670A9D}, a sample of 865 radio star candidates are obtained for the GLOSTAR B-configuration. 
These radio star candidates have effective sizes ranging from 0.51\arcsec to 3.04\arcsec, with a mean value of 0.83\arcsec. The in-band spectral index $\alpha$ of the radio star candidates ranges from -2.38 to 1.94, with a mean value of -0.41. As in \hii\ regions, from the 278 radio stars with $\sigma_{\alpha}<0.2$, there are 61\% (169/278) showing non-thermal emissions with $\alpha<-0.1$ and 39\% (109/278) showing thermal emissions with $\alpha>-0.1$. This is expected as both the thermal and non-thermal radio emission from radio stars have been observed \citep{Hoare2012PASP}.

Fig.\,\ref{fig:distr_glon_lat_RS} shows the Galactic distributions for radio star candidates in the GLOSTAR B-configuration catalog. The number of radio star candidates per 0.1 latitude bin is relatively flat and the source counts decrease near the edges of $b$ range possibly due to the higher noise level as discussed in Sec.\,\ref{sect:completeness}. 
The source counts of radio star candidates per 2\degr longitude bin are found to increase toward low longitudes, which is consistent with the increasing of the Galactic sources toward low longitudes as discussed in Sec.\,\ref{gal_distribution}. 
Given that there might be radio-emitting sources showing the same multiband emission properties, the nature of these radio star candidates are needed to be explored further.
 
\subsection{Variable sources}
\label{sect:variable_sources}
Variable radio sources are defined based on the ratios of peak flux densities between the GLOSTAR and CORNISH surveys for compact sources (i.e., $Y_{\rm factor}<2.0$), as outlined in \citet{Dzib2023AA670A9D}. 
A source is considered to be a variable source if (1) it shows a flux density ratio between GLOSTAR and CORNISH larger than 2.0; (2) it is detected in CORNISH at 7$\sigma$ but is not detected by GLOSTAR at 7$\sigma$; (3) it is a GLOSTAR source with a peak flux density higher than the CORNISH detection limit ($\rm 2.7\,mJy\,beam^{-1}$) but is not detected by CORNISH.  

Using the criteria listed above, we identified 245 variable radio sources in this work. 
Together with the 49 variable sources identified in \citet{Dzib2023AA670A9D} for the pilot region, there are a total of 294 variable sources in the GLOSTAR B-configuration catalog. Table\,\ref{tab:total_variable_sources} lists the source names, peak flux densities, and the suggested source types of all these variable radio sources. The majority of variable sources in the current work (76\%=186/245) are found to be extragalactic in origin or infrared quiet based on the classification of this work and CORNISH.
 The catalog of variable radio sources in this paper includes 12 \hii\ regions, 5 planetary nebulae, and 4 PDRs. 
The variability of \hii\ regions \citep[e.g.,][]{Yang2022MNRAS511280Y,Dzib2023AA670A9D} and planetary nebulae \citep[e.g.,][]{Cerrigone2011MNRAS4121137C,Suarez2015ApJ806105S,Yang2022MNRAS511280Y,Dzib2023AA670A9D} have been reported earlier, which are interesting targets for further exploration. 
 
\begin{table*}
\centering
\caption[]{ \it \rm Catalog of 245 variable radio sources identified by this work}
 \footnotesize
 \begin{tabular}
{p{3.cm}p{1.2cm}p{1.2cm}p{2.cm}p{1.2cm}p{1.2cm}p{2.cm}}
\hline
\hline
Source B-conf. &  \multicolumn{3}{c}{GLOSTAR}    &   \multicolumn{3}{c}{CORNISH}     \\
    \cmidrule(lr){2-4}\cmidrule(lr){5-7}
Gname & $S_{\rm peak}$ & $\sigma_{S_{\rm peak}}$ & classification & $S_{\rm peak}$ & $\sigma_{S_{\rm peak}}$ & classification \\

(1)  & (2)  & (3)  & (4) & (5)  & (6) & (7) \\
\hline
G010.3377$+$ 01.0601 &   $-$ &   $-$ &    $-$   & 10.54 & 1.04 & IR$-$Quiet \\
G010.3599$+$ 00.1307 & 13.26 & 0.73 & Egc & 5.79 & 0.63 & IR$-$Quiet \\
G010.5099$-$01.0018 & 11.18 & 0.62 & Egc & 3.44 & 0.47 & IR$-$Quiet \\
G010.5894$-$00.8981 &   $-$ &   $-$ &    $-$   & 3.49 & 0.52 & IR$-$Quiet \\
G010.8677$-$00.0052 &   $-$ &   $-$ &    $-$   & 4.14 & 0.55 & IR$-$Quiet \\
G011.0368$+$ 01.0899 &   $-$ &   $-$ &    $-$   & 10.4 & 1.07 & IR$-$Quiet \\
$\vdots$ & $\vdots$ & $\vdots$ & $\vdots$ & $\vdots$ & $\vdots$ & $\vdots$  \\
G014.8462$-$00.7751 & 3.65 & 0.22 & Radio$-$star &   $-$ &   $-$ &    $-$   \\
G014.8827$-$00.4943 & 4.22 & 0.24 & Egc &   $-$ &   $-$ &    $-$   \\
G014.9523$+$ 00.5942 & 3.72 & 0.22 & Radio$-$star &   $-$ &   $-$ &    $-$   \\
G014.9945$-$00.7486 & 4.74 & 0.35 & PDR &   $-$ &   $-$ &    $-$   \\
G015.0136$-$00.6969 & 4.08 & 0.54 & PDR &   $-$ &   $-$ &    $-$   \\
G015.0345$-$00.6771 & 50.55 & 2.86 & HII &   $-$ &   $-$ &    $-$   \\
$\vdots$ & $\vdots$ & $\vdots$ & $\vdots$ & $\vdots$ & $\vdots$ & $\vdots$  \\
G059.1808$-$01.0626 & 16.8 & 1.08 & Egc & 8.33 & 0.89 & IR$-$Quiet \\
G059.1910$-$00.0700 &   $-$ &   $-$ &    $-$   & 1.7 & 0.26 & IR$-$Quiet \\
G059.6786$-$01.0884 &   $-$ &   $-$ &    $-$   & 3.7 & 0.6 & Radio$-$Star \\
G059.8198$-$00.4796 &   $-$ &   $-$ &    $-$   & 1.75 & 0.28 & Radio$-$Galaxy  \\
G059.9401$+$ 01.0185 &   $-$ &   $-$ &    $-$   & 2.18 & 0.32 & IR$-$Quiet \\
\hline
\hline
\end{tabular}
\tablefoot{ Source names in column (1) are taken from GLOSTAR or CORNISH if no GLOSTAR detections.
Symbol $-$ refers to no measurements.
Only a small portion of the data is provided here. 
 The full table will be available in electronic form at the CDS via anonymous ftp to cdsarc.cds.unistra.fr (130.79.128.5)
or via https://cdsarc.cds.unistra.fr/cgi-bin/qcat?J/A+A/. }
\label{tab:total_variable_sources}
\end{table*}
\section{Conclusions}
\label{sect:conclusion}
The GLOSTAR survey covers 145 square degrees of the Galactic plane and observes both spectral lines and continuum in the 4-8 GHz using the VLA in B- and D-configurations and the Effelsberg 100-m telescope. In this paper, we present a catalog of continuum sources in the 68 square degrees of the Galactic plane\,($2\degs<\ell<28\degs,\,36\degs<\ell<40\degs$,\,$56\degs<\ell<60\degs$ $\&$ $|b|<1\degs$) observed with the VLA in B-configuration. 
The images have an angular resolution of 1\arcsec, a typical 1$\sigma$ noise level of $\rm\sim 0.08\,mJy\,beam^{-1}$, and are restricted to detect extended emission on angular scales smaller than 4\arcsec. 
The high and low reliability catalogs from this region are presented in Tables\,\ref{tab:total_7sigma} and \ref{tab:total_5to7sigma}, respectively.
   The main results are summarised below. 
\begin{enumerate}
\item  We obtained a high-reliability catalog of 5437 sources above 7$\sigma$ and a low-reliability catalog of 7917 sources with signal-to-noise ratios between 5$\sigma$ and 7$\sigma$, using the source extraction tool BLOBCAT. The 1$\sigma$ noise level of GLOSTAR is spatially varying and 95\% of the observing fields have $\rm  1\sigma<0.13\,mJy\,beam^{-1}$. Above the 7$\sigma$ detection limit, the catalog is typically complete to point sources at 95\%. 
The systematic positional uncertainty of the GLOSTAR B-configuration in this work is $\lesssim$0\farcs1. 

\item We further investigated the high-reliability catalog (i.e., 5437 sources with $ \rm S/N \ge 7\sigma$), with physical properties summarized in Table\,\ref{tab:summary_phy_param}.
We extracted the peak intensities from the 9 sub-bands from 4-8\,GHz, and determined the in-band spectral indices $\alpha$ for 5435 sources by fitting the peak flux density in each sub-band. The uncertainty of the spectral index is negatively correlated with the single-to-noise ratio (S/N).  
The mean value of the in-band spectral index is $\sim -0.6$, and 74\% of sources have $\alpha<-0.1$, indicating that the catalog is dominated by non-thermal radio emission. 
The spectral indices of compact sources between GLOSTAR and THOR are consistent.   
 
\item 
We classified all sources above $7\,\sigma$-threshold based on the presence or absence of counterparts/emissions in Galactic plane surveys at infrared and submillimeter wavelengths, as well as the SIMBAD database. 
We identified candidates of 251 \hii\ regions, 784 radio stars, 282 PNe, 4402 extragalactic sources, and 29 others (11 PDRs and 18 unclear). 
The consistency in classification between GLOSTAR and CORNISH survey is 100\% for \hii\ regions and PNe, and 98\% for extragalactic sources.

\item A significant fraction of candidates of the \hii\ region and PNe show a spectral index $\alpha<-0.1$ (or even $\alpha<-0.5$), suggesting that there are significant numbers of non-thermal emission sources corresponding to radio jets, winds, and outflows in the vicinity of young \hii\ regions and PNe.
 As expected, the majority of extragalactic candidates (77\%) show $\alpha<-0.1$, indicating non-thermal emission. 
The remaining 23\% showing $\alpha >-0.1$ are likely to be the gigahertz peak spectrum sources or high-frequency peakers. 
 For the above three source types, the $\alpha<-0.1$ group is more likely to be associated with clustered sources compared to the $\alpha>-0.1$ group.

\item We compared the measured flux densities between GLOSTAR and other radio surveys (CORNISH, MAGPIS, and THOR), and find a high level of agreement in the flux density and spectral index for compact sources detected in these surveys.  
We identified 245 variable radio sources listed in Table\,\ref{tab:total_variable_sources} by comparing the fluxes between the GLOSTAR and CORNISH survey, and most of these variable sources are found to be infrared and submillimeter quiet. 

\end{enumerate}

To date, the GLOSTAR survey using VLA in its B-configuration has the highest resolution ($\sim 1.0\arcsec$) and best sensitivity ($1\sigma \rm \sim 0.08\,mJy\,beam^{-1}$) for the Galactic plane in the C band (4-8\,GHz). 
In this work, we present a continuum catalog of 13354 sources $\geq 5\sigma$ and 5437 sources $\geq 7\sigma$ detection level. 
 From the high-reliability $7\sigma-$threshold catalog, we identified the largest sample of candidates of \hii\ regions, PNe, extragalactic sources, and variable sources in the Galactic plane. 
 It is worth noting that a significant fraction of radio emission associated with massive star formation regions are non-thermal, indicating that relativistic electrons commonly exist. 
 All the catalogs and data are available online at the GLOSTAR website \emph{https://glostar.mpifr-bonn.mpg.de.}  
      
\begin{acknowledgements}
We would like to thank the referee David J. Helfand for the helpful comments and suggestions on our manuscript. AYY acknowledges support from the National Natural Science Foundation of China (NSFC) grants No. 11988101, No. 11973013, and No. 12303031.
S.A.D. acknowledge the M2FINDERS project from the European Research Council (ERC) under the European Union's Horizon 2020 research and innovation programme
(grant No 101018682).
This research was partially funded by the ERC Advanced Investigator Grant GLOSTAR (247078). 
 M.\,R.\,R. is a Jansky Fellow of the National Radio Astronomy Observatory, USA. 
RD is a member of the International Max Planck Research School (IMPRS) for Astronomy and Astrophysics at the Universities of Bonn and Cologne. 
AY would like to thank the help of Philip Lucas and Read Mike when using the data of the UKIDSS survey. 
The UKIDSS survey was made with the UKIRT Wide Field Camera that was funded by the UK Particle Physics and Astronomy Research Council.
This work uses information from the GLOSTAR databases at 
\url{http://glostar.mpifr-bonn.mpg.de} supported by the MPIfR (Max-Planck-Institut für Radioastronomie), Bonn, which is based on observations with the Karl G. Jansky Very Large Array (VLA) of NRAO (The National Radio Astronomy Observatory is a facility of the National Science Foundation operated under cooperative agreement by Associated Universities, Inc.) and 100-m telescope of the MPIfR at Effelsberg. 
 It also made use of information from the ATLASGAL
database at 
\url{http://atlasgal.mpifr-bonn.mpg.de/cgi-bin/ATLASGAL_DATABASE.cgi}
supported by the MPIfR, Bonn, as well as information from the CORNISH
database at \url{http://cornish.leeds.ac.uk/public/index.php} which
was constructed with support from the Science and Technology 
Facilities Council of the UK.
This work has used data from GLIMPSE and MIPSGAL surveys of the \emph{Spitzer} Space Telescope, which is operated by the Jet Propulsion Laboratory, California Institute of Technology under a contract with NASA.
This publication also makes use of data products from the Wide-field Infrared Survey Explorer, which is a joint project of the University of California, Los Angeles, and the 
Jet Propulsion Laboratory/California Institute of Technology, funded 
by the National Aeronautics and Space Administration. 
This paper used the data products from the Hi-GAL survey of the \emph{Herschel} telescope which is an ESA space observatory with science instruments provided by European-led Principal Investigator consortia and with important participation from NASA. 
This research has made use of the SIMBAD database and the VizieR catalog, operated at CDS, Strasbourg, France.
This document was prepared using the collaborative tool Overleaf available at: 
\url{https://www.overleaf.com/}.




\end{acknowledgements}

\bibliographystyle{aa}
\bibliography{ref}
\clearpage

\begin{appendix}

\section{The pipeline logic flowchart}
\begin{figure*}
 \centering
 \begin{tabular}{cc}
    \includegraphics[width = 0.48\textwidth]{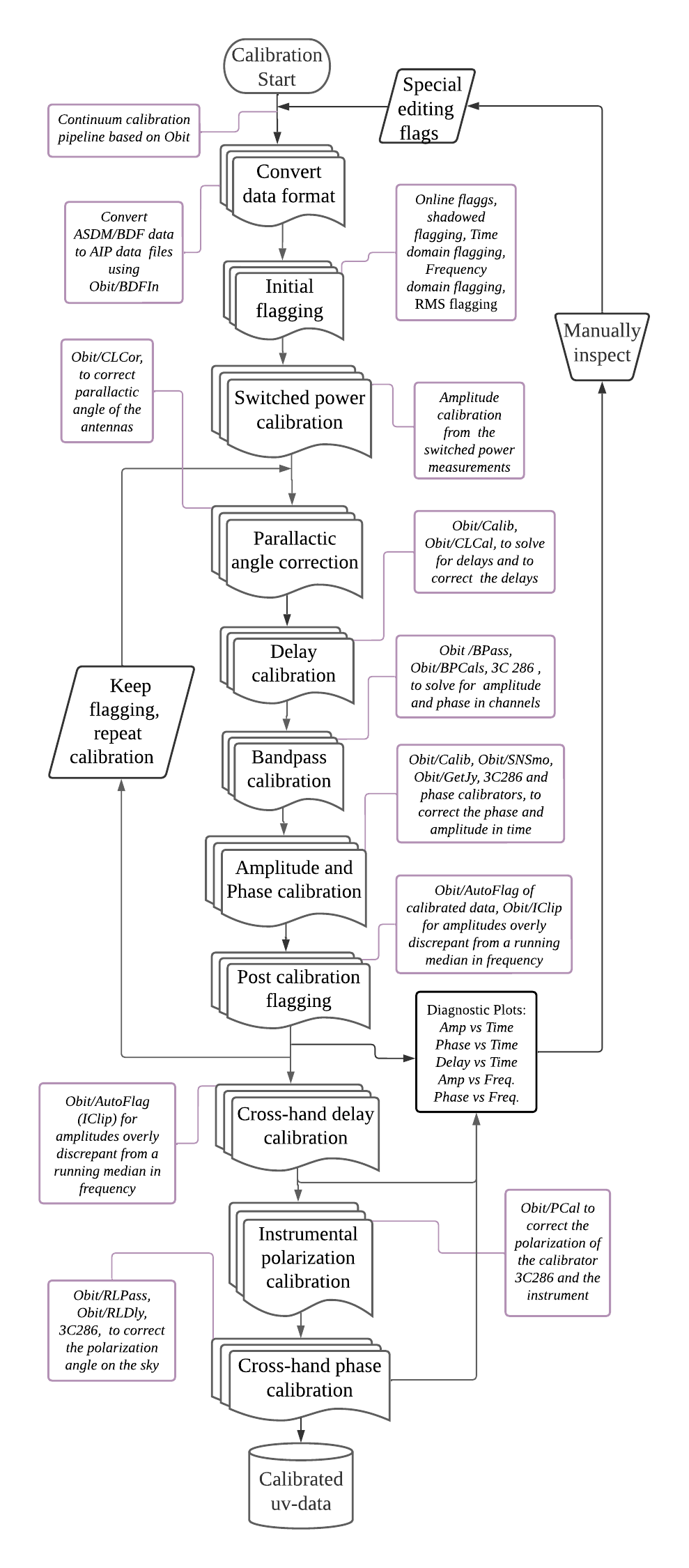} 
    &
      \includegraphics[width = 0.48\textwidth]{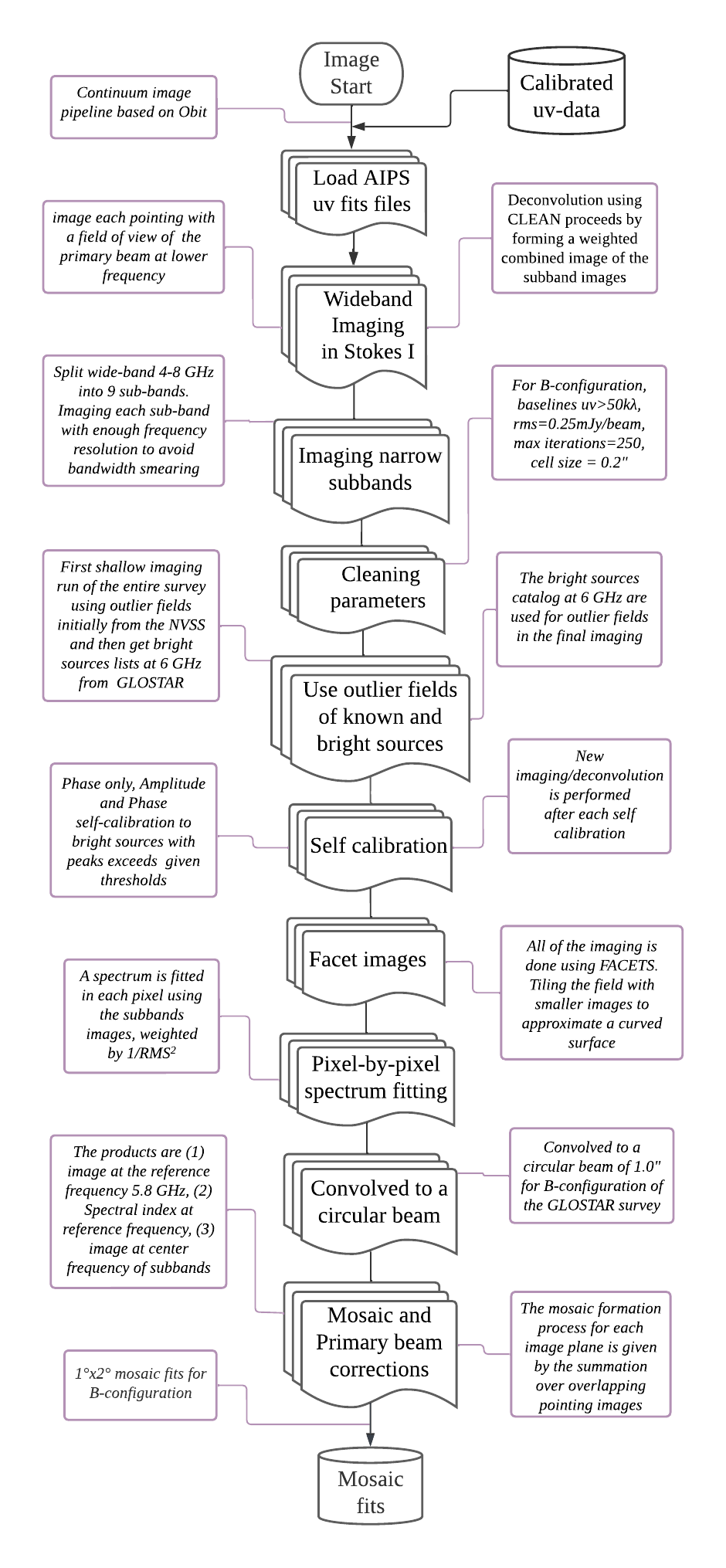} \\  
 \end{tabular}
  \caption{Flowchart illustration of the GLOSTAR calibration and imaging pipelines. 
  See the text of Sect.\,\ref{sect:calibration_and_image} and \citet{Brunthaler2021AA651A85B} for more details.}
 \label{fig:calibration_imaging_pipeline}
 \end{figure*}

In this section, we summarize the calibration and imaging steps for the continuum data reduction of the GLOSTAR survey, as discussed in Sect.\,\ref{sect:calibration_and_image} and as outlined in previous papers of GLOSTAR \citep[e.g.,][]{Medina2019AA627A175M,Brunthaler2021AA651A85B,Dzib2023AA670A9D}.  
The logic flowcharts of the calibration and imaging pipeline with the OBIT tasks and processes are shown in  Fig.\,\ref{fig:calibration_imaging_pipeline}.

\section{The low-reliability catalog}
As outlined in Sect.\,\ref{sect:source_cat_construction}, the low-reliability catalog refers to the GLOSTAR sources with a signal-to-noise ratio (S/N) between 5$\sigma$ and 7$\sigma$. 
We have identified 7917 sources between 5$\sigma$ and 7$\sigma$ in this work. 
Table A.1 lists basic parameters measured from low-reliability sources of the GLOSTAR B-configuration catalog in this work.
\begin{table*}[ht]
\centering
\caption[]{ \it \rm  The B-configuration catalog of sources with S/N in the range between 5$\sigma$ to 7$\sigma$ in this work. }
 \scriptsize
 \begin{tabular}{p{2.2cm}p{1.0cm}p{1.0cm}p{1.0cm}p{1.0cm}p{1.0cm}p{1.0cm}p{1.0cm}p{2.3cm}p{2.2cm}}
\hline
\hline
GLOSTAR B-conf. & $\ell$ & $b$ & $\rm S/N$ & $S_{\rm peak}$ & $\sigma_{S_{\rm peak}}$ & $S_{\rm int}$ & $\sigma_{S_{int}}$ & SIMBAD & GLOSTAR D-conf.  \\
Gname & $\degr$ & $\degr$ &   & \multicolumn{2}{c}{$mJy/beam$}   & \multicolumn{2}{c}{$mJy$}   &  & Gname    \\
(1) & (2) & (3) & (4) & (5) & (6) & (7) & (8) & (9) & (10)  \\ 
G001.9817+00.4085 & 1.98167 & 0.40845 & 5.3 & 0.65 & 0.13 & 0.65 & 0.13 & $-$ &  $-$  \\
G001.9820-00.5941 & 1.98205 & -0.59413 & 5.3 & 0.72 & 0.14 & 0.57 & 0.14 & $-$ &  $-$  \\
G001.9872-00.5155 & 1.98721 & -0.51553 & 5.3 & 0.62 & 0.13 & 0.8 & 0.13 & $-$ &  $-$   \\
G002.0020+00.4964 & 2.00204 & 0.49643 & 5.9 & 1.11 & 0.2 & 0.95 & 0.2 & $-$ &  $-$   \\
G002.0101+00.0865 & 2.01011 & 0.08655 & 5.1 & 0.59 & 0.13 & 0.58 & 0.13 & $-$ &  $-$  \\
G002.0206+00.1785 & 2.02059 & 0.17848 & 5.3 & 0.65 & 0.13 & 0.58 & 0.13 & $-$ &  $-$  \\
G002.0312+00.3970 & 2.03121 & 0.39701 & 5.4 & 0.64 & 0.13 & 0.59 & 0.13 & $-$ &  $-$  \\
G002.0359-00.2900 & 2.03593 & -0.28997 & 5.1 & 0.56 & 0.12 & 0.47 & 0.11 & $-$ &  $-$  \\
G002.0366+01.0173 & 2.0366 & 1.01729 & 5.6 & 0.88 & 0.16 & 0.6 & 0.16 & $-$ &  $-$  \\
G002.0493-01.0230 & 2.04934 & -1.02301 & 5.4 & 1.34 & 0.26 & 0.74 & 0.25 & $-$ &  $-$  \\
G002.0539-00.0130 & 2.0539 & -0.01297 & 6.5 & 0.66 & 0.11 & 0.57 & 0.11 & $-$ &  $-$  \\
G002.0553-01.0535 & 2.05531 & -1.05347 & 5.3 & 2.47 & 0.49 & 2.0 & 0.48 & $-$ &  $-$  \\
G002.0598+00.8321 & 2.05983 & 0.83209 & 5.1 & 0.48 & 0.1 & 0.52 & 0.1 & $-$ &  $-$  \\
G002.0651-00.2293 & 2.06506 & -0.22928 & 5.0 & 0.48 & 0.1 & 0.49 & 0.1 & $-$ &  $-$  \\
G002.0687+01.0083 & 2.06872 & 1.00826 & 5.4 & 0.71 & 0.14 & 0.52 & 0.13 & $-$ &  $-$  \\
G002.0770+00.4259 & 2.07703 & 0.42593 & 5.1 & 0.51 & 0.11 & 0.41 & 0.1 & $-$ &  $-$  \\
G002.0810+00.8817 & 2.08101 & 0.88168 & 5.1 & 0.42 & 0.09 & 0.37 & 0.09 & $-$ &  $-$  \\
G002.0861+00.9456 & 2.08613 & 0.94565 & 5.8 & 0.53 & 0.09 & 0.52 & 0.09 & $-$ &  $-$  \\
G002.0887-00.3291 & 2.08869 & -0.32914 & 6.9 & 0.66 & 0.1 & 0.68 & 0.1 & $-$ & G002.088-00.329  \\
\\
\hline
\hline
\end{tabular}
\tablefoot{ 
The description of each column is presented in Sect.\,\ref{sect:catalog_description}.
 Only a small portion of the data is provided here. 
 The full table of 7917 sources with $5\sigma\leq S/N<7\sigma$ in the B-configuration of this work will be available in electronic form at the CDS via anonymous ftp to cdsarc.cds.unistra.fr (130.79.128.5)
or via https://cdsarc.cds.unistra.fr/cgi-bin/qcat?J/A+A/. }
\label{tab:total_5to7sigma}
\end{table*}

 
\section{The MIR images of \hii\ region candidates}
\label{appendix_sect:MIR_HII_glimpse}
\begin{figure*}[!h]
\centering
\includegraphics[width=0.23\textwidth, trim= 0 0 0 0,clip]{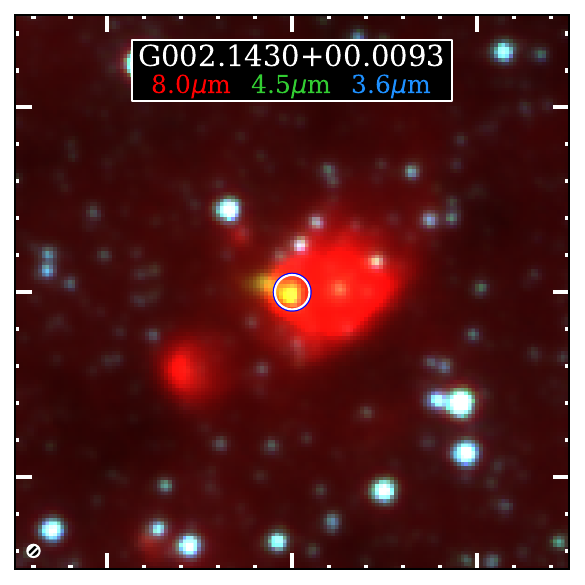}  \includegraphics[width=0.23\textwidth, trim= 0 0 0 0,clip]{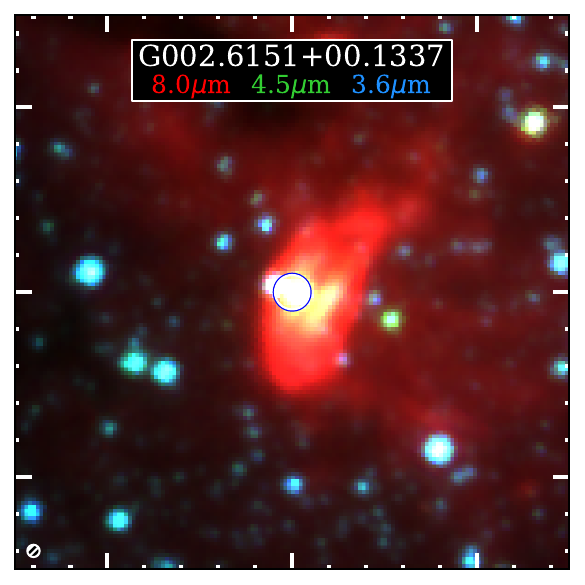}  \includegraphics[width=0.23\textwidth, trim= 0 0 0 0,clip]{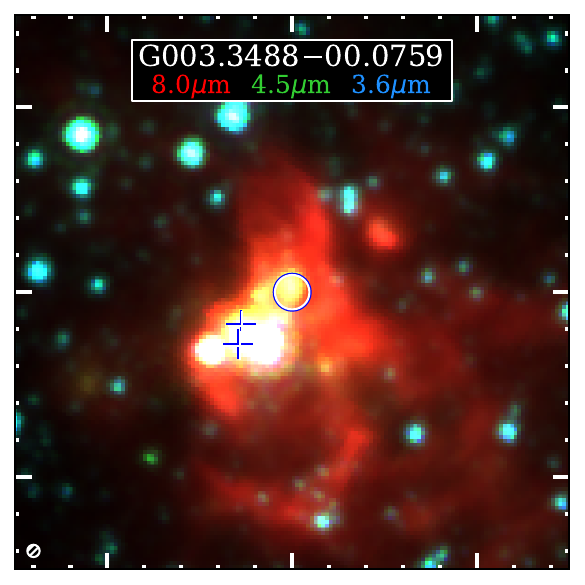}  \includegraphics[width=0.23\textwidth, trim= 0 0 0 0,clip]{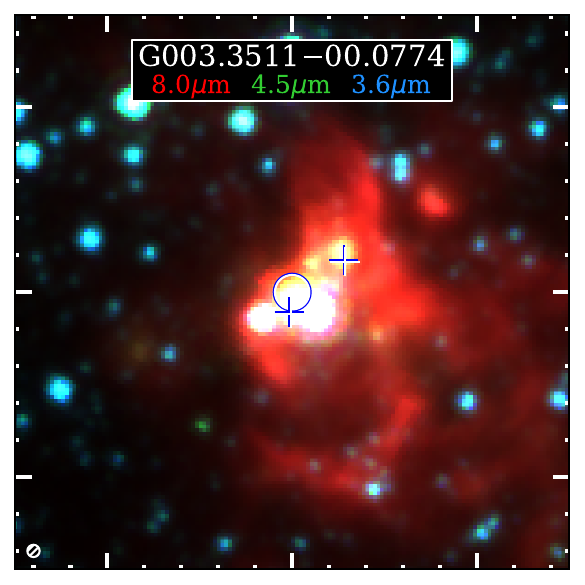} \\
\includegraphics[width=0.23\textwidth, trim= 0 0 0 0,clip]{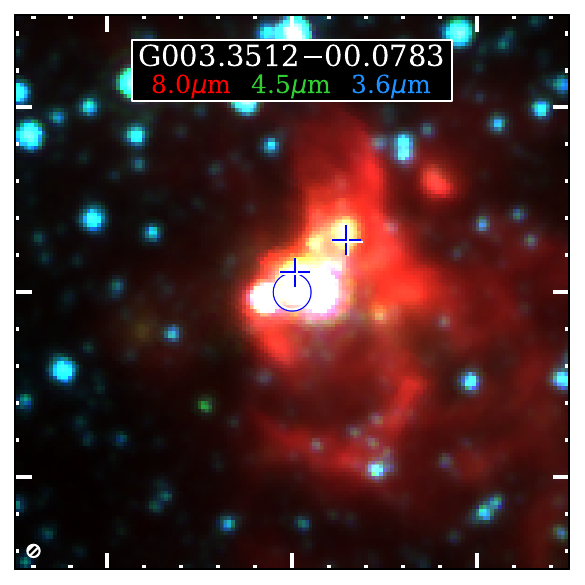}  \includegraphics[width=0.23\textwidth, trim= 0 0 0 0,clip]{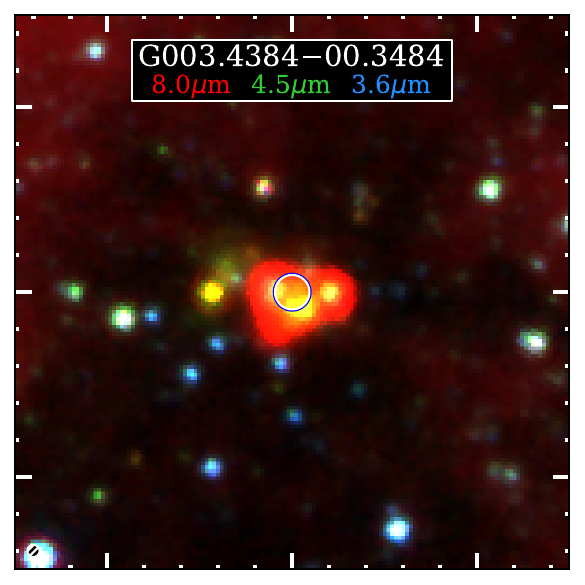}  \includegraphics[width=0.23\textwidth, trim= 0 0 0 0,clip]{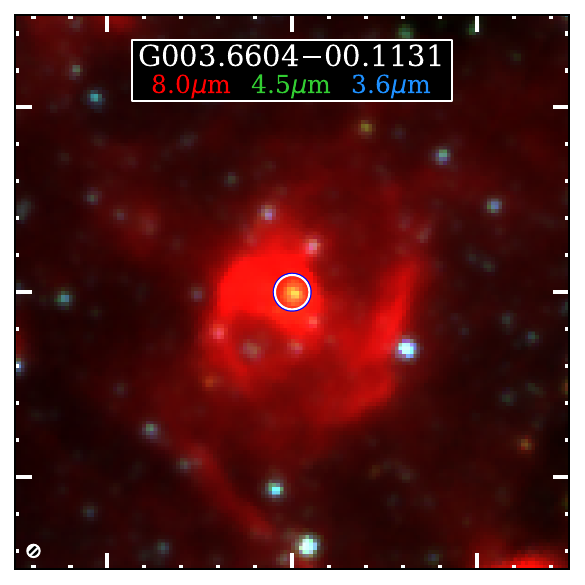}  \includegraphics[width=0.23\textwidth, trim= 0 0 0 0,clip]{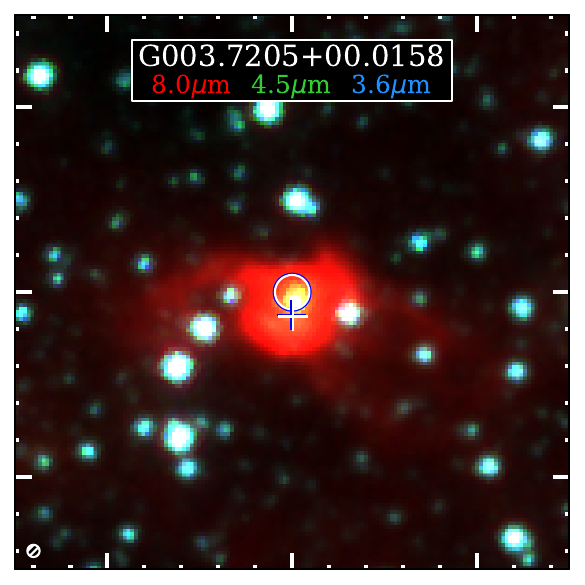} \\
\includegraphics[width=0.23\textwidth, trim= 0 0 0 0,clip]{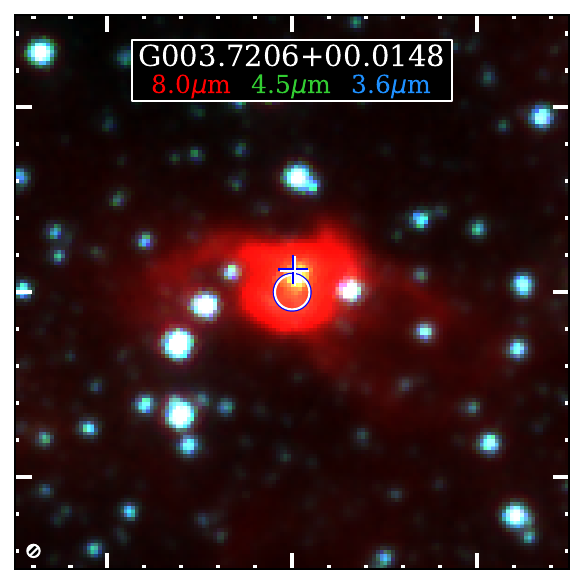}  \includegraphics[width=0.23\textwidth, trim= 0 0 0 0,clip]{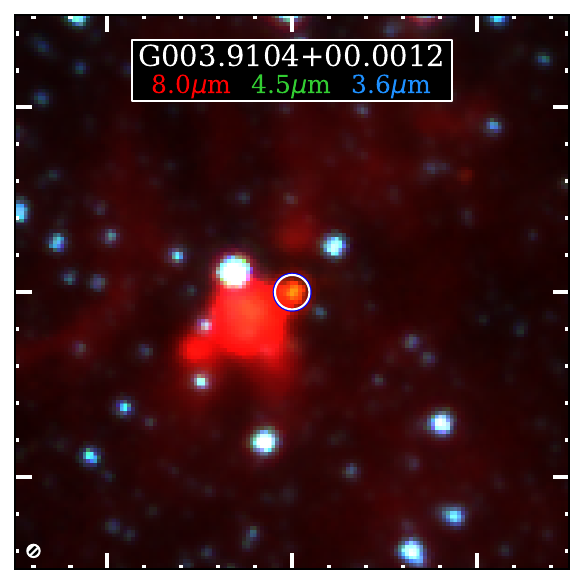}  \includegraphics[width=0.23\textwidth, trim= 0 0 0 0,clip]{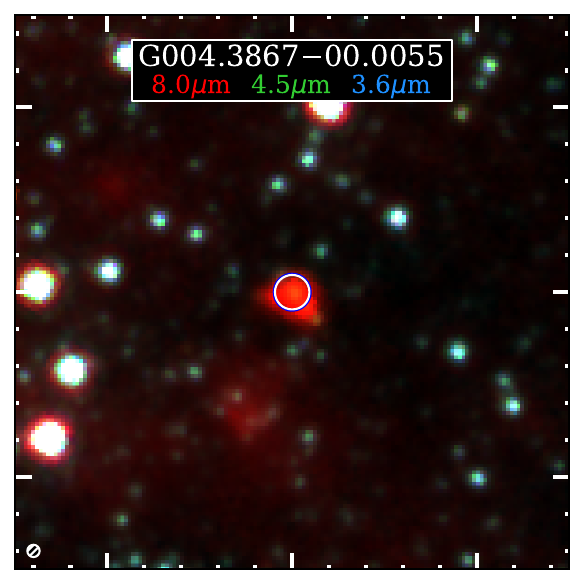}  \includegraphics[width=0.23\textwidth, trim= 0 0 0 0,clip]{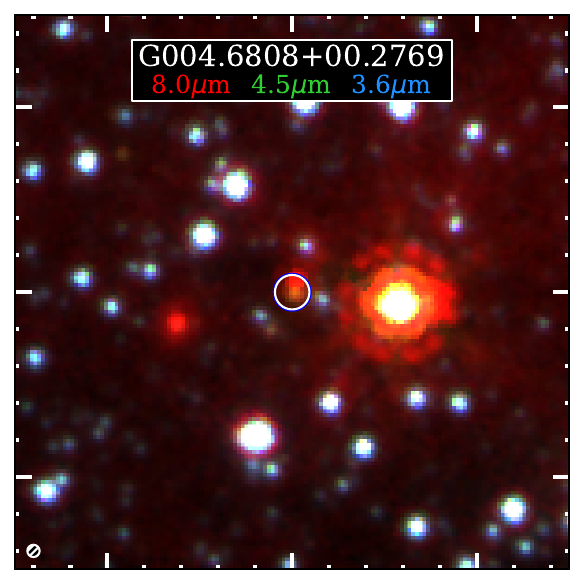} \\
\includegraphics[width=0.23\textwidth, trim= 0 0 0 0,clip]{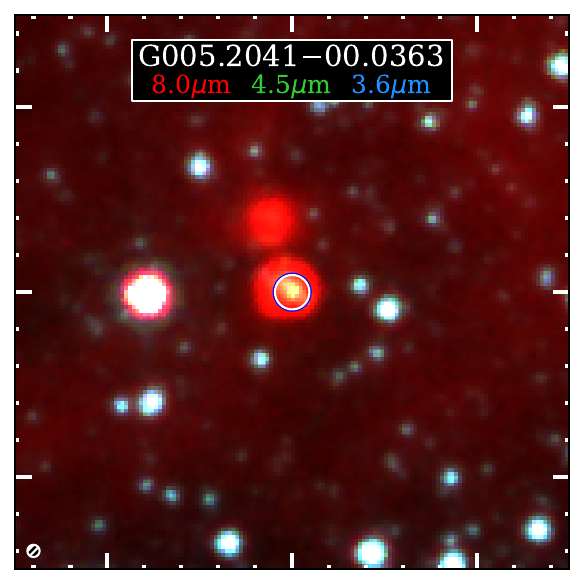}  \includegraphics[width=0.23\textwidth, trim= 0 0 0 0,clip]{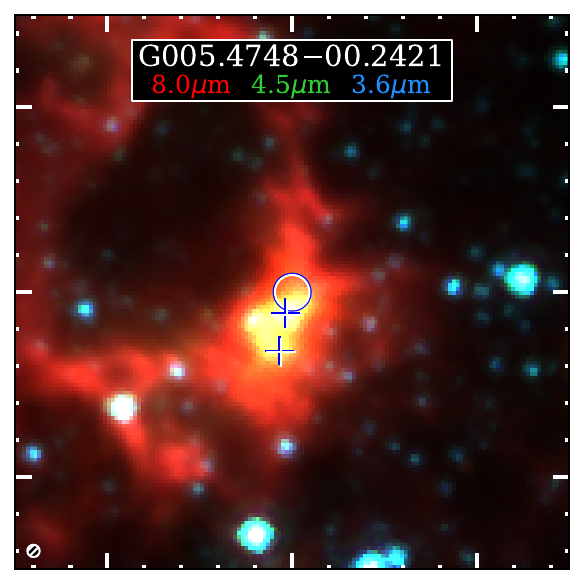}  \includegraphics[width=0.23\textwidth, trim= 0 0 0 0,clip]{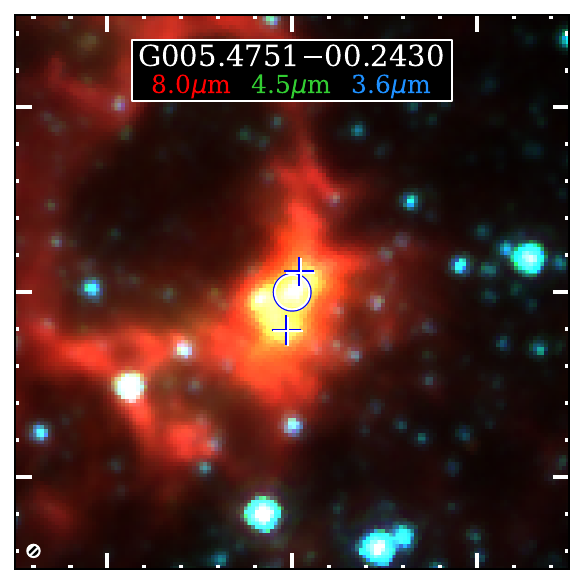}  \includegraphics[width=0.23\textwidth, trim= 0 0 0 0,clip]{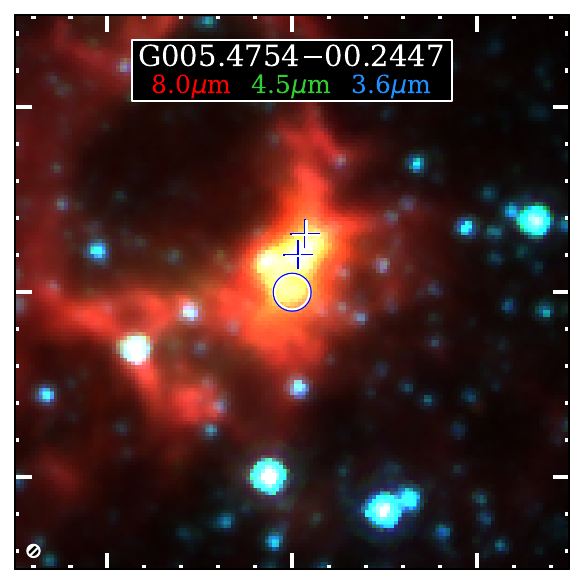} \\
\hspace{-1.0cm}
\includegraphics[width=0.28\textwidth, trim= 0 0 0 0,clip,valign=t]{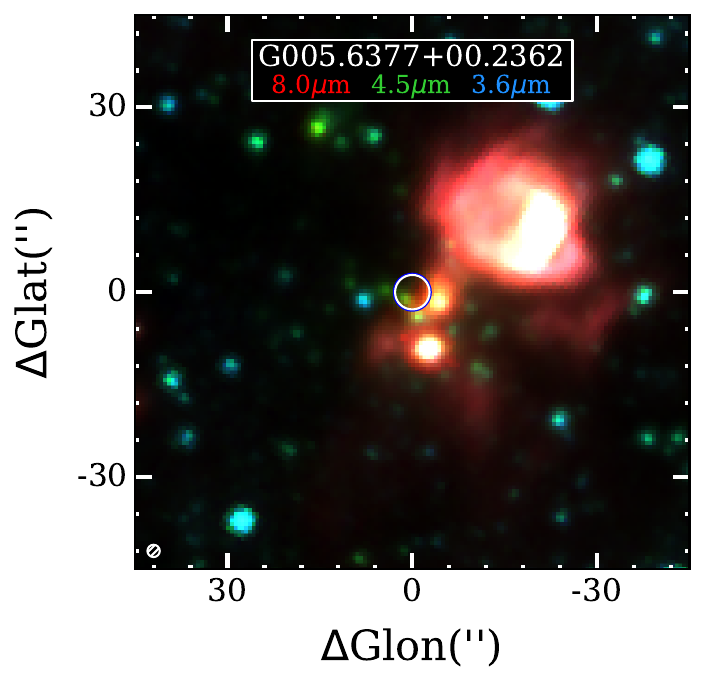}  \includegraphics[width=0.23\textwidth, trim= 0 0 0 0,clip,valign=t]{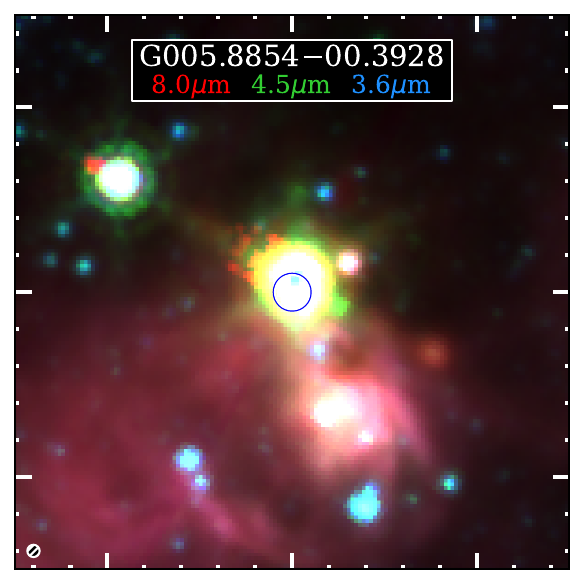}  \includegraphics[width=0.23\textwidth, trim= 0 0 0 0,clip,valign=t]{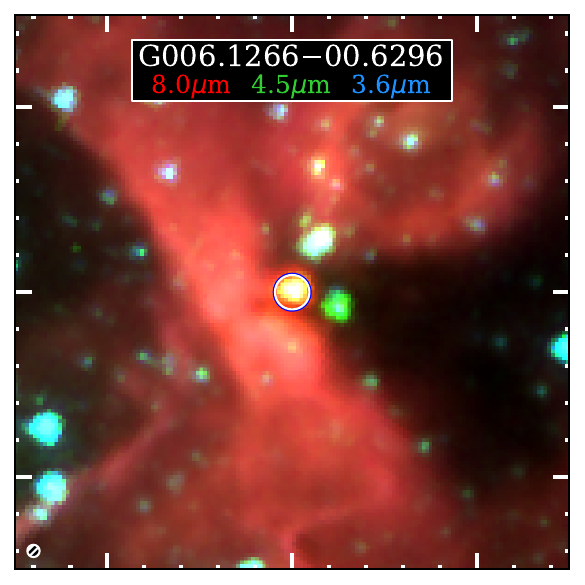}  \includegraphics[width=0.23\textwidth, trim= 0 0 0 0,clip,valign=t]{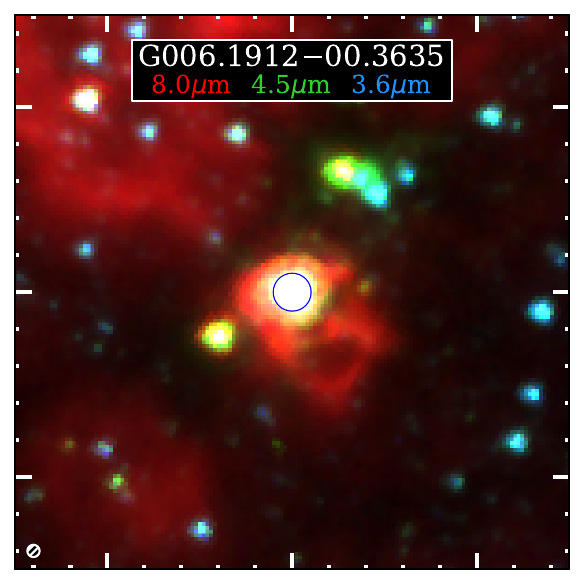} \\
\caption{Three colors images (red=8.0\,$\mu$m, green=4.5\,$\mu$m, and  blue=3.6\,$\mu$m) of the \hii\ region candidates identified in this work.
Each figure is  centered at the position of the identified \hii\ region candidate that is shown in blue-white circles. 
The images are shown in the boxes with a size of $90\arcsec\times90\arcsec$. 
The blue-white pluses refer to the positions of the fragmented \hii\ regions listed in Table~\ref{tab:over_resolved_source} or the nearby compact \hii\ regions. The beam size of 1\farcs0 is shown in the lower-left of each image. Only a small portion of the sample is presented here, and the MIR images for the full sample of 251 \hii\ region candidates are available in electronic form at the
Zenodo via \url{https://zenodo.org/uploads/8054107}. }
\label{fig:HII_MIR_images}
\end{figure*}

In Fig.\,\ref{fig:HII_MIR_images}, we present three colors images showing the MIR emission from the 251 the \hii\ region candidates identified by this work, based on data from the GLIMPSE survey. Only a small fraction of the MIR images is presented here and  the MIR images for the full sample will be available online.
Each figure is a $90\arcsec\times90\arcsec$ centered at the position of the \hii\ region candidate that is shown in blue-white circles, and the three colors are for red, green, and blue for 8.0\,$\mu$m, 4.5\,$\mu$m, and 3.6\,$\mu$m, respectively. 

%
%
\end{appendix}


\end{document}